\documentclass[11pt]{article}
\usepackage{amsmath,amssymb,amsfonts,amsthm,thmtools,bm,bbm,mathrsfs,mathabx,natbib,tocloft,comment,accents,geometry}
\usepackage[doublespacing]{setspace}

\usepackage{cancel,sectsty,color,array,hyperref,graphicx}
\usepackage[justification=centering]{caption}
\usepackage[nottoc,numbib]{tocbibind}
\hypersetup{pdfborder = {0 0 0},colorlinks=true,linkcolor=blue,urlcolor=blue,citecolor=blue}
\usepackage{appendix}
\usepackage[shortlabels]{enumitem}
\setlist[enumerate, 1]{1\textsuperscript{o}}
\usepackage{subfig}
\makeatletter
\def\@endtheorem{\endtrivlist}
\makeatother

\usepackage{pgf,tikz}
\usetikzlibrary{positioning}

\newtheorem{theorem}{Theorem}

\newtheorem{lemma}{Lemma}
\newtheorem{assumption}{Assumption}
\newtheorem*{assumption*}{Assumption}

\newtheorem{remark}{Remark}

\newtheorem{assumptionC}{$L^2$ completeness}

\newtheorem{assumptionIE}{Assumption}

\theoremstyle{definition}
\newtheorem*{example_tmp}{Example}
\newenvironment{example}
{ \begin{example_tmp}     }
	{ 
		\renewcommand{\qedsymbol}{$\square$}
		\qed 
	\end{example_tmp} 
}

\newcommand\inde{\protect\mathpalette{\protect\independenT}{\perp}}
\def\independenT#1#2{\mathrel{\rlap{$#1#2$}\mkern2mu{#1#2}}}

    \DeclareMathOperator*{\argmin}{arg\,min}

    \DeclareMathOperator*{\supp}{supp}

\newcommand{\BPi}{\bm{\Pi}}

\newcommand{\CA}{\mathcal{A}}

\newcommand{\CD}{\mathcal{D}}

\newcommand{\CX}{\mathcal{X}}

\newcommand{\CZ}{\mathcal{Z}}

\renewcommand\thmcontinues[1]{Continued}

\title{Treatment Effect Estimation with Noisy Conditioning Variables}
\author{Kenichi Nagasawa\thanks{The University of Warwick, Department of Economics. Email: kenichi.nagasawa@warwick.ac.uk Address: Department of Economics, University of Warwick, Coventry, CV4 7AL, United Kingdom. This paper is based on a portion of my job market paper, which was circulated under the title ``Identification and Estimation of Group-Level Partial Effects.'' I am grateful to my advisor Matias Cattaneo for advice and encouragement. I would like to thank Francesca Molinari and anonymous referees for valuable suggestions and thorough feedback that helped to substantially improve the paper. I also thank St\'{e}phane Bonhomme, Sebastian Calonico, Max Farrell, Yingjie Feng, Joachim Freyberger, Andreas Hagemann, Michael Jansson, Lutz Kilian, Xinwei Ma, Eric Renault, Roc\'{i}o Titiunik, Gonzalo Vazquez-Bare, and seminar/conference participants at various institutions for helpful comments.
}}
\date{\today}

\begin{document}

\begin{titlepage}
	\maketitle
	\thispagestyle{empty}
	\begin{abstract}
		I develop a new identification strategy for treatment effects when noisy measurements of unobserved confounding factors are available. I use proxy variables to construct a random variable conditional on which treatment variables become exogenous. The key idea is that, under appropriate conditions, there exists a one-to-one mapping between the distribution of unobserved confounding factors and the distribution of proxies. To ensure sufficient variation in the constructed control variable, I use an additional variable, termed excluded variable, which satisfies certain exclusion restrictions and relevance conditions. I establish asymptotic distributional results for semiparametric and flexible parametric estimators of causal parameters. I illustrate empirical relevance and usefulness of my results by estimating causal effects of attending selective college on earnings. 
	\end{abstract}
	\textit{Keywords:} average structural function, $L^2$-completeness, control functions, non-classical measurement errors.
\end{titlepage}

\section{Introduction}
In observational studies, controlling for confounding factors is crucial to identify causal effects of interest. The main challenge is that measuring these confounding elements are often difficult, if not impossible. A widely used approach to address this issue is to use proxy variables in place of the unobserved characteristics. For instance, to account for differences in unobserved worker's ability, researchers routinely include test scores in their regression controls. However, these measurements are generally contaminated with noise. In simple linear regression settings, it is well-known that using proxies as regressors induces attenuation bias. In nonparametric settings, \cite{Battistin-Chesher_2014_JoE} showed that average treatment effects are not identified under covariate measurement errors and that the direction of the bias cannot be determined a priori. Therefore, simply controlling for proxy variables does not lead to reliable inference on causal effects. In this paper, I provide a novel identification strategy for causal effects when important control variables are imperfectly measured. This new approach has a wide range of applications since coarse measurements are prevalent in practice \cite[e.g.,][]{Gillen-Snowberg-Yariv_2019_JPE}. 

In its simplest form, the identification problem of interest is captured in the model
\begin{equation*}
	Y = \beta_0 + \beta_1 D + \beta_2 X + \beta_2(X^*-X) + \epsilon, \qquad \mathbb{E}[\epsilon|D,X^*,X] =0
\end{equation*}
where $D$ is the treatment of interest and $X$ is a proxy variable for the unobserved confounding factor $X^*$. If $X^*$ were observed, one would estimate the equation $Y=\beta_0 + \beta_1 D+\beta_2X^*+\epsilon$. In practice, however, one regresses $Y$ on $D$ and $X$, using the error-ridden variable in place of $X^*$. The regression estimate using $X$ suffers from measurement error bias, and the treatment effect $\beta_1$ is not consistently estimated. A textbook solution to this problem is to use a repeated measurement as an instrumental variable (IV) for $X$. Denoting the second measurement of $X^*$ by $Z$, the two-stage least squares (2SLS) method is equivalent to estimating
\begin{equation}\label{eq:tsls_representation}
	Y = \beta_0 +\beta_1 D+ \beta_2 \mathbb{E}[X|D,Z] + \varepsilon,\qquad\mathbb{E}[\varepsilon|D,Z] =0
\end{equation}
where $\mathbb{E}[X|D,Z]$ is used as it is the fitted value in the first-stage equation. Here $\mathbb{E}[X|D,Z]$ plays the role of a control function \citep{HeckmanRobb1985}; see \citet[p.5356]{MATZKIN2007} for a definition of control functions.\footnote{Here I take the perspective of control function rather than IV because the control function approach extends to non-linear settings naturally whereas IV counterparts face some difficulties \citep{Blundell-Powell_2003}. For instance, nonparametric IV methods do not identify causal effects unless the unobserved heterogeneity is additively separable.}

The most novel contribution of this paper is to extend the above control function method to non-linear settings, including general nonparametric models. 
Specifically, I show that under appropriate assumptions, the random vector $V=(\mathbb{E}[X|D,Z],\dots,\mathbb{E}[X^k|D,Z])'$ for some $k\in\mathbb{N}$ is a valid control function in the sense that the treatment variable becomes conditionally independent of unobserved confounding factors given $V$. This is a natural extension of \eqref{eq:tsls_representation} since the 2SLS method controls for only the first conditional moment of proxy variables linearly whereas my identification approach also controls for the higher conditional moments in a nonparametric manner.
Accommodating non-linear models is especially important for treatment effect estimation because the linear model \eqref{eq:tsls_representation} imposes constant treatment effects, and as well-documented in the literature, ignoring treatment effect heterogeneity may produce misleading estimates.

The control function approach may be particularly appealing to applied researchers for its simplicity: the estimation procedure boils down to running regressions and computing sample averages. 
To facilitate implementation of the new control function method, I discuss semiparametric/flexible parametric modelling of the outcome equation, leveraging existing results in the control function literature. In addition, I develop a Lasso-based estimation procedure to flexibly choose regression specifications, building on \cite{Chernozhukov-Newey-Singh_2022_ecma}. With cross-validation, it is easy to choose Lasso tuning parameters, and I provide a closed-form variance estimator. I characterize a set of sufficient conditions for $\sqrt{n}$-consistency and asymptotic normality of the proposed estimator as well as consistency of the variance estimator.

To demonstrate the empirical relevance of the new results, I apply the control function method to estimating causal effects of attending selective college on post-college earnings, building on the empirical framework of \cite{Dale-Krueger_2002}. Their identification strategy controls for the set of colleges to which students applied and got admitted, and it is considered one of most credible identification strategies in non-experimental contexts \cite[e.g.,][for a recent application of this approach]{Chettyetal2020}. Yet, the method requires detailed information about college application processes, and it is not applicable in many settings due to data limitations. 
In my empirical application, I follow the main idea of Dale and Krueger but use the control function approach to overcome the data issue. Namely, I observe only partial information about college applications and use the noisy measures to construct control functions. My empirical results indicate that ignoring measurement errors in control variables lead to non-trivial differences in causal effect estimates.

\paragraph{Relation to existing literature}
The main alternatives to my control function approach are the integral equation approach of \cite{Deaner2021,Miao-Geng-TchetgenTchetgen} and the operator diagonalization method of \cite{HuSchennach2008}.

My identifying assumptions greatly overlap with those of the integral equation approach, yet there is one important difference: the integral equation approach depends on a difficult-to-verify high-level condition, which I do not impose. My method may be more appealing because of the difficulty in verifying this high-level condition in general. As discussed in the appendix, with a similar high-level condition, my method yields the same identification result as the integral equation approach, and thus, this high-level condition is the key difference between the two approaches.
Also, I exhibit a simple example where the key high-level condition of the integral equation approach fails whereas my identifying assumptions hold.

\cite{HuSchennach2008} pioneered the use of completeness conditions in nonparametric measurement error models and their methods have been successfully applied beyond measurement error contexts \citep[e.g.,][]{Arellano-Blundell-Bobhomme_2017,Bohhomme-Lamadon-Manresa_2018,Sasaki2015}. Building on their idea, I also use a completeness condition to formalize the notion of a proxy variable. \cite{HuSchennach2008} established powerful nonparametric identification results, and although the identifying assumptions are non-nested, their approach identifies a larger class of parameters than my approach e.g.,\ the distribution of unobserved confounding factors. 
On the other hand, my method has practical advantages because it can relax some of the identifying assumptions by exploiting additional structures on the outcome equation. 
Such relaxation of identifying assumptions does not seem to hold for the operator diagonalization technique (i.e.,\ identifying assumptions for a specific model are not weaker than for a general model). In addition, my control function method is easy to implement as estimation boils down to running regressions and computing sample averages.

This paper also contributes to the extensive literature on control functions \citep[for reviews, see][]{Blundell-Powell_2003,MATZKIN2007,Wooldridge2015JHR}. 
Many existing results construct a control variable from IV, whereas I use proxy variables.
Since the identifying assumptions are quite different, the new result in this paper complements existing results by expanding the scope of applicability of control function methods.
Many existing approaches use the invertibility of the first-stage equation to construct a control function, and this feature excludes the case of multi-dimensional unobserved heterogeneity with a scalar treatment variable.
Provided that appropriate proxies are available, my method allows for multiple unobserved heterogeneity and does not require the invertibility of the first-stage equation. This aspect is practically relevant as imposing the scalar restriction on unobserved confounding factors is not appealing in some settings. 
A common drawback of the control function method is that it has difficulty handling discrete endogenous variables. Although my method shares this feature to an extent, I discuss how additional assumptions help achieve the identification of causal parameters with a discrete endogenous variable.

As discussed below, my identification strategy is closely related to the method of \cite{AltonjiMatzkin2005}. In settings with group structure, \cite{ArkhangelskyImbens2019} provided a framework where the key condition of Altonji and Matzkin (the exchangeability condition) follows from model primitives. Similarly, I provide a framework where the exchangeability-type condition holds, but my model differs in not having an explicit group structure.

I motivated my identification strategy as a generalization of 2SLS estimation using repeated measurements. While existing literature on non-linear models with measurement errors is extensive \citep[for a review, see][and references therein]{SCHENNACH2020}, my approach is unique in the use of what I call excluded variable, which includes repeated measurements and IVs but also other types of variables. In addition, my approach is distinct from deconvolution methods.

\paragraph{Roadmap}
In the next section, I describe the econometric model and discuss the identification results. Semiparametric and flexible parametric estimation methods are developed in Section \ref{sec: estimation}, and I apply the results of this paper to estimating causal effects of attending selective college on post-college earnings in Section \ref{sec: empirical application}. Section \ref{sec:conclusion} concludes.

\section{Econometric model and identification results}
I employ the potential outcome notation. $\{Y(d):d\in\mathcal{D}\}$ denotes the set of potential outcomes, $\mathcal{D}$ is the set of possible treatment levels, $D$ is the realized treatment level, and $X^*$ represents unobserved confounding factors. The key identification challenge I focus on is endogeneity due to statistical dependence between $D$ and $X^*$. If $D$ and $X^*$ are independent (e.g.,\ in a randomized control trial), simple comparison of means for different treatment levels suffices for identification of treatment effects. For nonparametric outcome models, I focus on continuous treatment, but discrete cases are also discussed. $X$ denotes a proxy variable for $X^*$, and $Z$ is what I call excluded variable. A leading example of $Z$ is a repeated measurement as in the linear 2SLS case of \eqref{eq:tsls_representation}. Yet, there are other types of variables that can be used in the place of $Z$. Since $Z$ does not need to be a repeated measurement, I refer to $Z$ as excluded variable to emphasize this aspect. I will discuss examples of excluded variables in Section \ref{sec: excluded variable example}.

I note that the dimension of proxy variable $X$ should be at least as large as that of $X^*$ in general (precise assumptions below). An important consequence of this feature is that to allow for rich unobserved heterogeneity (e.g.,\ multi-dimensional, continuous $X^*$), the proxy variable has to exhibit sufficient variation (multi-dimensional, continuous $X$). Conversely, if one is willing to restrict variation in $X^*$, a discrete proxy $X$ suffices for my identification results below. Also, covariates without measurement errors can be introduced. For brevity, I do not consider covariates explicitly. The analysis below goes through by conditioning on the covariates throughout.

To ground discussion on concrete terms, I use the empirical setting of \cite{Dale-Krueger_2002} (DK henceforth) as a running example.
\begin{example}[\cite{Dale-Krueger_2002}]
	$D$ denotes college-level mean of SAT score (measure of college selectivity; a higher school-level mean indicates that a bigger fraction of high-achieving students attends the college), $Y(d)$ is the potential earnings at age 32, and $X^*$ represents student's unobserved ability. DK noted that $D$ and $X^*$ were likely to be correlated through college admission processes. In my empirical application, I use math and reading test scores at 12th grade for $X$ and SAT score for $Z$. 
\end{example}
The causal parameter I focus on is the average structural function (ASF) \citep{Blundell-Powell_2003}: 
\begin{equation*}
	\vartheta(d) = \mathbb{E}[Y(d)], \qquad d\in\mathcal{D} 
\end{equation*}
In DK's setting, $\vartheta(d)$ is the earnings function with respect to college selectivity measure $d$. As one varies school-level mean SAT score $d$, the distribution of unobserved ability is held constant, and thus, the change in $\vartheta(d)$ represents the ceteris-paribus effect of college selectivity on future earnings. In different empirical settings with a binary treatment $\mathcal{D}=\{0,1\}$, $\vartheta(1)-\vartheta(0)$ is the average treatment effect. 
As studied in the literature, other distributional features of the potential outcomes may be of interest e.g.,\ the quantile and distributional structural functions. Using my approach, these objects are identifiable under the same assumptions that identify the ASF. As the identification arguments are similar, I focus on the ASF.

\subsection{Nonparametric identification}\label{sec: nonparametric identification}
I denote statistical independence by $\inde$. For a random vector $A$, I write $F_A,f_A$ for the distribution function and density (with respect to some measure) of $A$, respectively. Let $\supp(A)$ be the support of $A$. Following are the identifying assumptions I impose on econometric models.
\begin{assumption}\label{ASSM: conditional independence outcome}
	$Y(d)\inde (D,Z)|X^*$.
\end{assumption}
\begin{assumption}\label{ASSM: proxy exclusion}
	$X\inde (D,Z)|X^*$.
\end{assumption}
\begin{assumption}\label{ASSM: L2 completeness}
	For any $b\in L^2(F_{X^*})$, the joint distribution $F_{XX^*}$ satisfies
	\begin{equation*}
		\mathbb{E}[b(X^*)|X]=0 \quad F_{X}\text{-almost surely} \qquad\Rightarrow \qquad b(X^*)=0 \quad F_{X^*}\text{-almost surely}.
	\end{equation*}
\end{assumption}
\begin{assumption}\label{ASSM: regularity}
	$\mathbb{E}[|Y(d)|^2]<\infty$ for all $d\in\CD$. Let $\lambda,\lambda_1,\lambda_2,\lambda_3$ be products of Lebesgue and/or counting measures. The joint distribution of $(X,X^*,D,Z)$ is absolutely continuous with respect to the product measure $\lambda\times\lambda_1\times\lambda_2\times\lambda_3$. 
	$\mathbb{E}[|\frac{f_{XX^*}(X,X^*)}{f_{X}(X)f_{X^*}(X^*)}|^2]<\infty$ and $ \frac{f_{X^*|DZ}(X^*|D,Z)}{f_{X^*}(X^*)}\leq C <\infty$ with probability one. 
	If $X$ has continuous elements, let $X=(X_1',X_2')'$ where $X_1$ has a continuous distribution, and $f_{X|DZ}(x_1,X_2|D,Z)$ is continuous in $x_1$ with probability one. If $D$ is continuously distributed, $\mathbb{E}[Y(d)|D,Z]$ is continuous in $d$ with probability one.
\end{assumption}
\begin{assumption}\label{ASSM: support condition}
	Let $\CX=\supp(X)$ and $\CZ_d=\supp(Z|D=d)$.
	For each $\tilde{d}\in \CD$, there exists a function $\phi_{\tilde{d}}$ such that the event $\{f_{X|DZ}(x |D, Z) = f_{X|DZ}(x|\tilde{d}, \phi_{\tilde{d}}(D,Z))$ for all $x\in\CX\}$ has probability one and $\mathbb{P}[\phi_{\tilde{d}}(D,Z)\in \CZ_{\tilde{d}}]=1$.
\end{assumption}
To interpret Assumption \ref{ASSM: conditional independence outcome}, note that it is implied by $Y(d)\inde D|X^*$ and $Y(d)\inde Z|D,X^*$. The first part states that once conditional on $X^*$, the treatment variable becomes exogenous i.e.,\ $X^*$ represents the confounding factors. The other condition is that $Z$ satisfies an exclusion restriction in the sense that $Z$ does not affect the outcome given the treatment and confounding factors. This exclusion restriction on $Z$ is distinct from the standard IV exclusion restriction because the restriction needs to hold only conditional on confounding factors $X^*$. In my setting, excluded variables $Z$ are allowed to be correlated with confounding factors, which is not the case for IVs.

Assumption \ref{ASSM: proxy exclusion} states that given the ``correctly measured'' variable, its noisy measurement is independent of other variables. Intuitively, it requires that measurement errors in $X$ be orthogonal to treatment and excluded variables conditional on confounding factors. This type of restriction is common in the measurement error literature \citep[e.g.,\ Assumption 2 in][]{HuSchennach2008}.
As a visual aid to understand Assumptions \ref{ASSM: conditional independence outcome}-\ref{ASSM: proxy exclusion}, Figure \ref{fig:dag} provides directed acyclical graphs that are compatible with the imposed conditional independence restrictions. 

To assess the plausibility of these assumptions in empirical settings, it is important to specify some elements of the treatment assignment mechanism as well as the outcome determination process. I illustrate this point using DK's empirical setting.
\begin{example}[continued]
	To motivate their empirical strategy, DK considered the following simple model of college admission. Colleges assign each student an ``admission score,'' which is a linear combination of variables such as SAT score and high school GPA i.e.,\ $\text{admission score}=\sum_{j=1}^J\gamma_j \tilde{X}_j$ where $\tilde{X}_j$'s are student characteristics admission offices observe (but econometricians may not observe $\tilde{X}_j$'s). A college admits students if and only if the score is above the college's threshold. Since the information used to compute the admission score reflects student's academic ability, we can view it as $\tilde{X}_j = X^* + \tilde{\epsilon}_j$ where $X^*$ denotes student's academic ability and $\tilde{\epsilon}_j$ is a noise in the measurement. Then, DK's model of college admission can be written as
	\begin{equation*}
		\text{College } l \text{ admits student } i \text{ if and only if } \gamma X^*_i + \epsilon_{il} > C_l
	\end{equation*}
	where $C_l$ is the college $l$'s threshold score, $\gamma=\sum_{j=1}^J\gamma_j$, and the idiosyncratic term $\epsilon_{il}$ is a linear combination of $\tilde{\epsilon}_j$'s and other idiosyncratic shocks that may vary across colleges. 
	In addition, DK considered the simple model of earnings determination
	\begin{equation}\label{eq: DK simple outcome eq}
		Y = D\theta_0 +  X^*\gamma_0 +\varepsilon,\qquad \mathbb{E}[\varepsilon|D,X^*]=0
	\end{equation}
	where $\theta_0,\gamma_0$ denote parameters. The unobserved ability $X^*$ is present in both the admission equation and the outcome equation, which is the source of endogeneity in this model. The identification result below allows for more general models than the linear model \eqref{eq: DK simple outcome eq}.
\end{example}
In this model of college admission, Assumption \ref{ASSM: conditional independence outcome} is plausible. 
Holding $X^*$ constant, the remaining variation in the college admission process is due to $\epsilon$'s, which represents a sum of idiosyncratic shocks. Therefore, given $X^*$, $D$ becomes orthogonal to the unobserved term $\varepsilon$ in the outcome equation \eqref{eq: DK simple outcome eq}, impliying $Y(d)\inde D|X^*$. To assess the other condition $Y(d)\inde Z|D,X^*$, note $Z$ corresponds to an SAT score. Since SAT is a measurement of academic ability, once conditioning on the college selectivity and academic ability, SAT scores in high school are unlikely to have predictive power on post-college earnings, satisfying the exclusion restriction.

To evaluate Assumption \ref{ASSM: proxy exclusion}, we think of test scores as a noisy measurement of academic ability. One way to express this is that $X=X^*+U_x,Z=X^*+U_z$ where $U_x,U_z$ denote measurement errors. Then, Assumption \ref{ASSM: proxy exclusion} requires that $U_x$, residual variation in test scores, is independent of $(D,U_z)$ conditional on $X^*$. The requirement that $U_x\inde U_z|X^*$ is plausible as test scores may fluctuate around its mean (ability) for idiosyncratic reasons e.g.,\ whether students were feeling sick on the day of test taking. Regarding $U_x\inde D|X^*$, in my empirical application, academic tests were administered by an agency of the U.S.\ Department of Education solely for the survey purpose, and they were not used for college admission. Thus, once conditional on the ability $X^*$, these test scores are likely to be independent of college admission processes. Note that when evaluating Assumptions \ref{ASSM: conditional independence outcome}-\ref{ASSM: proxy exclusion}, the functional forms in DK's model did not play any role. The above reasoning is equally valid for nonparametric models.

Assumption \ref{ASSM: L2 completeness} formalizes the idea that the proxy variable $X$ has a strong relationship with $X^*$. It is the $L^2$-completeness of the conditional distributions of $X^*$ given $X$. Completeness conditions can be thought of as a generalization of the IV rank condition in linear models to nonparametric settings \citep{NeweyPowell2003ECMA}. Since this is a rank condition, the dimension of $X$ should be at least as large as that of $X^*$ in general. In the literature, there are several known sufficient conditions for completeness \citep[e.g.,][]{Andrews2017JoE,DHaultfoeuille2011ET,Hu-Schennach-Shiu2017}. For instance, if researchers are willing to impose the measurement error structure such as $X=\chi(X^*+\eta)$ where $\chi$ is  invertible and $X^*\inde \eta$, then primitive sufficient conditions for completeness exist (see Lemma 4 in the appendix). In a panel data setting, \cite{Wilhelm2015} discussed justifications for completeness assumptions using past observations as proxies. In my empirical application, test scores are a strong indicator of academic ability, and the rank condition seems reasonable.

A key aspect of Assumptions \ref{ASSM: conditional independence outcome}-\ref{ASSM: L2 completeness} is that $X^*$ includes important unobserved confounding factors and that we observe a proxy variable for each element of $X^*$. In DK's setting, one might argue that colleges look at not only academic ability but also some personality trait (e.g.,\ ``grit'') that is reflected in essays and recommendation letters. If the non-cognitive skill also affects future earnings, then we need to modify the identifying assumptions by including the non-cognitive ability as part of $X^*$. What this means in practice is that researchers need to observe a proxy variable for the non-cognitive skill in addition to a proxy for academic ability. In my empirical application, I use survey questions intended to measure psychological attributes as a noisy measurement of non-cognitive skills. Since I use two-dimensional proxy variables for the two-dimensional ability, Assumptions \ref{ASSM: conditional independence outcome}-\ref{ASSM: L2 completeness} continute to hold even if the non-cognitive skill is important part of the selection mechanism.

Assumption \ref{ASSM: regularity} is a collection of regularity conditions. It imposes that outcome has a finite second moment, some ratios of densities are bounded, and some functions of the data generating process are continuous. These requirements are standard and mild. The formulation handles discrete and continuous treatment variables in a unified framework.

Assumption \ref{ASSM: support condition} is a common support condition for the control function. Theorem \ref{thm:conditional independence} below states that under Assumptions \ref{ASSM: conditional independence outcome}-\ref{ASSM: regularity}, the stochastic process $\mathfrak{V}=\{f_{X|DZ}(x|D,Z):x\in\CX\}$ is a valid control function in that the potential outcome is conditionally independent of the treatment variable given $\mathfrak{V}$. As well-known in the control function literature, a common support condition is crucial to identify the ASF with a general nonparametric outcome equation \citep{Blundell-Powell_2003,ImbensNewey2009}. See Assumption 2 of \cite{ImbensNewey2009} and their discussion. Informally speaking, to identify causal effects, one needs to be able to vary $D$ while holding constant $\mathfrak{V}$. Since $\mathfrak{V}$ is a function of $D$, this is possible only if the excluded variable $Z$ can ``undo'' the effect of $D$ on $\mathfrak{V}$, which is the content of Assumption \ref{ASSM: support condition}. 

Generally, the common support condition requires the large support of $Z$ and some structure on the treatment equation. To illustrate this point, consider the following example:
\begin{equation*}
	D = Z + X^* + \eta,\qquad Z\inde (X^*,\eta).
\end{equation*}
Then, Assumption \ref{ASSM: support condition} holds with $\phi_{\tilde{d}}(d,z)= \tilde{d}-d+z$ if the support of $Z$ is the entire real line. This claim holds because there exists a bijection between $f_{X|DZ}(\cdot|d,z)$ and $f_{X^*|DZ}(\cdot|d,z)$ by Assumptions \ref{ASSM: proxy exclusion}-\ref{ASSM: regularity}. The conditional distribution of $X^*$ is determined by the value $D-Z$, and $Z$ having a large support ensures that the effect of $D$ on the conditional distribution on $X^*$ can be undone by a change in $Z$. This example resembles the type of restrictions imposed by \cite{ImbensNewey2009}. Although my setting shares common elements with the existing control function literature, a key distinction is that my approach accommodates multi-dimensional unobserved heterogeneity in the treatment equation provided that appropriate proxy variables exist. I elaborate on this point in Section \ref{sec: connection with literature}.

The common support condition may be more stringent for a discrete treatment variable. With a binary treatment, Assumption \ref{ASSM: support condition} translates to $0<\mathrm{Pr}[D=1|\mathfrak{V}]<1$ with probability one, the well-known overlap condition. This overlap assumption may fail to hold if the treatment assignment follows a threshold-crossing model $D=\mathbbm{1}\{Z'\delta \geq X^*\}$. An intuitive explanation for this failure is that the conditional distributions of $X^*$ differ considerably across the treated and control groups; the conditional support of $X^*$ for $D=1$ is $(-\infty,Z'\delta]$ while the conditional support for $D=0$ is $(Z'\delta,\infty)$. The difficulty in handling discrete endogenous variables is not specific to my approach but generally shared by control function methods \citep[see e.g.,][]{Wooldridge2015JHR}.  Fortunately, the conditional independence result below holds without the overlap assumption, and one can replace Assumption \ref{ASSM: support condition} with alternative restrictions to achieve the identification of causal effects. One possibility is to rely on identification-at-infinity type arguments. Another approach is to impose additional structures on the outcome equation, which I discuss in Section \ref{sec: semiparametric parametric outcome}.

The following is the main identification result of this paper. A formal proof is presented in the appendix, and the main ideas are sketched in Section \ref{sec:proof_sketch}.
\begin{theorem}\label{thm:conditional independence}
	Suppose Assumptions \ref{ASSM: conditional independence outcome}, \ref{ASSM: proxy exclusion}, \ref{ASSM: L2 completeness}, and \ref{ASSM: regularity} hold. Then,
	\begin{equation*}
		Y(d)\inde D|\mathfrak{V} 
	\end{equation*}
	where the sigma-field generated by $\mathfrak{V}$ is the smallest sigma-field that makes the mapping $(D,Z)\mapsto \int g(x)f_{X|DZ}(x|D,Z)d\lambda(x)$ measurable for any non-negative measurable map $g$.
	In addition, if Assumption \ref{ASSM: support condition} holds, the ASF is identified by
	\begin{equation*}
		\vartheta(d) = \mathbb{E}[\mu(d,\mathfrak{V})]
	\end{equation*}
	where $\mu(D,\mathfrak{V})=\mathbb{E}[Y|D,\mathfrak{V}]$.
\end{theorem}
The statement above focuses on the ASF, but as standard in the literature, other parameters of interest such as the quantile structural function can be identified using the same argument. Note that the conditional independence result $Y(d)\inde D|\mathfrak{V}$ holds without Assumption \ref{ASSM: support condition}, and it is possible to replace the common support condition with restrictions on the outcome equation as discussed in the next section.

\begin{example}[continued]
	Theorem \ref{thm:conditional independence} implies that under Assumptions \ref{ASSM: conditional independence outcome}-\ref{ASSM: support condition}, researchers can specify a quite general model of the earnings, not just an additive linear model like \eqref{eq: DK simple outcome eq}, to identify the causal effects of college selectivity. For instance, in the empirical application below, I specify the random coefficient model
	\begin{equation*}
		Y = \varepsilon_1 + \varepsilon_2 D ,\qquad (\varepsilon_1,\varepsilon_2)\inde D|X^*
	\end{equation*}
	where the unobserved random variables $(\varepsilon_1,\varepsilon_2)$ can be arbitrarily correlated with the unobserved worker's ability $X^*$. This random coefficient specification can be of empirical interest as the model allows for the treatment effect to vary across different levels of unobserved worker's ability. Such heterogeneity is likely to be important in practice, yet a simple model like \eqref{eq: DK simple outcome eq} does not accommodate such feature of data.
\end{example}

In devising an estimation method based on Theorem \ref{thm:conditional independence}, the potentially high dimension of $\mathfrak{V}$ appears to pose a challenge. However, the sigma-field generated by $\mathfrak{V}$ is coarser than the sigma-filed generated by $(D,Z)$. Thus, intuitively, the effective dimension of $\mathfrak{V}$ is at most the dimension of $(D,Z)$ and potentially smaller. To make a heuristic argument, suppose that a single-index restriction holds for the conditional distribution of $X$ i.e.,\ $f_{X|DZ}(x|d,z)= f(x,d'\gamma+z'\delta)$ for some fixed function $f$ and vectors $\gamma,\delta$. If  $f$ is continuously differentiable in the second argument and $\partial f(x_1,u)/\partial u\neq 0$ for some $x_1$, then there is a bijection $T:f_{X|DZ}(x_1|d,z)\rightarrow d'\gamma +z'\delta$. In turn, we have a representation $f_{X|DZ}(x|D,Z) = f(x,T(f_{X|DZ}(x_1|D,Z)))$ for all $x\in \CX$. Thus, conditioning on $f_{X|DZ}(x_1|D,Z)$ is equivalent to conditioning on $\mathfrak{V}$. Generalizing this argument to multi-index cases can be done (locally) via the inverse function theorem.

To formalize the above idea, I impose the following conditions. For simplicity, I assume that all elements of $(D,Z)$ are continuous, but when there are discrete elements, I fix values of discrete elements and consider each probability event separately.
\begin{assumption}\label{ASSM: control function dimension}
	$(D,Z)$ is continuously distributed.
	Mapping $(d,z)\mapsto F_{X|DZ}(x|d,z)$ is continuously differentiable for almost all $x$. For $(d,z)\in\supp(D,Z)$, there exist $x_1,\dots,x_l$ and an open set containing $(d,z)$ such that the derivative of $(d,z)\mapsto (F_{X|DZ}(x_1|d,z),\dots,F_{X|DZ}(x_l|d,z))'$ has a constant rank on the open set and $(x_1,\dots,x_l)$ is maximal in the sense that for any $x_{l+1}$, the rank of derivative of $(d,z)\mapsto(F_{X|DZ}(x_1|d,z),\dots,F_{X|DZ}(x_{l+1}|d,z))'$ does not increase.
\end{assumption}
For the last part of the assumption, the maximal set of points is finite as the derivative is a $l\times (\mathrm{dim}(D)+\mathrm{dim}(Z))$ matrix and its rank cannot be larger than $\mathrm{dim}(D)+\mathrm{dim}(Z)$. As the single-index example above indicates, the maximal set may be a singleton, which formalizes the idea that $\mathfrak{V}$ has a dimension smaller than $(D,Z)$. The following theorem shows that the sigma-field generated by $\mathfrak{V}$ is equivalent to the sigma-field of a finite-dimensional random vector.
\begin{theorem}\label{thm: finite-dimensional V}
	Suppose Assumption \ref{ASSM: control function dimension} holds and that $\supp(D,Z)$ is compact. Then, there exist points $\{x_{1},\dots,x_{k}\}$ such that with $V = (F_{X|D,Z}(x_{1}|D,Z),\dots,F_{X|D,Z}(x_{k}|D,Z) )'$, the sigma-field generated by $V$ equals the one generated by $\mathfrak{V}$. In particular,
	\begin{equation*}
		\mathbb{E}[Y(d)|\mathfrak{V}]=  \mathbb{E}[Y(d)|V] \quad \text{with probability one}.
	\end{equation*}
\end{theorem}
This theorem operationalizes the control function result of Theorem \ref{thm:conditional independence} by providing a method to compute conditional expectations given $\mathfrak{V}$. Elements of $V$ have the form $\int g(x) f_{X|DZ}(x|D,Z)d\lambda(x)$ where $g$ is some $X$-measurable function. In Theorem \ref{thm: finite-dimensional V}, I take $g(\cdot) = \mathbbm{1}\{\cdot\leq x\}$ for some $x$. This specific choice is not essential, and there are many other possibilities. For example, one can use monomial transformation i.e.,\ $g(x) = x^j$ and have $V = (\mathbb{E}[X|D,Z],\dots, \mathbb{E}[X^k|D,Z])'$. In the introduction, I used this choice to preview the identification result.
In theory, any choice of $g$'s is valid provided that continuous differentiability and the maximal rank condition as in Assumption \ref{ASSM: control function dimension} hold. In the sequel, let
\begin{equation*}
	B(x)=(g_1(x),\dots,g_k(x))'
\end{equation*}
be a choice of transformations of $x$ and set $V=\mathbb{E}[B(X)|D,Z]$. In the empirical application, I use $g_{\ell}(x)=x^{\ell}$. Ex ante, the dimension $k$ is not known to researchers, and a practical approach is to include a sufficient number of elements in the control functions and use a model selection method to choose relevant regressors. I develop this method formally in Section \ref{sec: semiparametric estimation} where I use Lasso to select relevant control functions.

Another consequence of Theorem \ref{thm: finite-dimensional V} is that the estimation problem is well-posed, as opposed to ill-posed, under regularity conditions. 
Well-posedness refers to the situation where estimand of interest can be represented as a unique solution to some estimation equation and the solution varies continuously as some feature of the underlying data generating process changes \citep[][]{CarrascoFlorensRenault2007}. 
If an estimation problem is not well-posed, it is ill-posed.
A well-known example of ill-posedness is nonparametric IV estimation: the parameter of interest (a structural function) can be recovered by computing an inverse of an integral operator, but this inverse is known to be discontinuous, leading to ill-posedeness.
Ill-posedness of estimation problems may lead to slower convergence rates, and obtaining reliable estimates might be difficult unless the sample size is large.
In my identification argument, I use a completeness condition, which guarantees a unique solution of an integral operator, but I do not need to compute the inverse of such integral operator. Instead, I construct a control function $V$. As Theorem \ref{thm: finite-dimensional V} shows that this control function is finite-dimensional, it fits into the framework of \cite{HahnRidder2013} and they provided sufficient conditions under which causal parameters are continuous (even differentiable) in the control function. 

The assumption of compact support of $(D,Z)$ may be at odds with Assumption \ref{ASSM: support condition} as the common support condition often requires a large support of $Z$. In the empirical application below, I replace Assumption \ref{ASSM: support condition} with restrictions on the outcome equation to maintain the compact support assumption. Yet, it is possible to relax the compact support condition while retaining the conclusion of Theorem \ref{thm: finite-dimensional V}.
An alternative assumption to the compact support of $(D,Z)$ is that the conditional proxy distribution satisfies index restrictions
\begin{equation*}
	F_{X|DZ}(x|D,Z) = F(x,\iota_1(D,Z),\dots, \iota_k(D,Z)) 
\end{equation*}
for some fixed (but unknown) functions $F,\iota_1,\dots,\iota_k$ where $\iota_l$ is real-valued for $l=1,\dots,k$, $\iota_l(D,Z)$ is continuously distributed and has a compact support for each $l=1,\dots, k$, and $F$ is continuously differentiable with respect to values of $(\iota_1,\dots,\iota_k)$. Here the dimension $k$ is unknown. Then, with the rank condition as in Assumption \ref{ASSM: control function dimension}, a finite-dimensional representation of the control function holds. Note that compact support of $\iota_l(D,Z)$ does not contradict with Assumption \ref{ASSM: support condition}.

In the next section, I discuss how  additional structures on the outcome equation can replace the common support condition. As demonstrated below, this approach maintains quite flexible specifications of econometric models while relaxing some of the identifying assumptions.

\subsection{Semiparametric/parametric outcome equation}\label{sec: semiparametric parametric outcome}
The existing work on control function methods has recognized that the common support condition may fail in some empirical applications \citep[e.g.,][]{Chernozhukov-etal2019,Florens-Heckman-Meghir-Vytlacil_2008}. Building on the available results, I consider additional structures on the outcome equation, which can substantially relax the support condition.

\paragraph{Random coefficient model}
I consider a version of the random coefficient model
\begin{equation}\label{eq: random coefficient model}
	Y =\varepsilon_1 +\varepsilon_2 D,\qquad (\varepsilon_1,\varepsilon_2)\inde D|X^*
\end{equation}
with $D\in\mathbb{R}$.
As discussed above, this model allows for heterogeneous treatment effects.
The causal parameter of interest I consider is the average of the random slope on the treatment variable:
\begin{equation*}
	\theta_0 = \mathbb{E}[\varepsilon_2].
\end{equation*}
This parameter represents the average marginal effect of the treatment $D$. Under Assumptions \ref{ASSM: conditional independence outcome}-\ref{ASSM: regularity} and \ref{ASSM: control function dimension}, the model \eqref{eq: random coefficient model} yields
\begin{equation*}
	\mathbb{E}[Y|D,V] = \mu_1(V) + \mu_2(V)D
\end{equation*}
where $\mu_1,\mu_2$ are unknown functions. \cite{NeweyStouli2020} showed that if the conditional variance of $D$ given $V=v$ is positive for almost all $v$, then $\theta_0$ is identified under regularity conditions. This new identifying condition is substantially weaker than the common support condition: while the common support condition demands that $\supp(D|V)$ equals $\supp(D)$ almost surely, the new condition only requires $\supp(D|V)$ has more than one point almost surely. Therefore, with the random coefficient specification \eqref{eq: random coefficient model}, an excluded variable $Z$ may have discrete variation and one can still achieve the identification of the causal parameter $\theta_0$.\footnote{Let $\nu(D,Z)=\mathbb{E}[B(X)|D,Z]=V$. For $v\in\supp(V)$, if there exist two distinct points $(d_1,z_1),(d_2,z_2)\in\supp(D,Z)$ such that $v=\nu(d_1,z_1)=\nu(d_2,z_2)$, then the conditional variance of $D$ given $V=v$ is positive.}

With a binary treatment $D$, the model \eqref{eq: random coefficient model} places few restrictions on the outcome determination process, and one needs to impose additional structures to relax the overlap condition. One possibility is to impose that $\mu_1(v),\mu_2(v)$ are real analytic and the conditional support of $V$ given $D$ contains an open set. This identification argument was used by \cite{Arellano-Bonhomme_2017}.  A nice feature of this approach is that the assumptions encompass the cases where $\mu_1(v),\mu_2(v)$ are polynomial functions of $v$, a widely used specification in empirical studies. 

\paragraph{Binary outcome}
The random coefficient model above is natural for a continuous outcome variable but may not be appropriate for limited dependent variables. For binary outcomes, \cite{Blundell-Powell_2004} considered flexible yet tractable outcome models. Suppose that $D\in\mathbb{R}^{\dim(D)}$, $\dim(D)\geq 2$, has at least one element that is continuously distributed and that
\begin{equation*}
	Y =\mathbbm{1}\{D'\gamma_0    \geq \varepsilon\}, \qquad \varepsilon\inde D|X^*
\end{equation*} 
where $\gamma_0$ is a vector of coefficient parameters. Here, one can think of $D$ containing additional covaraites as well as treatment variables of interest. Under Assumptions \ref{ASSM: conditional independence outcome}-\ref{ASSM: regularity} and \ref{ASSM: control function dimension}, the above model implies
\begin{equation*}
	\mathbb{E}[Y|D,V] = F_{\varepsilon|V}(D'\gamma_0 |V)
\end{equation*}
where $V=\mathbb{E}[B(X)|D,Z]$ and $F_{\varepsilon|V}(\cdot|v)$ is the conditional CDF of $\varepsilon$ given $V=v$. Existing studies have established sufficient conditions for the identification of coefficient parameters and the ASF. For instance, Proposition 4.1 of \cite{Jochmans2013} states that if the first element of $\gamma_0$ is non-zero and the first element of $D$ has everywhere positive Lebesgue density conditional on $(D_2,V)$, where $D_2\in\mathbb{R}^{\dim(D)-1}$ is the remaining elements of $D$, and the conditional support of $D$ given $V$ is not contained in a proper linear subspace of $\mathbb{R}^{\dim(D)}$, then the coefficient vector $\gamma_0$ is identified up to scale. With $\gamma_0$ identified, the ASF can be computed by averaging $\mathbb{E}[Y|D,V] = F_{\varepsilon|V}(u|V)$ over $V$ fixing $u=D'\gamma_0$. Here, the index structure mitigates the possible curese of dimensionality.

I should note that with a binary endogenous variable following a threshold crossing model, the full-rank condition on $D$ given $V$ may fail and additional structures on $F_{\varepsilon|V}(e|v)$ may be necessary to achieve the identification of the coefficient vector. This aspect is related to the results of \cite{Blundell-Powell_2004}, where they assumed that endogenous variables have a continuous distribution.

\paragraph{Flexible parametric model}
Let $q(d,v)$ be a vector of some transformations of $(d,v)$ e.g.,\ $q(d,v) = (1,d,v,d*v)'$. If desired, higher-order polynomial transformations may be included in $q(d,v)$. I postulate that the outcome equation satisfy
\begin{equation}\label{eq:flexible_parametric_specification}
	\mathbb{E}[Y|D,V] = \Lambda\big( q(D,V)'\gamma_0 \big)
\end{equation}
where $\Lambda$ is a known, strictly monotonic link function and $\gamma_0$ is the parameter to be estimated. This model encompasses a parametric version of the models considered above. For the random coefficient model, suppose
\begin{equation*}
	Y =  \varepsilon_1 + \varepsilon_2 D,\qquad \mathbb{E}[\varepsilon_1|D,V] = p(V)'\gamma_1,\quad \mathbb{E}[\varepsilon_2|D,V] = p(V)'\gamma_2
\end{equation*}
where $p(v)$ is a vector of transformations of $v$ and the conditional independence $\varepsilon_l\inde D|V$, $l=1,2$ follows under Assumptions \ref{ASSM: conditional independence outcome}-\ref{ASSM: regularity} and \ref{ASSM: control function dimension}. Relative to the above semiparametric model, the additional restriction here is $\mathbb{E}[\varepsilon_l|V]=p(V)'\gamma_l$ for $l=1,2$. 
This random coefficient model fits into \eqref{eq:flexible_parametric_specification}, where $\Lambda$ equals the identity function, $q(D,V) = (p(V)',Dp(V)')'$, and $\gamma_0=(\gamma_1',\gamma_2')'$.

For the case of a binary outcome, one may consider the outcome equation  
\begin{equation*}
	Y =\mathbbm{1}\{\gamma_1' D \geq \varepsilon \}
\end{equation*}
and using Theorem \ref{thm:conditional independence}, $\mathbb{E}[Y|D,V] = F(\gamma_1'D |V)$ where $F$ is the conditional distribution of $\varepsilon$ given $V$. For tractability, we may assume that the conditional distribution of $\varepsilon$ given $V$ is a location family i.e.,\ $F(\,\cdot\,|V) = \Lambda(\,\cdot\,-p(V)'\gamma_2)$ and specify $\Lambda$ as the normal/logit CDF, which gives rise to the specification \eqref{eq:flexible_parametric_specification} with $q(D,V)=(D',p(V)')'$ and $\gamma_0=(\gamma_1,\gamma_2')'$.

Under \eqref{eq:flexible_parametric_specification}, identification of the ASF holds if the parameter $\gamma_0$ is identified. In turn, the coefficients $\gamma_0$ is identified if the matrix $\mathbb{E}[q(D,V)q(D,V)']$ is non-singular. This full-column rank condition is weaker than the support invariance condition, and \cite{Chernozhukov-etal2019} and \cite{NeweyStouli2020} provided sufficient conditions for the non-singularity of $\mathbb{E}[q(D,V)q(D,V)']$.

To see what types of restrictions imply \eqref{eq:flexible_parametric_specification}, consider the binary outcome model
\begin{equation*}
	Y=\mathbbm{1}\{\gamma_1'D\geq \varepsilon\},\qquad \varepsilon\inde (D,Z)|X^*.
\end{equation*}
Suppose that $F_{\varepsilon|X^*}(e|x^*)=F(e+\delta'x^*)$ for some function $F$ and coefficient vector $\delta\in\mathbb{R}^{\dim(X^*)}$ and  that $f_{X^*|DZ}(x^*|d,z) = f(x^*-\iota(d,z))$ for some fixed functions $f,\iota$. Also, $X=X^*+U_x$ with $U_x\inde (X^*,D,Z)$ and $\mathbb{E}[U]=0$. Then, $V=\mathbb{E}[X|D,Z]=\iota(D,Z)$ and \eqref{eq:flexible_parametric_specification} holds with
\begin{equation*}
	\Lambda(\cdot) = \int F(\cdot + \delta'u ) f(u)du, \qquad q(D,V)= (D',V')'.
\end{equation*}
With $\dim(X^*)=1$, if $F$ and $f$ are standard normal CDF and density respectively, then $\Lambda(y) = F(y/\sqrt{1+\delta^2})$, implying a probit model.
The non-singularity of $\mathbb{E}[q(D,V)q(D,V)']$ holds in this case if for each $d\in\CD$, there exist $z_1,z_2$ in $\CZ_d$ such that $\iota(d,z_1)\neq\iota(d,z_2)$.

\subsection{Examples of excluded variables other than a repeated measurement}\label{sec: excluded variable example}
As discussed above, a repeated measurement is a leading example for excluded variables $Z$. Yet, there exist other types of variables that qualify as excluded variables. For $Z$ to be an excluded variable, it needs to satisfy the exclusion restriction Assumption \ref{ASSM: conditional independence outcome}. In addition, excluded variables should be correlated with $X^*$ and/or $D$ to help satisfy Assumption \ref{ASSM: support condition} in the nonparametric model or appropriate full-rank conditions for semiparametric/parametric models.

One example of excluded variables is an IV for the treatment variable. Suppose $D=h(Z,\eta)$ where $h$ is a non-stochastic function and $\eta$ is some unobserved heterogeneity. By the standard IV exclusion restriction, $(Y(d),X^*,\eta)\inde Z$ is plausible, and we may additionally impose $Y(d)\inde \eta|X^*$. Then, $Y(d)\inde (D,Z)|X^*$ (Assumption \ref{ASSM: conditional independence outcome}) follows. If $h(z,\eta)$ is non-trivial function in $z$, $Z$ is correlated with $D$, which helps to satisfy the support condition.  

In DK's setting, one may use educational attainment of worker's parents as excluded variables. The exclusion restriction (i.e.,\ $Y(d)\inde Z|D,X^*$) is plausible as potential wage levels are unlikely to be affected by parents' education level once you control for ability and own education as well as other observed characteristics. The relevance condition (i.e.,\ $Z$ correlated with $X^*$ and/or $D$) is also plausible because parents' education is likely to influence the probability of college attendance (e.g.,\ having college-graduate parents makes going to college a norm). Note that educational attainment of worker's parents may not satisfy the IV exclusion restriction (i.e.,\ $(Y(d),X^*)\inde Z$) as it is potentially correlated with worker's unobserved ability through human capital formation. This example demonstrates that there exist empirically relevant excluded variables other than repeated measurements and IVs.

\subsection{Connection with existing results}\label{sec: connection with literature}
\paragraph{Integral equation approach}
My identifying assumptions are closely related to those used by \cite{Deaner2021,Miao-Geng-TchetgenTchetgen}. Both my method and the integral equation approach impose Assumptions \ref{ASSM: conditional independence outcome}, \ref{ASSM: proxy exclusion}, and \ref{ASSM: L2 completeness}. The main difference is Assumption \ref{ASSM: support condition}. In the integral equation approach, instead of Assumption \ref{ASSM: support condition}, the identifying assumptions are
\begin{itemize}
	\item For any $b\in L^2(F_{X^*})$ and $d\in\CD$, $\mathbb{P}[\mathbb{E}[b(X^*)|D,Z]=0|D=d]=1$ implies $\mathbb{E}[b(X^*)|D]=0$ almost surely;
	\item Let $\{\pi_{j,d}\in\mathbb{R},v_{j,d}\in L^2(F_{X|D=d}),u_{j,d}\in L^2(F_{Z|D=d})\}_{j\geq 1}$ be a singular system associated with the linear operator $g\mapsto \mathbb{E}[g(X)|Z,D=d]$. Assume
	\begin{equation}\label{eq: high-level condition integral equation}
		\sum_{j=1}^{\infty} \pi_{j,d}^{-2} \mathbb{E}\big[\mathbb{E}[Y|D,Z]u_{j,d}(Z)\big\vert D=d\big]^2 <\infty.
	\end{equation}
\end{itemize}
See the appendix for details.
The first condition turns out to be implied by Assumption \ref{ASSM: support condition}.\footnote{This implication was pointed out by a referee. I thank them for very helpful feedback.} Then, the key difference lies in the second condition, which imposes some high-level restrictions on the outcome equation. 

An important advantage of the control function approach is that it avoids the high-level assumption \eqref{eq: high-level condition integral equation}, which may be difficult to verify in practice. For example, consider the following simple model:
\begin{equation*}
	Y = \mathbbm{1}\{\beta D  \geq X^* \},\quad X = X^*+U_x, \quad	Z = X^*+U_z ,
\end{equation*}
\begin{equation*}
	\left[ \begin{array}{c} X^*\\ D  \end{array}\right] \sim \mathrm{Normal}\left(\left[\begin{array}{c} 0\\ 0\end{array}\right], \left[\begin{array}{cc} \sigma_x^2 & \sigma_x\sigma_d\rho  \\ \sigma_x\sigma_d\rho& \sigma_d^2	\end{array}\right] \right),
\end{equation*}
$(X^*,D)\inde (U_x,U_z)$, $U_x\inde U_z$, and $U_x,U_z$ both follows the standard normal distribution. In this model, for any value of $\rho \in (-1,1)$, the high-level condition \eqref{eq: high-level condition integral equation} fails while Assumptions \ref{ASSM: conditional independence outcome}-\ref{ASSM: control function dimension} hold. See the appendix for derivations. In this specific example, explicit calculation of singular values and associated orthonormal functions is possible due to the normality assumption. In more general settings, it seems difficult to provide easy-to-verify sufficient conditions for the finiteness of the infinite sum unless one imposes specific structures on the outcome equation.

If one is willing to impose the high-level condition \eqref{eq: high-level condition integral equation}, then the integral equation approach produces an estimation equation that is linear in nuisance parameters, which may facilitate estimation.
Also, it has no difficulties with discrete endogenous variables. 
Yet, as the above example suggests, verifying \eqref{eq: high-level condition integral equation} needs to be done on a case-by-case basis, and it is not easy to do so. The control function method applies to a general nonparametric outcome equation if Assumption \ref{ASSM: support condition} holds, and in Section \ref{sec: semiparametric parametric outcome}, I demonstrate that even when the common support condition does not hold, the control function method can effectively deal with non-linear outcome models.
In addition, my method can leverage available results from the control function literature to provide a variety of modeling devices that facilitate implementations of my identification result.

\paragraph{Operator diagonalization method}
The identifying assumptions of my approach and those of the operator diagonalization method \citep{HuSchennach2008} are non-nested. In particular, Assumption \ref{ASSM: support condition} does not imply and is not implied by a completeness condition on $(X^*,Z)$ with respect to $Z$ (conditional on $D$), although Assumption \ref{ASSM: support condition} implies a weaker version of completeness (see the appendix Section C).

The nonparametric identification results of \cite{HuSchennach2008} are more general than Theorem \ref{thm:conditional independence} in the sense that they identify a larger class of parameters, including the distribution of $X^*$.
Also, I assume that the treatment variable $D$ does not have a measurement error issue, whereas the operator diagonalization method can handle measurement errors in the treatment variable if there exists an appropriate proxy for mismeasured treatment.

The main advantage of my control function approach lies in practical implementation. It can effectively use additional structures on the outcome equation to relax some of the identifying assumptions. This point is practically relevant because in the majority of empirical applications, sample sizes are not large enough to warrant fully nonparametric estimation. In addition, control function methods are easy to implement with the help of cross validation as shown in Section \ref{sec: semiparametric estimation}, whereas the operator diganalization method relies on sieve estimation methods with multiple tuning parameters.

To be concrete, compare a general nonparametric model with the random coefficient model discussed above. Whereas identification in a general nonparametric model requires $X$ and $Z$ to satisfy completeness-like conditions for $X^*$, the additional structure afforded by random coefficient models enables identification with only completeness condition on $X$ and very mild relevance condition for $Z$. In particular, discrete $Z$ is accommodated, even when the unobserved heterogeneity $X^*$ is continuously distributed.
The operator diagnozliation method seems to require the same strong identifying assumptions in models with additional restrictions as in a general nonparametric model, because it does not exploit specific features of those models.

\paragraph{Control function method in triangular models}
Most constructions of control functions rely on strict monotonicity of the first-stage equation. To be specific, consider the model
\begin{equation*}
	D = h(Z,\eta)
\end{equation*}
where $h$ is a fixed function and $\eta$ is an unobserved variable. \cite{ImbensNewey2009} (IN henceforth) showed that if $Z\inde (Y(d),\eta)$ and $h(z,\cdot)$ is invertible in the second argument for almost all $z$, then $F_{D|Z}(D|Z)$ is a valid control function. 

For the sake of comparison, I consider how IN's approach might be used to construct a valid control function (i.e.,\ $Y(d)\inde D|F_{D|Z}(D|Z)$) in the measurement error problem of this paper. 
The main difference in the identifying assumptions lies in (i) the scalar restriction on the unobserved heterogeneity and (ii) the exclusion restriction on $Z$.

Suppose $Z$ satisfies the IV exclusion restriction in the sense that $(Y(d),X^*)\inde Z$. Then, IN's approach produces a valid control function if $X^*$ enters into the first-stage equation as a scalar e.g.,\ there exists some  real-valued function $\iota$ such that
\begin{equation*}
	D= h(Z,\iota(X^*,U))
\end{equation*}
where $U$ is another unobserved term independent of other variables, and $h(z,\cdot)$ is strictly monotonic in its second argument. In my empirical setting, $X^*$ is two-dimensional (cognitive and non-cognitive ability), so the scalar restriction is indeed a substantive restriction.  In the simple model of college admission considered above, the scalar restriction and strict monotonicity may be plausible as admission decision is based on the scalar admission score. In other settings, the scalar restriction might be violated \cite[e.g.,][Section 2.1]{Florens-Heckman-Meghir-Vytlacil_2008}. My approach can relax this scalar restriction, provided that appropriate proxy variables are available, because it does not rely on the invertibility of the first-stage equation.
On the other hand, if the invertibility holds, IN does not require a proxy variable as it exploits the available structure.

Another important distinction is the exclusion restriction on $Z$.
Whereas my approach allows for statistical dependence between $X^*$ and $Z$, IN requires $Z\inde X^*$. 
In fact, correlation between $X^*$ and $Z$ helps with the common support condition for my approach.
This distinction is empirically relevant.
In DK's setting, a valid IV is not readily available because in observational settings, it is difficult to find an instrument that is independent of unobserved ability $X^*$ and yet affects college attendance.
Thus, IN's approach may not be applicable in DK's setting.
My approach remains valid if measurement errors in $Z$ (SAT score) are orthogonal to the outcome conditional on the unobserved ability, which seems plausible as I argued in Section \ref{sec: nonparametric identification}.

\paragraph{Group-level correlated random effect models}
Theorem \ref{thm:conditional independence} has a close connection with the identification results of \cite{AltonjiMatzkin2005}. To explain, it is helpful to use the notation
\begin{equation*}
Y(d) = \mathscr{Y}(d,\varepsilon),\qquad D\inde \varepsilon|X^*
\end{equation*}
where $\mathscr{Y}$ is an unknown non-stochastic function and $\varepsilon$ is an unobserved heterogeneity. Altonji and Matzkin based their identification results on finding pairs $(d_1,z_1),(d_2,z_2)$ such that
\begin{equation}\label{eq:unobs_hetero_dist}
f_{\varepsilon|DZ}(\cdot|d_1,z_1) = f_{\varepsilon|DZ}(\cdot|d_2,z_2)
\end{equation}
where $f_{\varepsilon|DZ}$ is the conditional density of $\varepsilon$ given $(D,Z)$ (see their equation (1.4)). In proving Theorem \ref{thm:conditional independence}, I show that \eqref{eq:unobs_hetero_dist} is implied by
\begin{equation*}
f_{X|DZ}(\cdot|d_1,z_1) = f_{X|DZ}(\cdot|d_2,z_2).
\end{equation*}
Thus, my identification strategy uses the proxy distribution to find pairs $(d_1,z_1),(d_2,z_2)$ such that Altonji and Matzkin's exchangeability condition holds. In this sense, my paper provides a framework in which the exchangeability condition \eqref{eq:unobs_hetero_dist} follows from model primitives. 

\cite{ArkhangelskyImbens2019} also provided a framework that implies a version of exhcnageability condition, and they presented additional identification results. They focused on settings where observational units belong to groups and there exists a group-level unobserved heterogeneity. My model does not have an explicit group structure, and what my identification strategy does is to form groups based on the value of $V$. That is, two observations belong to the same group if they have the same value of $V$. Similar to the setup in Arkhangelsky and Imbens, the treatment assignment becomes exogenous within groups, and treatment effects can be identified using this induced group structure.

\subsection{Proof sketch of the conditional independence result}\label{sec:proof_sketch}
I sketch the identification argument using a simple setting. I focus on the case where $D,X,X^*$ are all discrete. Specifically, the supports of $X$ and $X^*$ are $\CX=\{x_1,\dots,x_L\}$ and $\CX^*=\{x_1^*,\dots,x_L^*\}$ for some $L$. Note that in this special case, Assumption \ref{ASSM: L2 completeness} reduces to the full-column rank of the matrix
\begin{equation*}
	\bm{\Pi} =\left[\begin{array}{ccc}
		\mathrm{Pr}[X=x_1|X^*=x_1^*] & \dots & \mathrm{Pr}[X=x_1|X^*=x_L^*] \\
		\vdots &\ddots & \\
		\mathrm{Pr}[X=x_L|X^*=x_1^*] & \dots & \mathrm{Pr}[X=x_L|X^*=x_L^*]\end{array}\right]. 
\end{equation*}
Define
\begin{equation*}
	V= \big[	\mathrm{Pr}[X=x_1|D,Z] \ \dots \ \mathrm{Pr}[X=x_L|D,Z] \big]'
\end{equation*}
which is the conditional distribution of the proxy variable $X$. Now I show that $V$ is a valid control function in the sense that $Y(d)\inde D|V$. To verify this claim, it suffices to show $X^*\inde D|V$ since Assumption \ref{ASSM: conditional independence outcome} imply $\mathrm{Pr}[Y(d)\leq y|D,V] = \mathbb{E}[\mathrm{Pr}[Y(d)\leq y|X^*]|D,V]$ and $X^*\inde D|V$ implies the desired result.

The law of total probabilities and Assumption \ref{ASSM: proxy exclusion} imply 
\begin{align*}
	\mathrm{Pr}[X=x|D,Z] &= \sum_{l=1}^L \mathrm{Pr}[X=x|X^*=x_l^*,D,Z]\mathrm{Pr}[X^*=x_l^*|D,Z] \\
	&= \sum_{l=1}^L \mathrm{Pr}[X=x|X^*=x_l^*]\mathrm{Pr}[X^*=x_l^*|D,Z],
\end{align*}
and stacking this equation for different values of $x\in\{x_1,\dots,x_L\}$,
\begin{equation*}
	V =	\bm{\Pi}U,\qquad U=\big[ \mathrm{Pr}[X^*=x_1^*|D,Z]\ \dots \ \mathrm{Pr}[X^*=x_L^*|D,Z]\big]'
\end{equation*}
By the full-rank condition,
\begin{equation}\label{eq:U=PiV}
	U = \big(\bm{\Pi}'\bm{\Pi} \big)^{-1}\bm{\Pi}'V.
\end{equation} 
This equality indicates that if two groups of workers have the same conditional distribution of test scores, then they also have the same conditional distribution of unobserved ability since $\BPi$ is non-stochastic. This in turn implies that the conditional proxy distribution has the balancing property. To substantiate this last claim, for any $l\in\{1,\dots,L\}$,
\begin{align*}
	\mathrm{Pr}[X^*=x_l|D,V] &= \mathbb{E}\big[\mathrm{Pr}[X^*=x_l|D,Z]\big|D,V\big] = \mathbb{E}[e_l'U|D,V] \\
	&= \mathbb{E}[e_l'\big(\bm{\Pi}'\bm{\Pi} \big)^{-1}\bm{\Pi}'V |D,V]\\
	&= \mathbb{E}[e_l'\big(\bm{\Pi}'\bm{\Pi} \big)^{-1}\bm{\Pi}'V |V] \\
	&= \mathbb{E}[\mathrm{Pr}[X^*=x_l|D,Z] |V]= \mathrm{Pr}[X^*=x_l|V]
\end{align*}
where $e_l\in\mathbb{R}^L$ is the unit vector whose $l$th element is unity, the first equality holds as $V$ is a function of $(D,Z)$, the third equality follows from \eqref{eq:U=PiV}, the fourth equality is by $\BPi$ being non-random, and the fifth equality applies \eqref{eq:U=PiV} again. The conclusion $\mathrm{Pr}[X^*=x_l|D,V] = \mathrm{Pr}[X^*=x_l|V]$ establishes the desired result $X^*\inde D|V$.

\section{Estimation}\label{sec: estimation}
In applications, it is important to allow for covariates. 
Denoting additional covariates by $S$, Assumptions \ref{ASSM: conditional independence outcome} and \ref{ASSM: proxy exclusion} may be modified to $Y(d)\inde (D,Z)|X^*,S$ and $X\inde (D,Z,S)|X^*$, respectively. 
Then, I include additional covariates in the control function regression and the outcome regression i.e,\ $V=\mathbb{E}[B(X)|D,Z,S]$ and $\mathbb{E}[Y|D,V,S]$.
To lighten the notation, I again make covariates implicit, but including them does not change estimation results below.

\subsection{Semiparametric estimation}\label{sec: semiparametric estimation}
Theorem \ref{thm: finite-dimensional V} indicates that instead of conditioning on $\mathfrak{V}$, it suffices to condition on a finite-dimensional vector $V=\mathbb{E}[B(X)|D,Z]$ with some choice of $B(x) = (g_1(x),\dots,g_k(x))'$. While $k$ is guaranteed to be finite, it is not known a priori. Therefore, I develop practical estimation procedures that automatically select relevant control functions using Lasso. To this goal, I build on the recent developments in estimation methods using Neyman-orthogonal moments. The treatment here mostly follows \cite{Chernozhukov-Newey-Singh_2022_ecma} (henceforth CNS) and I refer interested readers to the paper for details. For concreteness, I focus on the random coefficient model $Y=\varepsilon_1 + \varepsilon_2 D$ as considered in Section \ref{sec: semiparametric parametric outcome} with $D\in\mathbb{R}$, where the parameter of interest is $\theta_0=\mathbb{E}[\varepsilon_2]$. Although the discussion in this section focuses on this specific model, the estimation method below can be easily extended to causal parameters in other models.

Let $\mu_0(d,v) = \mathbb{E}[Y|D=d,V=v]=\mu_{0,1}(v) + \mu_{0,2}(v)d$ and $V=\nu_0(D,Z)$ with $\nu_0(D,Z)=\mathbb{E}[B(X)|D,Z]$ where $B(x)$ potentially contains redundant elements, which may be discarded via a model selection procedure. A moment condition that identifies $\theta_0$ is
\begin{equation*}
	\theta_0 = \mathbb{E}[ \mu_0(D+1,V) - \mu_0(D,V)] \equiv \mathbb{E}[m(W,\mu_0,\nu_0)]
\end{equation*}
where $W=(Y,D,Z',B(X)')'$ is a data observation and $m(w,\mu,\nu)=\mu(d+1,\nu(d,z))-\mu(d,\nu(d,z))$ is the moment function. Having characterized the moment function $m$, a simple approach to estimate $\theta_0$ is to first estimate $\mu_0,\nu_0$ and form a sample analogue of $\mathbb{E}[m(W,\mu_0,\nu_0)]$. Yet, such plug-in approach is known to suffer from bias arising from noises in the first-stage estimates of $(\mu_0,\nu_0)$, especially when one uses Lasso or other machine learning methods to select regressors. Neyman-orthogonal estimation equations can be used to address this issue.

To describe Neyman-orthogonal moments, note that for any square-integrable function $\mu$, 
\begin{equation*}
	\mathbb{E}[m(W,\mu,\nu_0)] = \mathbb{E}\bigg[ \bigg(\frac{f_{DV}(D-1,V)}{f_{DV}(D,V)}-1\bigg)\mu(D,V) \bigg]
\end{equation*}
holds with some regularity conditions. The function $f_{DV}(D-1,V)/f_{DV}(D,V)-1$ plays the role of a Riesz representer as discussed in CNS. Let $\Gamma$ be a function class that we use to estimate $\mu_0$, and denote by $\alpha_0$ the $L^2$ projection of a Riesz representer onto $\Gamma$. The exact form of $\Gamma$ will be specified below. Using the pathwise derivative calculation of \cite{HahnRidder2013} and \cite{Newey1994a} with the assumptions below, one can show that
\begin{align*}
	\psi(W,\mu,\nu,\alpha) &= m(W,\mu,\nu) + \alpha(D,\nu(D,Z))[Y-\mu(D,\nu(D,Z))] \\
	&\quad + \bigg[\frac{\partial}{\partial V'} m(W,\mu,\nu) - \alpha(D,\nu(D,Z))\frac{\partial}{\partial V'} \mu(D,\nu(D,Z))\bigg] [B(X) - \nu(D,Z)]
\end{align*}
is an orthogonal score function for the parameter $\theta_0$. $\psi$ being an orthogonal score means that (i) $\mathbb{E}[\psi(W,\mu_0,\nu_0,\alpha_0)]=\theta_0$ and (ii) for a path $\{\mu_t,\nu_t,\alpha_t:t\in [0,\epsilon),\epsilon>0\}$,
\begin{equation*}
	\frac{\partial}{\partial t}\mathbb{E}[\psi(W,\mu_t,\nu_t,\alpha_t)]\big\vert_{t=0} = 0
\end{equation*}
holds. This second property suggests that the score function is insensitive to the first-stage estimation errors in $(\mu_0,\nu_0,\alpha_0)$. Then, one can form an estimator of $\theta_0$ by first estimating $(\mu_0,\nu_0,\alpha_0)$ and plugging them into $\psi$ to form a sample analogue of $\mathbb{E}[\psi(W,\mu_0,\nu_0,\alpha_0)]$.

Following CNS, I use cross-fitting. Let $\{I_l\}_{l=1}^L$ be a $L$-fold random partition of the sample. The number of partition $L$ is fixed (a common choice is $L=5,$ or $10$), and the size of each partition should be similar. Estimation involves multiple steps. In the first step, I estimate the control function $V=\nu_0(D,Z)$. Any nonparametric/machine learning method may be used to construct estimates of $V$, and I write $\widehat{V}_{ll'}=\widehat{\nu}_{ll'}(D,Z)$ for the estimate of $V$ using observations not in $I_l,I_{l'}$. Here, all the elements in $\mathbb{E}[B(X)|D,Z]$ are estimated, some of which may be redundant and discarded later. In the second step, I estimate $(\mu_0,\alpha_0)$ using Lasso. Let $p(v)=(p_1(v),\dots,p_K(v))'$ be a vector of approximating functions (e.g.,\ polynomial splines) whose dimension $K=K_n$ depends on the sample size. Define
\begin{equation*}
	\widehat{\Omega}_l = \frac{1}{n-n_l}\sum_{l'\neq l}\sum_{i\in I_{l'}}  \left[
	\begin{array}{cc}
		p(\widehat{V}_{i,ll'})p(\widehat{V}_{i,ll'})' & p(\widehat{V}_{i,ll'})p(\widehat{V}_{i,ll'})'D_i\\
		p(\widehat{V}_{i,ll'})p(\widehat{V}_{i,ll'})'D_i & p(\widehat{V}_{i,ll'})p(\widehat{V}_{i,ll'})'D_i^2
	\end{array}
	\right]
\end{equation*}
\begin{equation*}
	\widehat{M}^{\mu}_l = \frac{1}{n-n_l}\sum_{l'\neq l}\sum_{i\in I_{l'}} \left[
	\begin{array}{c}
		p(\widehat{V}_{i,ll'})Y_i\\
		p(\widehat{V}_{i,ll'})D_i Y_i
	\end{array}
	\right],\qquad \widehat{M}^{\alpha}_l= \frac{1}{n-n_l}\sum_{l'\neq l}\sum_{i\in I_{l'}} \left[
	\begin{array}{c}
		\mathbf{0}\\
		p(\widehat{V}_{i,ll'})
	\end{array}
	\right]
\end{equation*}
where $n_l$ is the size of $I_l$. Then, the Lasso estimators of $(\mu_0,\alpha_0)$ are given by $\widehat{\mu}_l(d,v) = (p(v)',p(v)'d) \widehat{\rho}_l$, $\widehat{\alpha}_l(d,v) = (p(v)',p(v)'d) \widehat{\pi}_l$ where
\begin{equation*}
	\widehat{\rho}_l =\argmin_{\rho} \big\{ \rho'\widehat{\Omega}_l\rho - 2\rho'\widehat{M}^{\mu}_l  + 2 \kappa_n \|\rho\|_1 \big\},\qquad  \widehat{\pi}_l =\argmin_{\pi} \big\{ \pi'\widehat{\Omega}_l\pi - 2\pi'\widehat{M}^{\alpha}_l  + 2 \kappa_n \|\pi\|_1 \big\},
\end{equation*}
$\kappa_n$ is a sequence of penalty terms that shrinks to zero, and $\|\cdot\|_1$ is the $L^1$ norm on $\mathbb{R}^{2K}$. Finally, an estimator of the causal parameter is formed by
\begin{equation*}
	\widehat{\theta}_n = \frac{1}{n}\sum_{l=1}^L\sum_{i\in I_l} \psi(W_i,\widehat{\mu}_l,\widehat{\nu}_l,\widehat{\alpha}_l)
\end{equation*}
where $\widehat{\nu}_l$ is an estimate of $\nu_0$ using observations not in $I_l$. Also, one can estimate the asymptotic variance of $\widehat{\theta}_n$ by
\begin{equation*}
	\widehat{\Psi}_n = \frac{1}{n}\sum_{l=1}^L \sum_{i\in I_l} \big[ \psi(W_i,\widehat{\mu}_l,\widehat{\nu}_l,\widehat{\alpha}_l) - \widehat{\theta}_n\big]^2.
\end{equation*}
The proposed procedure allows researchers to be agnostic about which elements of the control function $V$ to be included in the regression. This aspect can be practically important as there is potential bias-variance trade-off. When the effective dimension of $V$ is low, including redundant elements may lead to noisier estimates in finite samples, whereas omitting relevant controls can lead to biased estimates.  The automatic selection using Lasso may offer efficiency gains while ensuring that the bias does not arise from omitting important controls.

To present a formal result on the asymptotic distribution of $\widehat{\theta}_n$, I define additional notations.
Write $\|\cdot\|_{\ell}$ for the $\ell$-norm $\ell=1,2$ on the Euclidean space.  Let $q(d,v) =  (p(v)'\ dp(v)')'$ be the vector of approximating functions and define $\Gamma$ to be the $L^2$ closure of linear combinations of $\{q_l(d,v): l\in\mathbb{N}\}$. 
For $u=(u_1,\dots,u_l)\in\mathbb{Z}_{\geq0}^l$, write $|u|=\sum_{\ell=1}^ku_{\ell}$ and $\partial^u f(x)=\partial^{|u|}f(x)/\partial^{u_1}x_1\dots\partial^{u_l}x_l$. 
Below $C>1$ denotes a positive constant independent of the sample size and represents a different number at different places.
\begin{assumption}\label{ASSM: estimation DGP}
	Observations $\{Y_i,D_i,Z_i,X_i\}_{i=1}^n$ are a random sample. $\{Y(d):d\in \mathcal{D}\}\inde (D,Z)|X^*$.
	$\CD$ is bounded. $\mu_0$ is twice continuously differentiable and $\alpha_0$ is Lipschitz continuous. $\mu_0,\alpha_0$, and derivatives of $\mu_0$ are bounded. Let $\varepsilon=Y-\mu_0(D,V)$ and $\eta = B(X)-V$. With probability one, $\mathbb{E}[\varepsilon^2|D,Z] + \mathbb{E}[\|\eta\|_2^2|D,Z]\leq C<\infty$. Also, $\mathbb{E}[\varepsilon^4] +\mathbb{E}[\|\eta\|_2^4]<\infty$.
\end{assumption}
This assumption imposes regularity conditions on the underlying data generating process, which are mild and standard in the literature. One non-standard aspect is that it strengthens Assumption \ref{ASSM: conditional independence outcome} to $\{Y(d):d\in \mathcal{D}\}\inde (D,Z)|X^*$. Although it is a strictly stronger condition mathematically, in many econometric models used in practice, it holds when Assumption \ref{ASSM: conditional independence outcome} does. This conditional independence restriction implies $\mathbb{E}[\varepsilon|D,Z]=0$ (Lemma A.3 in the appendix), which is crucial for the form of the asymptotic variance \citep{HahnLiaoRidderShi}.

The next assumption imposes that the first-stage estimate of the control function is consistent and converges at a certain rate in $L^2$ norm.
\begin{assumption}\label{ASSM: estimation first-stage}
	The first-stage estimator $\widehat{\nu}_n$ satisfies $\int \|\widehat{\nu}_n(d,z) - \nu_0(d,z)\|_2^2 dF_{DZ}(d,z) = O_{\mathbb{P}}(\tilde{\omega}_n^2)$ where $\tilde{\omega}_n=o(1)$ is the convergence rate of $\widehat{\nu}_n$.  
\end{assumption}
Define
\begin{equation*}
	\Xi_{a,n}=\max_{1\leq l\leq K_n}\max_{|u|=a}\sup_{v}|\partial^u p_l(v)|\qquad a=0,1,2,
\end{equation*}
where $\partial^0 p(u) = p(u)$. Then, let
\begin{equation*}
	\omega_{n} =  \Xi_{1,n} \tilde{\omega}_n.
\end{equation*}
This rate plays an important role in the rest of assumptions. With specific $p(\cdot)$, bounds on $\Xi_{a,n}$ are available in the literature. For instance, with polynomial splines, $\Xi_{a,n}=K_n^{1/2+a}$ \citep[][]{NEWEY1997}.

The next assumption needs some additional notation: given a set of indices $J\subset\{1,\dots,L\}$ and $v\in\mathbb{R}^L$, let $\#|J|$ be the cardinality of $J$, $v_J$ be the $\#|J|\times 1$ subvector of $v$ whose elements are $\{v_l:l\in J\}$, and $v_{J^c}$ be the subvector of $v$ that consists of elements not in $v_J$.
\begin{assumption}\label{ASSM: estimation basis functions}
	The basis function $p(\cdot)$ is twice continuously differentiable. The eigenvalues of $\Omega=\mathbb{E}[q(D,V)q(D,V)']$ are bounded above uniformly in $K$.
	For any $s=O(\omega_n^{-2})$, with probability approaching one,
	\begin{equation*}
		\min_{\substack{J\subset\{1,\dots,2K\}\\ \#|J|\leq s} }\min_{\|\rho_{J^c}^{}\|_1\leq 3 \|\rho_J^{}\|_1} \frac{\rho'\widehat{\Omega}\rho}{\rho_J'\rho_J^{}}\geq c,\qquad \min_{\substack{J\subset\{1,\dots,2K\}\\ \#|J|\leq s} }\min_{\|\rho_{J^c}^{}\|_1\leq 3 \|\rho_J^{}\|_1} \frac{\rho'\Omega\rho}{\rho_J'\rho_J^{}}\geq c
	\end{equation*}
	where $c>0$ is independent of $K$.
\end{assumption}
This condition is a sparse eigenvalue condition commonly used in the Lasso literature (see Assumption 3 of CNS).
\begin{assumption}\label{ASSM: estimation approximation error}
	Let $\bar{\rho},\bar{\pi}$ be least squares coefficients of projecting $\mu_0,\alpha_0$ onto $q(D,V)$, respectively. i.e.,\ $\Omega\bar{\rho}=\mathbb{E}[q(D,V)Y]$ and $\Omega \bar{\pi} = \mathbb{E}[q(D,V)\alpha_0(D,V)]$.
	There exists $\xi >1/2$ such that for each positive integer $s$, there exist $\widetilde{\rho},\widetilde{\pi} \in\mathbb{R}^{2K_n}$ satisfying that the number of non-zero elements in $\widetilde{\rho},\widetilde{\pi}$ are bounded by $s$ and
	\begin{align*}
		\|\bar{\rho} - \widetilde{\rho}\|_2 + \|\bar{\pi}-\widetilde{\pi}\|_2 \leq C s^{-\xi}.
	\end{align*}
	Let $\bar{\mu}(d,v)=q(d,v)'\bar{\rho}$ and $\bar{\alpha}(d,v)=q(d,v)'\bar{\pi}$. Also, assume
	\begin{align*}
		&\mathbb{E}[|\mu_0(D,V)-\bar{\mu}(D,V)|^2] +\mathbb{E}[|\alpha_0(D,V)-\bar{\alpha}(D,V)|^2] =o(n^{-1/2}),\\
		&\mathbb{E}[\|\partial \{\mu_0(D,V)-\bar{\mu}(D,V)\}/\partial V\|_2^2] + \mathbb{E}[\|\partial^2 \{\mu_0(D,V)-\bar{\mu}(D,V)\}/\partial D\partial V\|_2^2]=o(1).
	\end{align*} 
\end{assumption}
Assumption \ref{ASSM: estimation approximation error} imposes that $\mu_0,\alpha_0$ admit sparse approximations. I follow \cite{Bradicetal} in the formulation of this condition (their Assumptions 3 and 8). One aspect that differs from CNS is that I need to estimate the derivative of $\mu_0$. To accommodate this change, I impose that $p(\cdot)$ can approximate smooth functions and their derivatives in the $L^2$ norm. This approximation property holds for many of standard choices of approximating functions e.g.,\ polynomial splines.

\begin{theorem}\label{thm: semiparametric estimation}
	Assumptions \ref{ASSM: estimation DGP}, \ref{ASSM: estimation first-stage}, \ref{ASSM: estimation basis functions}, and \ref{ASSM: estimation approximation error} hold. In addition, suppose $\max_{1\leq l\leq L}n/n_l=O(1)$, $\omega_n = o(\kappa_n)$, $\omega_n^{2\xi/(2\xi+1)}\Xi_{1,n}(\kappa_n/\omega_n) + \Xi_{2,n}\tilde{\omega}_n=o(1)$, and $\omega_n^{2\xi/(2\xi+1)}(\kappa_n/\omega_n) + (\Xi_{0,n}\Xi_{2,n})^{1/2}\tilde{\omega}_n=o(n^{-1/4})$. Then,
	\begin{equation*}
		\sqrt{n}\big(\widehat{\theta}_n -\theta_0\big) \rightsquigarrow \mathrm{Normal}(0,\Psi_0)
	\end{equation*}
	where $\Psi_0= \mathrm{Var}[ \psi(W,\mu_0,\nu_0,\alpha_0)]$ and $\rightsquigarrow$ denotes convergence in distribution. Also, the variance estimator $\widehat{\Psi}_n$ converges in probability to $\Psi_0$.
\end{theorem}
The theorem states that the estimator $\widehat{\theta}_n$ is $\sqrt{n}$-consistent and asymptotically normal under the rate restrictions on the first-stage estimator, the Lasso penalty term, and the sup norm on the $\partial^u p(v)$ with $|u|\leq 2$. To illustrate, suppose we use polynomial splines as approximating functions and $\tilde{\omega}_n=n^{-\kappa_1}$, $K_n = n^{\kappa_2}$ for some $\kappa_1,\kappa_2>0$. Then, the hypothesis of Theorem \ref{thm: semiparametric estimation} implies $2\kappa_1-3\kappa_2 > \frac{1}{2} + \frac{1}{4\xi}$. The restriction is somewhat stringent on the rate at which the number of approximation terms can grow (i.e.,\ $K_n$) because both the first and second stages are non-parametric in $V$. It might be possible to relax rate restrictions and to obtain a better finite-sample distribution approximation by carefully analyzing the higher-order terms of the estimator \citep[e.g.,\ in the spirit of][]{CattaneoJanssonMa2019}. I leave such analysis for future work.

\subsection{Flexible parametric approach}\label{sec:flexible_parametric_estimation}
In this section, I consider a flexible parametric estimation procedure based on the model \eqref{eq:flexible_parametric_specification} in Section \ref{sec: semiparametric parametric outcome}. Recall that \eqref{eq:flexible_parametric_specification} posits $\mathbb{E}[Y|D,V]=\Lambda(q(D,V)'\gamma_0)$ where $\Lambda(\cdot),q(\cdot)$ are specified by researchers and $\gamma_0$ is to be estimated. Let $d_v$ be the dimension of $B(x)$ and $d_q$ be some positive integer. For the control function $V$, I specify the model
\begin{equation*}
	\mathbb{E}[B(X)|D,Z] = Q(D,Z) \delta_{0},
\end{equation*}
where $Q:\CZ\CD\to\mathbb{R}^{ d_v\times d_q}$ is a matrix-valued transformation of $(D,Z)$ and $\delta_{0}\in\mathbb{R}^{d_q}$ is the parameter to be estimated. As a baseline, one may use $Q(D,Z)=I\otimes q_2(D,Z)$ with $I$ being the identity matrix, $q_2(D,Z)=(1,D',Z')$, and $\otimes$ denoting the Kronecker product. Researchers can include higher-order polynomial terms to enhance flexibility. Note that this is a reduced form equation, and as long as the model has good predictive power, the procedure is expected to work reasonably well. 

For implementation, first estimate $\delta_{0}$ by least squares and form $\widehat{V}_i = Q(D_i,Z_i)\widehat{\delta}_{n}$. Then, estimate $\gamma_0$ in $\mathbb{E}[Y|D,V]=\Lambda(q(D,V)'\gamma_0)$ by (non-linear) regression of $Y$ on $q(D,\widehat{V})$. Finally, the estimator for the ASF is formed by
\begin{equation*}
	\widehat{\vartheta}_n(d) = \frac{1}{n}\sum_{i=1}^n \Lambda\big(q(d,\widehat{V}_i)'\widehat{\gamma}_n \big).
\end{equation*}

For inference, it is useful to have a closed-form variance estimator. Let $\dot{\Lambda}$ be the first derivative of $\Lambda$,  
\begin{equation*}
	\widehat{q}_i =q(D_i,\widehat{V}_i),\quad \widehat{\varepsilon}_i = Y_i - \Lambda(\widehat{q}_i'\widehat{\gamma}_n), \quad Q_i=Q(D_i,Z_i),\quad \widehat{\eta}_i = B(X_i) - Q_i\widehat{\delta}_n,
\end{equation*}
\begin{equation*}
	\widehat{\Gamma}_1 = \frac{1}{n}\sum_{i=1}^n Q_i' Q_i,\quad \widehat{\Gamma}_2 = \frac{1}{n}\sum_{i=1}^n \big|\dot{\Lambda}(\widehat{q}_i'\widehat{\gamma}_n)\big|^2 \widehat{q}_i\widehat{q}_i',  \quad\widehat{\Gamma}_3 =  -\frac{1}{n}\sum_{i=1}^n|\dot{\Lambda}(\widehat{q}_i'\widehat{\gamma}_n)|^2\widehat{q}_i\widehat{\gamma}_n'\partial q(D_i,\widehat{V}_i)Q_{i},
\end{equation*}
\begin{equation*}
	\widehat{c}_1(d) = \frac{1}{n}\sum_{i=1}^n\dot{\Lambda}(q(d,\widehat{V}_i)'\widehat{\gamma}_n)q(d,\widehat{V}_i)',\qquad \widehat{c}_2(d) =\frac{1}{n}\sum_{i=1}^n\dot{\Lambda}(q(d,\widehat{V}_i)'\widehat{\gamma}_n)\widehat{\gamma}_n'\partial q(d,\widehat{V}_i) Q_i,
\end{equation*}
and $\partial q$ be the derivative of $q$ with respect to $V$. Then,
\begin{equation*}
	\widehat{\Psi}_n(d)=\frac{1}{n}\sum_{i=1}^n \big[ \Lambda(q(d,\widehat{V}_i)'\widehat{\gamma}_n) - \widehat{\vartheta}_n(d) +\widehat{c}_1(d)\widehat{\Gamma}_2^{-1}\widehat{q}_i\dot{\Lambda}(\widehat{q}_i'\widehat{\gamma}_n)\widehat{\varepsilon}_i + \{\widehat{c}_1(d)\widehat{\Gamma}_2^{-1}\widehat{\Gamma}_3 + \widehat{c}_2(d)\}\widehat{\Gamma}_1^{-1}Q_i'\widehat{\eta}_i \big]^2
\end{equation*}
is an estimator for the asymptotic variance of $\sqrt{n}(\widehat{\vartheta}(d)-\vartheta(d))$. Note that the variance estimator does not require additional nuisance parameter estimation.

Since the asymptotic distributional theory for $\widehat{\vartheta}_n(d)$ is well-established \citep[e.g.,][]{NeweyMcfadden1994}, I relegate the discussion of the asymptotic theory to the appendix. Under the assumptions stated there, the ASF estimator $\widehat{\vartheta}_n(d)$ is asymptotically normal and the variance estimator is consistent.

\section{Empirical application}\label{sec: empirical application}
I apply the new results of this paper to studying causal effects of attending selective college on earnings. I build on the empirical framework of DK, but my approach has an advantage that data requirements are less demanding. Specifically, DK's main identification strategy relies on detailed information about college admission processes, which is not available in my setting.  My control function method overcomes the data limitations, and my empirical result suggests that measurement errors indeed have non-trivial impact on causal effect estimates.

As discussed in Section \ref{sec: nonparametric identification}, DK's empirical framework posits that a college computes an admission score for each student and admits students if and only if their admission score is above the college's cutoff. That is,
\begin{equation*}
	\text{College} \text{ admits a student} \text{ if and only if } X^*_1 +X^*_2 + \epsilon > C
\end{equation*}
where $C$ denotes the cutoff level, $X^*_1 +X^*_2 + \epsilon$ denotes the admission score, $X_1^*$ represents academic ability, $X_2^*$ is non-cognitive skill, and $\epsilon$ is an idiosyncratic term.

DK pointed out that students who applied to and got admitted to the same set of colleges must have similar admission scores e.g.,\ if students got admitted to college 1 but rejected from college 2 with cutoffs $C_1<C_2$, their admission scores would lie in the interval $(C_1,C_2]$. Using this insight, DK controlled for average SAT scores of colleges to which students applied and got admitted in order to account for the non-random selection. Therefore, for DK's empirical strategy, it is crucial to observe college application choices and admission results.

However, such detailed information is not often available. In my dataset, the National Longitudinal Study of 1972 (NLS72), some information about college admission process is observed, but there are two issues regarding the measurement. First, many missing observations on admission results make it difficult to control for the mean SAT score of colleges to which students got admitted. In fact, DK also analyzed the NLS72 dataset as a robustness check to their main analysis, but they only controlled for the mean SAT score of colleges to which students applied, what they call self-revelation models. In their empirical result, self-revelation models yielded similar estimates to their main specifications, yet theoretically, not conditioning on admission results might induce selection bias. Another issue is that the survey elicited only up to three schools to which students applied, but a non-negligible fraction of respondents applied to four or more colleges. In the final sample I use for analysis, $18.6\%$ ($210$ observations) said they applied to more than three colleges.
Thus, the variable of mean SAT score of colleges to which students applied may have non-negligible measurement errors.

The control function method developed in this paper can address these data limitations by using partial information as proxy variables. In particular, I use math and reading test scores at 12th grade as proxies for academic ability $X_1^*$, a survey question that intended to measure psychological traits as a proxy for non-cognitive skill $X_2^*$, and the error-ridden measurement of the average SAT score of colleges to which students applied as another proxy variable. For the excluded variable $Z$, I use an SAT test score, another survey question as a measure of psychological traits, and whether worker's parents attended college. I impose Assumptions \ref{ASSM: conditional independence outcome}-\ref{ASSM: regularity}, \ref{ASSM: control function dimension} as well as the outcome equation
\begin{equation*}
	Y=\varepsilon_1 + \varepsilon_2 D,\qquad (\varepsilon_1,\varepsilon_2)\inde D |X^*.
\end{equation*}
The outcome model strikes the balance between flexibility and practicality. Importantly, the random coefficient model allows for heterogeneous treatment effects that can vary across the unobserved ability levels.
The plausibility of the substantive assumptions \ref{ASSM: conditional independence outcome}-\ref{ASSM: L2 completeness} have been discussed in Sections \ref{sec: nonparametric identification} and \ref{sec: excluded variable example}.
One caveat I should note is that the proxy for non-cognitive ability is a discrete variable, and thus, I impose that the unobserved non-cognitive skill has only moderate variation.

Table \ref{table:summary_statistics} presents summary statistics of variables used in the analysis. The sample size is $1130$. Earnings, the outcome of interest, were measured when survey respondents were around 32 years old. The average of school mean of SAT (the treatment variable) is $1029$ with standard deviation $137$. In the College and Beyond survey analyzed by DK, the mean of the same variable was $1194$ with standard deviation $92.8$. Thus, workers in my sample attended less selective colleges on average relative to those in the College and Beyond survey. This feature of the sample could be important as DK found null effects of college selectivity on future earnings among workers who attended highly selective colleges while other studies found positive, sometime large, effects for workers with relatively low ability \citep[e.g.,][]{Hoekstra2009,Zimmerman2014}.

Table \ref{table:restults} presents the estimation results for the average effect on log earnings of attending college with $100$ higher mean SAT score. To better understand the performance of my proposed method, I also estimated the log earnings regression with the proxy and excluded variables as control regressors using the Lasso-based method of CNS. This second approach is prone to measurement errors as test scores are noisy measurements of ability. 
The control function estimate indicates that earnings increases by $7.9\% \approx 100*(\exp(0.076)-1)$ when a worker attended college with $100$ higher mean SAT score, whereas the ``na\"{i}ve'' estimate indicates the increase is $3.7\%$. Thus, the control function estimate of the causal effect is more than twice the measurement-error prone estimate. As expected, the control function estimate has a larger standard error with a wider confidence interval. Using the influence function representations, I tested the null hypothesis of the two estimators converging to the same value, and the $p$-value for the two-sided alternative was $0.083$. Thus, my empirical result indicates that measurement errors in control variables have non-trivial impact on causal effect estimates.

\section{Conclusion}\label{sec:conclusion}
I developed a new identification strategy for causal effects such as the average structural functions by exploiting proxy variables for unobserved confounding factors. This new approach identifies causal parameters with relatively weak assumptions by exploiting additional structures on the outcome equation. I developed a Lasso-based method to flexibly choose regression specifications for my control function method. As illustrated through an empirical application, measurement errors in covariates can have non-negligible impact on causal estimates. My method can be used to apply the main idea of \cite{Dale-Krueger_2002} under less demanding data requirements.



\vspace{100pt}

\begin{figure}[h]
	\centering
	\caption{Independence Assumptions via DAG}\label{fig:dag}
	\scalebox{1.2}{%
		\begin{tikzpicture}
			\node[text centered] (z) {$Z$};
			\node[below right = 0.5 of z, text centered] (xstr) {$X^*$};
			\node[below = 1 of xstr, text centered] (y) {$Y$};
			\node[left=1 of xstr, text centered] (d) {$D$};
			\node[right =2 of z,text centered] (x) {$X$};
			
			\draw[-,dotted, line width= 1] (z) --  (d);
			\draw[<-,line width= 1] (z) --  (xstr);
			\draw [->, line width= 1] (d) -- (y);
			\draw[->,line width= 1] (xstr) --(d);
			\draw[->,line width= 1] (xstr) -- (y);
			\draw[->,line width=1] (xstr) -- (x);
			\draw[-,dotted,line width=1] (y) -- (x);
	\end{tikzpicture}}\hspace{25pt}	 \scalebox{1.2}{%
		\begin{tikzpicture}
			\node[text centered] (z) {$Z$};
			\node[below right = 0.5 of z, text centered] (xstr) {$X^*$};
			\node[below = 1 of xstr, text centered] (y) {$Y$};
			\node[left=1 of xstr, text centered] (d) {$D$};
			\node[right =2 of z,text centered] (x) {$X$};
			
			\draw[->, line width= 1] (z) --  (d);
			\draw[-,dotted,line width= 1] (z) --  (xstr);
			\draw [->, line width= 1] (d) -- (y);
			\draw[->,line width= 1] (xstr) --(d);
			\draw[->,line width= 1] (xstr) -- (y);
			\draw[->,line width=1] (xstr) -- (x);
			\draw[-,dotted,line width=1] (y) -- (x);
	\end{tikzpicture}}{\footnotesize\begin{flushleft}
	\textbf{Notes}.
	Arrows represent causal effects and dotted lines mean that two variables may have a causal relationship with unspecified direction of effects. 
\end{flushleft}}
\end{figure}

\begin{table}[h]
	\renewcommand{\arraystretch}{1.2}
	\centering
	
	\caption{Summary statistics}
	{\resizebox{0.8\columnwidth}{!}{ 
			\label{table:summary_statistics}
\begin{tabular}{lcccc}
\hline\hline
\multicolumn{1}{l}{}&\multicolumn{1}{c}{Mean}&\multicolumn{1}{c}{Std. dev.}&\multicolumn{1}{c}{Min}&\multicolumn{1}{c}{Max}\tabularnewline
\hline
{\bfseries Outcome}&&&&\tabularnewline
~~Log earnings&  10.13&  0.58&  8.52&  11.58\tabularnewline
\hline
{\bfseries Treatment}&&&&\tabularnewline
~~School mean of SAT&1029.26&137.76&611.00&1440.00\tabularnewline
\hline
{\bfseries Proxy variables}&&&&\tabularnewline
~~Math score&   1.00&  0.64& -1.30&   1.70\tabularnewline
~~Ave. SAT of colleges applied to&1042.09&124.06&640.00&1440.00\tabularnewline
~~Personality trait score 1&   2.48&  0.37&  1.00&   3.00\tabularnewline
\hline
{\bfseries Excluded variables}&&&&\tabularnewline
~~SAT score&1034.17&194.82&460.00&1550.00\tabularnewline
~~Reading score&   0.86&  0.76& -1.60&   2.10\tabularnewline
~~Personality trait score 2&   2.27&  0.41&  1.00&   3.00\tabularnewline
~~Father attended college&   0.39&  0.49&  0.00&   1.00\tabularnewline
~~Mother attended college&   0.24&  0.43&  0.00&   1.00\tabularnewline
\hline
{\bfseries Covariates}&&&&\tabularnewline
~~Female&   0.41&  0.49&  0.00&   1.00\tabularnewline
~~Racial minority&   0.12&  0.32&  0.00&   1.00\tabularnewline
~~Log family income&   9.56&  0.65&  7.31&  10.31\tabularnewline
\hline
\end{tabular}
 } }
	{\footnotesize\begin{flushleft}\textbf{Notes}. The observations are from the National Longitudinal Survey 1972. The sample size is $1130$. The sample construction followed DK by focusing on individuals who attended four-year college in the fall of 1972 and with earnings greater than $\$5000$. Observations were dropped if any of the variables had a missing value. 	\end{flushleft}}
\end{table}

\begin{table}[h]
	\renewcommand{\arraystretch}{1.2}
	\centering
	
	\caption{Estimates of the causal effect of attending selective college}
	{\resizebox{0.6\columnwidth}{!}{ 
			\label{table:restults}
\begin{tabular}{lcc}
\hline\hline
\multicolumn{1}{l}{}&\multicolumn{1}{c}{Control Function}&\multicolumn{1}{c}{Noisy Regressors}\tabularnewline
\hline
Estimate&0.076&0.036\tabularnewline
Std. Err.&(0.028)&(0.015)\tabularnewline
95\% CI.&[0.022 0.131]&[0.007 0.064]\tabularnewline
\hline
\end{tabular}
 } }
	{\footnotesize\begin{flushleft}
			\textbf{Notes}. For estimation, I follow the procedure described in Section \ref{sec: semiparametric estimation}. The first-stage estimation of the control function is done using Lasso, and I use monomial transformation of proxy variables. For the second-step estimation, I use cubic splines as a approximation basis with $K_n=50$. For the lasso penalty term, I use $\kappa_n = c(\log n)^{1/2} n^{-1/3}$ where the constant $c$ is chosen via cross-validation, as suggested by CNS. I also followed a recommendation by \cite{Chernozhukov-etal2018} and obtained multiple estimates by iterating procedures with different sample splits to incorporate the uncertainty from random sample splitting into the standard error computation.
	\end{flushleft}}
\end{table}

\newpage
\hspace{10pt}
\newpage

\begin{appendices}

\section{Identification results}

\subsection{Proof of Theorem \ref{thm:conditional independence}}
\begin{remark}
	For the proof below, consider the linear operator $\Pi:L_2(F_{X^*})\mapsto L_2(F_{X})$, $\Pi(g)(x) = \mathbb{E}[g(X^*)|X=x]$. Assumption \ref{ASSM: regularity} implies that $\Pi$ is compact \citep[see e.g.,][p.5659]{CarrascoFlorensRenault2007}. By Theorem 15.16 of \cite{Kress2014}, there exist non-negative singular values $\{\tau_j\}_{j\geq 1}$ and orthonormal sequences $\{\phi_j\}_{j\geq 1}\subset L_2(F_{X^*})$, $\{\varphi_j\}_{j\geq1}\subset L_2(F_{X})$ for the operator $\Pi$. Theorem 15.18 of \cite{Kress2014} implies that for any function $g\in L_2(F_{X^*})$ in the orthogonal complement of the null space of $\Pi$, we have
	\begin{equation}\label{eq: remark}
		g(x^*) = \sum_{j=1}^{\infty} \frac{\mathbb{E}[h(X)\varphi_j(X)]}{\tau_j}\phi_j(x^*),\qquad h(x) = \Pi(g)(x)
	\end{equation}
	with $\sum_{j=1}^{\infty} \tau_j^{-2}\mathbb{E}[h(X)\varphi_j(X)]^2<\infty$.\hfil\qed
\end{remark}
In the sequel, the equal sign involving random variables is understood as equality with probability one.

By the law of iterated expectations and Assumption \ref{ASSM: proxy exclusion}, 
\begin{align*}
	\frac{f_{X|DZ}(x|D,Z)}{f_{X}(x)} &= \int \frac{f_{X|X^*DZ}(x|x^*,D,Z)}{f_{X}(x)} f_{X^*|DZ}(x^*|D,Z)d\lambda_1(x^*)\\
	&= \int \frac{f_{X|X^*}(x|x^*)}{f_{X}(x)} f_{X^*|DZ}(x^*|D,Z)d\lambda_1(x^*)\\
	&= \Pi\bigg(\frac{f_{X^*|DZ}(\cdot|D,Z)}{f_{X^*}(\cdot)}\bigg) (x).
\end{align*}
Since Assumptions \ref{ASSM: L2 completeness} and \ref{ASSM: regularity} imply that $f_{X^*|DZ}(\cdot|d,z)/f_{X^*}(\cdot)$ lies in the orthogonal complement of the null space of $\Pi$, \eqref{eq: remark} implies
\begin{align*}
	\frac{f_{X^*|DZ}(x^*|D,Z)}{f_{X^*}(x^*)} &= \sum_{j=1}^{\infty} \tau_j^{-1}\int \frac{f_{X|DZ}(x|D,Z)}{f_{X}(x)} \varphi_j(x)f_X(x) d\lambda(x) \phi_j(x^*)\\
	&= \sum_{j=1}^{\infty} \tau_j^{-1}\int f_{X|DZ}(x|D,Z) \varphi_j(x)d\lambda(x)  \phi_j(x^*).
\end{align*}
This equation implies, for any $y\in\mathbb{R}$,
\begin{align*}
	&\mathrm{Pr}[Y(d)\leq y|D,Z]\\
	&= \int \mathrm{Pr}[Y(d)\leq y|X^*=x^*,D,Z]f_{X^*|DZ}(x^*|D,Z)d\lambda_1(x^*)\\
	&= \int \mathrm{Pr}[Y(d)\leq y|X^*=x^*]f_{X^*|DZ}(x^*|D,Z)d\lambda_1(x^*)\\
	&= \int \mathrm{Pr}[Y(d)\leq y|X^*=x^*] \bigg(\sum_{j=1}^{\infty} \tau_j^{-1}\int \varphi_j(x)f_{X|DZ}(x|D,Z) d\lambda(x)  \phi_j(x^*)\bigg)f_{X^*}(x^*)d\lambda_1(x^*)
\end{align*}
where the first equality is by the law of iterated expectations, and the second equality follows from Assumption \ref{ASSM: conditional independence outcome}.
Lemma \ref{lemma: interchange} below implies that the infinite sum and the outer integral are exchangeable, and thus,
\begin{align*}
	\mathrm{Pr}[Y(d)\leq y|D,Z] &=\sum_{j=1}^{\infty} \tau_j^{-1}\mathbb{E}[\mathrm{Pr}[Y(d)\leq y|X^*]\phi_j(X^*)]\int \varphi_j(x)f_{X|DZ}(x|D,Z) d\lambda(x) \\
	&\equiv \sum_{j=1}^{\infty} \frac{\varsigma_j(d)}{\tau_j}\int \varphi_j(x)\mathfrak{V}(x) d\lambda(x),\quad \varsigma_j(d)=\mathbb{E}[\mathrm{Pr}[Y(d)\leq y|X^*]\phi_j(X^*)]
\end{align*}
where $\sum_{j=1}^{\infty}\tau_j^{-2}|\int\varphi_j(x)\mathfrak{V}(x)d\lambda(x)|^2 <\infty$ by \eqref{eq: remark}. Then, $\mathrm{Pr}[Y(d)\leq y|D,Z]$ is $\sigma(\mathfrak{V})$-measurable, and by $\sigma(\mathfrak{V})\subseteq\sigma(D,Z)$, $\mathrm{Pr}[Y(d)\leq y|D,Z]= \mathrm{Pr}[Y(d)\leq y|\mathfrak{V}]$. Thus, $\mathrm{Pr}[Y(d)\leq y|D,\mathfrak{V}]=\mathbb{E}[\mathrm{Pr}[Y(d)\leq y|D,Z]|D,\mathfrak{V}]=\mathbb{E}[\mathrm{Pr}[Y(d)\leq y|\mathfrak{V}]|D,\mathfrak{V}] = \mathrm{Pr}[Y(d)\leq y|\mathfrak{V}]$, which is the desired conditional independence result.

To prove the second part of the theorem, first note $\mathbb{E}[Y|D=d,\mathfrak{V}]=\mathbb{E}[Y(d)|D=d,\mathfrak{V}] = \mathbb{E}[Y(d)|\mathfrak{V}]$ where the second equality holds by the conditional independence result above. Also, $\mathbb{E}[Y(d)|\mathfrak{V}]=\mathbb{E}[Y(d)|D,Z]$ by the argument above. If $D$ has a continuous distribution, then by continuity of $\mathbb{E}[Y(d)|D,Z]$ in $d$, $\mathbb{E}[Y|D=d,\mathfrak{V}]$ is continuous in $d$ with probability one. Thus, the ASF will be identified via
\begin{equation*}
	\vartheta(d) = \mathbb{E}[\mathbb{E}[Y|D=d,\mathfrak{V}]]
\end{equation*}
provided that the expectation is well-defined. Let $\mu$ be a $\sigma(D,\mathfrak{V})$-measurable function such that $\mathbb{E}[Y|D,\mathfrak{V}]=\mu(D,\mathfrak{V})$. Then, we have
\begin{equation*}
	\mathbb{E}[\mathbb{E}[Y|D=d,\mathfrak{V}]] = \mathbb{E}[ \mu(d,f_{X|DZ}(\cdot|D,Z)) ].
\end{equation*}
For this integral to be feasible, it suffices to have $\{ f_{X|DZ}(\cdot|\tilde{d},\tilde{z}):(\tilde{d},\tilde{z}) \in \supp(D,Z)\}= \{f_{X|DZ}(\cdot|d,z):z\in \CZ_d\}$ where $\CZ_d$ is the conditional support of $Z$ given $D=d$, and this condition is implied by Assumption \ref{ASSM: support condition}.\hfill$\qedsymbol$

\begin{lemma}
	Under the assumptions of Theorem, $D\inde X^*|\mathfrak{V}$.
\end{lemma}
\begin{proof}
	Similar to the above calculations on $\mathrm{Pr}[Y(d)\leq \cdot|D,Z]$, we have
	\begin{equation*}
		\mathrm{Pr}[X^*\leq a|D,Z] = \sum_{j=1}^{\infty} \tau_j^{-1} \mathbb{E}[\mathbbm{1}\{X^*\leq a\}\phi_j(X^*)] \int \varphi_j(x) \mathfrak{V}(x) d\lambda(x),\quad \forall a\in\mathbb{R}^{\dim(X)}
	\end{equation*}
	where $\dim(\cdot)$ is the dimension of its argument.
	Thus, $\mathrm{Pr}[X^*\leq a|D,Z]$ is $\sigma(\mathfrak{V})$-measurable and $\mathrm{Pr}[X^*\leq a|D,Z]=\mathrm{Pr}[X^*\leq a|\mathfrak{V}]$. Thus, $\mathrm{Pr}[X^*\leq a|D,\mathfrak{V}]=\mathbb{E}[\mathrm{Pr}[X^*\leq a|D,Z]|D,\mathfrak{V}] = \mathbb{E}[\mathrm{Pr}[X^*\leq a|\mathfrak{V}]|D,\mathfrak{V}]=\mathrm{Pr}[X^*\leq a|\mathfrak{V}]$.
\end{proof}

\begin{lemma}\label{lemma: interchange}
	For a transformation $\mathscr{T}$, suppose $\mathbb{E}[\mathscr{T}(Y(d))^2]<\infty$. Under Assumptions \ref{ASSM: proxy exclusion}, \ref{ASSM: L2 completeness}, and \ref{ASSM: regularity},
	\begin{equation*}
		\int \mathbb{E}[\mathscr{T}(Y(d))|X^*=x^*] \bigg(\sum_{j=1}^{\infty} \tau_j^{-1}e_j(d,z) \phi_j(x^*)\bigg)f_{X^*}(x^*)d\lambda_1(x^*) = \sum_{j=1}^{\infty} \tau_j^{-1} e_j(d,z) \varsigma_j(d) 
	\end{equation*}
	where $e_j(d,z) = \int f_{X|DZ}(x|d,z) \varphi_j(x) d\lambda(x)$ and $\varsigma_j(d)=\mathbb{E}[\mathbb{E}[\mathscr{T}(Y(d))|X^*]\phi_j(X^*)]$.
\end{lemma}
\begin{proof}
	Note that the infinite sum $\sum_{j=1}^{\infty} \tau_j^{-1} e_j(d,z) \phi_j$ converges in $L_2(F_{X^*})$ as discussed around \eqref{eq: remark}. Since $\mathbb{E}[\mathscr{T}(Y(d))|X^*]\in L_2(F_{X^*})$,
	\begin{align*}
		&\int \mathbb{E}[\mathscr{T}(Y(d))|X^*=x^*] \bigg(\sum_{j=1}^{\infty} \tau_j^{-1}e_j(d,z) \phi_j(x^*)\bigg)f_{X^*}(x^*)d\lambda(x^*)\\
		& = \lim_{N\to\infty} \int \mathbb{E}[\mathscr{T}(Y(d))|X^*=x^*] \bigg(\sum_{j=1}^{N} \tau_j^{-1} e_j(d,z) \phi_j (x^*)\bigg)f_{X^*}(x^*)d\lambda(x^*).
	\end{align*} 
	For each $N$,
	\begin{equation*}
		\int \mathbb{E}[\mathscr{T}(Y(d))|X^*=x^*] \bigg(\sum_{j=1}^{N} \tau_j^{-1} e_j(d,z) \phi_j(x^*)\bigg)f_{X^*}(x^*) d\lambda(x^*) = \sum_{j=1}^{N}\tau_j^{-1} e_j(d,z) \varsigma_j(d) 
	\end{equation*}
	and this infinite sum converges as $\sum_{j=1}^{\infty}\tau_j^{-2}e_j(d,z)^2<\infty$ and $\sum_{j=1}^{\infty}\varsigma_j(d)^2<\infty$. The latter inequality follows from $\mathbb{E}[\mathscr{T}(Y(d))|X^*]\in L_2(F_{X^*})$ and $\{\phi_j\}_{j\geq1}$ being an orthonormal sequence in $L_2(F_{X^*})$. Then, the desired conclusion follows.
\end{proof}
\begin{lemma}\label{lemma: E[Y|D,Z] measurable wrt V}
	Suppose that Assumptions \ref{ASSM: proxy exclusion}, \ref{ASSM: L2 completeness}, \ref{ASSM: regularity} hold, $(d,x^*)\mapsto \mathbb{E}[Y(d)|X^*=x^*]$ is measurable, and that $\{Y(d):d\in\CD\}\inde (D,Z)|X^*$. Then, $\mathbb{E}[Y|D,Z]$ is $(\mathfrak{V},D)$-measurable.
\end{lemma}
\begin{proof}
	Let $c$ be $c(Y(\cdot),d)=Y(d)$ i.e.,\ mapping that takes a function and its argument to the value of the function. Then, $Y=c(Y(\cdot),D)$ and
	\begin{align*}
		\mathbb{E}[Y|D=d,Z] &= \mathbb{E}[c(Y(\cdot),d)|D=d,Z] =\mathbb{E}[\mathbb{E}[c(Y(\cdot),d)|X^*]|D=d,Z ]\\
		&= \int \mathbb{E}[Y(d)|X^*=x^*] f_{X^*|DZ}(x^*|d,Z)d\lambda_1(x^*)
	\end{align*}
	where the second equality follows from $Y(\cdot)\inde (D,Z)|X^*$ and I used $\mathbb{E}[c(Y(\cdot),d)|X^*]=\mathbb{E}[Y(d)|X^*]$. Arguing as in the proof of Theorem \ref{thm:conditional independence},
	\begin{equation*}
		\mathbb{E}[Y|D=d,Z] = \sum_{j=1}^{\infty} \tau_j^{-1}\mathbb{E}[ \mathbb{E}[Y(d)|X^*]\phi_j(X^*) ] \int f_{X|DZ}(x|d,Z) \varphi_j(x)d\lambda(x).
	\end{equation*}
	Letting $\varsigma_{y,j}(d) =  \mathbb{E}[ \mathbb{E}[Y(d)|X^*]\phi_j(X^*) ]$,
	\begin{equation*}
		\mathbb{E}[Y|D,Z] = \sum_{j=1}^{\infty} \tau_j^{-1}\varsigma_{y,j}(D) \int \mathfrak{V}(x) \varphi_j(x)d\lambda(x),
	\end{equation*}
	which proves the desired result.
\end{proof}
The following is a slight modification of a known sufficient condition for bounded completeness.
\begin{lemma}
	Let $X=\chi(X^*+\eta)$ where $X^*\inde \eta$ and $\chi$ is invertible. Suppose $X^*$ and $\eta$ are continuously distributed, and the characteristic function of $\eta$ is non-zero everywhere. Then, the family of distributions $\{f_{X^*|X}(\cdot|x):x\in\CX\}$, where $\CX$ is the support of $X$, is bounded complete.
\end{lemma}
\begin{proof}
	For any bounded function $b$, $\mathbb{E}[b(X^*)|X] = \mathbb{E}[b(X^*)|X^*+\eta]$ almost surely because $\chi$ is invertible. Thus, without loss of generality, assume $\chi$ is the identity function. Since $f_{X^*X}(y,x) = f_{X^*}(y)f_{\eta}(x-y)$,	\begin{equation*}
		\mathbb{E}[b(X^*)|X] = \int b(y)f_{X^*}(y)f_{\eta}(X-y)dy = \int \tilde{b}(y) f_{\eta}(y-X)dy 
	\end{equation*}
	where $\tilde{b}(y) = b(-y)f_{X^*}(-y)$. By Theorem 2.1 of \cite{Mattner1993}, $\int \tilde{b}(y) f_{\eta}(y-X)dy =0$ almost surely implies $\tilde{b}(y)=0$ almost surely. This in turn implies $b(y)=0$ for $y$ such that $f_{X^*}(y) >0 $. Thus, bounded completeness holds.
\end{proof}

\subsection{Proof of Theorem \ref{thm: finite-dimensional V}}
Let $\mathfrak{V}_2$ be the sigma field generated by the collection of maps $\{(D,Z)\mapsto \mathrm{Pr}[X\leq a|D,Z] : a\in \mathbb{R}^{\dim(X)}\}$ where $\dim(X)$ is the dimension of $X$. I first show $\mathfrak{V}_2\supseteq\sigma(\mathfrak{V})$, which implies $\mathfrak{V}_2=\sigma(\mathfrak{V})$.

For any $X$-measurable function $g\geq 0$, there exists a sequence of simple functions $g_n$ such that the sets used to construct $g_n$ are rectangles and $g_n(x)\uparrow g(x)$ for almost all $x$. Then, $\int g_n(x)f_{X|DZ}(x|D,Z)d\lambda(x)\mapsto \int g(x) f_{X|DZ}(x|D,Z)d\lambda(x)$ almost surely by $f_X$ dominating $f_{X|DZ}(\cdot|d,z)$ for almost all $d,z$ and the monotone convergence theorem. Since $\int g_n(x)f_{X|DZ}(x|D,Z)\lambda(x)$ is $\mathfrak{V}_2$-measurable, so is $\int g(x) f_{X|DZ}(x|D,Z)d\lambda(x)$, which implies $\mathfrak{V}_2\supseteq\sigma(\mathfrak{V})$.

Now I show that $\mathfrak{V}_2=\sigma(V)$ where $V=(F_{X|DZ}(x_1|D,Z),\dots,F_{X|DZ}(x_k|D,Z))'$ for $\{x_1,\dots, x_k\}$ specified below.
One inclusion is immediate, and it suffices to show $\mathfrak{V}_2\subseteq\sigma(V)$.
For each $(d,z)$ in the support of $(D,Z)$, pick $(\tilde{x}_1,\dots, \tilde{x}_l)$ satisfying the conditions in Assumption \ref{ASSM: control function dimension} and let $F(D,Z)=(F_{X|DZ}(\tilde{x}_1|D,Z),\dots,F_{X|DZ}(\tilde{x}_l|D,Z))'$. Pick a point $a\in \mathbb{R}^{\dim(X)}$, let $R(d,z) = \mathrm{Pr}[X\leq a|D=d,Z=z]$, and consider the map $G(d,z) = (F(d,z)',R(d,z))'$. The derivative of $G$ has rank less than or equal to $l$ by the hypothesis. Without loss of generality, assume the rank of $\partial G(d,z)/\partial (d,z)'$ is $l$.

The following argument is adapted from a proof of the constant rank theorem \cite[e.g., Theorem 4.12 of][]{lee2003introduction}.
For exposition, write $\tilde{Z}=(D,Z)$ and partition $\tilde{Z}=(\tilde{Z}_1,\tilde{Z}_2)$ so that $\tilde{Z}_1\in\mathbb{R}^{l}$. Without loss of generality, assume that the first $l$ columns of the derivative of $F$ are linearly independent. The map $H:\tilde{Z}\mapsto (F(\tilde{Z}),\tilde{Z}_2)$ has derivative of full rank on some neighborhood of $(d,z)$. By the inverse function theorem, on some open set $U$ containing $(d,z)$, $H$'s inverse exists and is continuously differentiable. Write the inverse as $H^{-1}(\tilde{v},\tilde{z}_2) = (H^1(\tilde{v},\tilde{z}_2),H^b(\tilde{v},\tilde{z}_2))$ where $H^1(\cdot)\in\mathbb{R}^l$. 
Since $\tilde{Z}= H^{-1}(F(\tilde{Z}),\tilde{Z}_2)$, we have $H^b(\tilde{v},\tilde{z}_2)=\tilde{z}_2$. Also, $F\circ H^{-1}(\tilde{v},\tilde{z}_2) = \tilde{v}$ and we have $G\circ H^{-1}(\tilde{v},\tilde{z}_2) = (\tilde{v},\tilde{R}(\tilde{v},\tilde{z}_2))$ where $\tilde{R}(\tilde{v},\tilde{z}_2)=R(H^1(\tilde{v},\tilde{z}_2),\tilde{z}_2)$. Since the rank of the Jacobian of $G\circ H^{-1}$ is $l$, we see that $\tilde{R}$ does not depend on $\tilde{z}_2$ i.e.,\ $G\circ H^{-1}(\tilde{v},\tilde{z}_2) = (\tilde{v},\tilde{R}(\tilde{v}))$.

Now recalling that $(F(D,Z),\tilde{Z}_2) = H(D,Z)$, we have $G(D,Z) = (F(D,Z),\tilde{R}(F(D,Z)))$. Thus, on the neighborhood $U$, $\mathrm{Pr}[X\leq a|D,Z] =\tilde{R}(F(D,Z))$. Note that the choice of the neighborhood $U$ does not depend on the choice of $a$ in $\mathrm{Pr}[X\leq a|D,Z]$. Thus, the above argument shows that for any point $a$, there exists some function $\tilde{R}$ satisfying the equality on $U$.

By the compactness of the support, we can pick a finite number (say $L\in\mathbb{N}$) of $(d,z)'s$ whose associated open sets (denoted by $U_1,\dots, U_L$) cover the support of $(D,Z)$. Collect associated $x$ points and relabel them to form $(x_1,\dots, x_k)$. Define $V=(F_{X|DZ}(x_1|D,Z),\dots,F_{X|DZ}(x_k|D,Z))'\equiv \nu(D,Z)$. 
Fix $a\in\mathbb{R}^{\dim(X)}$. The above argument shows that on each $U_{\ell}$, $\mathrm{Pr}[X\leq a|D,Z] =\tilde{R}(V)$ for some measurable function $\tilde{R}$. It remains to show that it is possible to patch together the functions on $\cup_{\ell=1}^L U_\ell$ preserving $V$-measurability.

If $\nu(U_\ell)\cap \nu(U_{\ell'})$ for some $\ell\neq \ell'$ has positive probability, take $\tilde{R}$ on $U_{\ell}$ and we have  $\mathrm{Pr}[X\leq a|D,Z]\mathbbm{1}\{V\in \nu(U_\ell)\cap \nu(U_{\ell'})\} = \tilde{R}(V)\mathbbm{1}\{V\in \nu(U_\ell)\cap \nu(U_{\ell'})\}$ is $V$-measurable. On $\nu^{-1}( \{\cup_{\ell\neq \ell'}\nu(U_\ell)\cap \nu(U_{\ell'})\}^c )$, $\nu$ is injective, and we can write $\mathrm{Pr}[X\leq a|D,Z]\mathbbm{1}\{V\in \{\cup_{\ell\neq \ell'}\nu(U_\ell)\cap \nu(U_{\ell'})\}^c\}$ as some measurable function of $V$. 

Therefore, $\mathrm{Pr}[X\leq a|D,Z]$ can be expressed as a sum of $V$-measurable functions. Since the choice of $a$ is arbitrary, $\mathfrak{V}_2\subseteq \sigma(V)$.\hfill\qed

\section{Estimation}

\subsection{Proof of Theorem \ref{thm: semiparametric estimation}}
The proof builds on the arguments of \cite{Chernozhukov-Newey-Singh_2022_ecma} (henceforth CNS).
Let $\widehat{\mu}_2(v) = \widehat{\mu}(d+1,v)-\widehat{\mu}(d,v)$, which is the estimate of the coefficient on $D$ i.e., $\mu_0(D+d,V)-\mu_0(D,V)=\mu_{0,2}(V)d$, and 
\begin{align*}
	\varphi(W,\mu,\nu,\alpha) &= \alpha(D,\nu(D,Z))\big[Y-\mu(D,\nu(D,Z)) \big]\\
	&\quad + \left[\frac{\partial m(W,\mu,\nu)}{\partial V'} - \alpha(D,\nu(D,Z))\frac{\partial\mu(D,\nu(D,Z))}{\partial V'} \right][B(X)-\nu(D,Z)].
\end{align*}
Following CNS, use the decomposition
\begin{align*}
	\psi(W,\widehat{\mu}_l,\widehat{\nu}_l,\widehat{\alpha}_l) &= \psi(W,\mu_0.\nu_0,\alpha_0) \\
	&\quad + \widehat{\mu}_{l,2}(\widehat{\nu}_l(D,Z)) - \mu_{0,2}(V)\\
	&\quad + \varphi(W,\widehat{\mu}_l,\widehat{\nu}_l,\alpha_0) - \varphi(W,\mu_0,\nu_0,\alpha_0)\\
	&\quad + \varphi(W,\mu_0,\nu_0,\widehat{\alpha}_l) - \varphi(W,\mu_0,\nu_0,\alpha_0)\\
	&\quad + \varphi(W,\widehat{\mu}_l,\widehat{\nu}_l,\widehat{\alpha}_l) -  \varphi(W,\widehat{\mu}_l,\widehat{\nu}_l,\alpha_0) -\varphi(W,\mu_0,\nu_0,\widehat{\alpha}_l) + \varphi(W,\mu_0,\nu_0,\alpha_0)\\
	&\equiv \psi(W,\mu_0,\nu_0,\alpha_0)  + R_{1l} + R_{2l} + R_{3l} +R_{4l} .
\end{align*}
Lemmas \ref{remainder lemma R1 R2}, \ref{remainder lemma R3}, \ref{remainder lemma R4}, and Assumption \ref{ASSM: estimation first-stage} imply $\sqrt{n}\mathbb{E}[R_{1l}+R_{2l}|I_{l}^c]=o_{\mathbb{P}}(1)$, $\mathbb{E}[R_{3l} |I_{l}^c]=0$,  $\sqrt{n}\mathbb{E}[R_{4l}|I_{l}^c]=o_{\mathbb{P}}(1)$, and $\mathbb{E}[R_{jl}^2|I_{l}^c]=o_{\mathbb{P}}(1)$ for $j=1,2,3,4$. Then, with $n_l/n$ being uniformly bounded away from zero,
\begin{equation*}
	\frac{1}{\sqrt{n}}\sum_{l=1}^L\sum_{i\in I_l} \psi (W_i,\widehat{\mu}_l,\widehat{\nu}_l,\widehat{\alpha}_l)= \frac{1}{\sqrt{n}}\sum_{i=1}^n \psi(W_i,\mu_0,\nu_0,\alpha_0) + o_{\mathbb{P}}(1).
\end{equation*}

For consistency of the variance estimator, note
\begin{align*}
	\big[ \psi(W,\widehat{\mu}_l,\widehat{\nu}_l,\widehat{\alpha}_l)-\widehat{\theta}_n\big]^2 &=  \big[ \psi(W,\mu_0,\nu_0,\alpha_0)-\theta_0\big]^2 + \big[\psi(W,\widehat{\mu}_l,\widehat{\nu}_l,\widehat{\alpha}_l) -\psi(W,\mu_0,\nu_0,\alpha_0)\big]^2\\
	&\quad + \big[\widehat{\theta}_n-\theta_0\big]^2 - 2\big[\psi(W,\widehat{\mu}_l,\widehat{\nu}_l,\widehat{\alpha}_l) -\psi(W,\mu_0,\nu_0,\alpha_0)\big]\big[\widehat{\theta}_n-\theta_0\big]\\
	&\quad +2 \big[ \psi(W,\widehat{\mu}_l,\widehat{\nu}_l,\widehat{\alpha}_l)-\psi(W,\mu_0,\nu_0,\alpha_0)\big]\big[ \psi(W,\mu_0,\nu_0,\alpha_0)-\theta_0\big] \\
	&\quad - 2 \big[\widehat{\theta}_n-\theta_0\big]\big[ \psi(W,\mu_0,\nu_0,\alpha_0)-\theta_0\big]
\end{align*}
and by Lemmas \ref{remainder lemma R1 R2}-\ref{remainder lemma R4} and $\widehat{\theta}_n=\theta_0 +o_{\mathbb{P}}(1)$,
\begin{align*}
	\widehat{\Psi}_n &= \frac{1}{n}\sum_{i=1}^n \big[ \psi(W,\mu_0,\nu_0,\alpha_0)-\theta_0\big]^2 + o_{\mathbb{P}}(1) = \Psi_0 + o_{\mathbb{P}}(1).
\end{align*}

In the sequel, expectations are computed with respect to $\{W_i\in I_l\}$ conditional on observations in $I_l^c$. Thus, I treat $(\widehat{\mu}_l,\widehat{\nu}_l,\widehat{\alpha}_l)$ as fixed. Analysis is done for each fixed $l=1,\dots,L$, so I drop the subscript $l$ to save space. In the next subsection, I derive asymptotic properties of $(\widehat{\mu},\widehat{\alpha})$ given the first-stage estimator $\widehat{\nu}$. Using these results, I compute $\mathbb{E}[R_{jl}|I_l^c],\mathbb{E}[R_{jl}^2|I_l^c]$ for $j=1,2,3,4$ in the following section.
\subsubsection{Lasso estimators}
Recall $\mu_0(d,v) = \mu_{0,1}(v) + d\mu_{0,2}(v)$ and $\widehat{\mu}_2(v) = \widehat{\mu}(d+1,v)-\widehat{\mu}(d,v)$. In this section, I prove the following.
\begin{lemma}\label{lemma: convergence rate of muhat}
	\begin{equation*}
		\int \big|\widehat{\mu}(d,\widehat{\nu}(d,z))-\mu_0(d,\nu_0(d,z))\big|^2 dF_{DZ}(d,z)= O_{\mathbb{P}}\big( (\kappa_n/\omega_n)^2\omega_n^{4\xi/(2\xi+1)} \big),
	\end{equation*}
	\begin{equation*}
		\int \big|\widehat{\mu}_2(\widehat{\nu}(d,z))-\mu_{0,2}(\nu_0(d,z))\big|^2 dF_{DZ}(d,z)= O_{\mathbb{P}}\big( (\kappa_n/\omega_n)^2\omega_n^{4\xi/(2\xi+1)} \big),
	\end{equation*}
	\begin{equation*}
		\int \big|\widehat{\alpha}(d,\widehat{\nu}(d,z))-\alpha_0(d,\nu_0(d,z))\big|^2 dF_{DZ}(d,z)= O_{\mathbb{P}}\big( (\kappa_n/\omega_n)^2\omega_n^{4\xi/(2\xi+1)} \big),
	\end{equation*}
	and for $l=1,\dots, \dim(V)$, if $\Xi_{1,n}(\kappa_n/\omega_n)\omega_n^{2\xi/(2\xi+1)}=o(1)$ and $\Xi_{2,n}\tilde{\omega}_n=o(1)$, then
	\begin{equation*}
		\int \Big|\frac{\partial}{\partial v_l}\big\{\widehat{\mu}(d,\widehat{\nu}(d,z))-\mu_0(d,\nu_0(d,z))\big\} \Big|^2dF_{DZ}(d,z) =o_{\mathbb{P}}(1),
	\end{equation*}
	\begin{equation*}
		\int \Big|\frac{\partial}{\partial v_l}\big\{\widehat{\mu}(d,\widehat{\nu}_2(d,z))-\mu_{0,2}(d,\nu_0(d,z))\big\} \Big|^2dF_{DZ}(d,z) =o_{\mathbb{P}}(1).
	\end{equation*}
	\hfill\qed
\end{lemma}

Analysis of $\widehat{\mu}$ and $\widehat{\alpha}$ is identical, so I focus on $\widehat{\mu}$.
Define $M^{\mu}= \mathbb{E}[q(D,Z)Y]$. Let $s_0=\lfloor C\omega_n^{-2/(2\xi+1)}\rfloor$ and let $\widetilde{\rho}$ be the vector as defined in the assumption \ref{ASSM: estimation approximation error} for $s=s_0$.  Then,
\begin{equation}\label{eq: sparse approximation}
	\|\bar{\rho}-\widetilde{\rho}\|_2^2 \leq C s_0^{-2\xi} \leq C \omega_n^{2\xi/(2\xi+1)}.
\end{equation}
Let $J_0$ be the indices of nonzero elements of $\widetilde{\rho}$. Note $|J_0|=s_0$.
Also, define
\begin{equation*}
	\rho_{*} \in \argmin_{\rho} \bigg\{ (\rho-\bar{\rho})'\Omega(\rho-\bar{\rho}) + 2 \omega_n \sum_{j\in J_0^c} |\rho_j|\bigg\}
\end{equation*} 
and $J_{*}$ be the set of indices of nonzero elements of $\rho_{*}$.

\paragraph{Proof of Lemma \ref{lemma: convergence rate of muhat}}
I present a proof for $\widehat{\mu}$. The same argument goes through for $\widehat{\mu}_2$ and $\widehat{\alpha}$.
By adding and subtracting $q(d,\nu_0(d,z))'\widehat{\rho}$,
\begin{align}\nonumber
	\int \big|\widehat{\mu}(d,\widehat{\nu}(d,z))-\mu_0(d,\nu_0(d,z))\big|^2 dF_{DZ}(d,z)&\leq 2 \int \big|\widehat{\mu}(d,\widehat{\nu}(d,z))-\widehat{\mu}(d,\nu_0(d,z))\big|^2 dF_{DZ}(d,z)\\
	\label{eq: add and subtract}
	&\quad +2 \int \big|\widehat{\mu}(d,v)-\mu_0(d,v)\big|^2 dF_{DV}(d,v)
\end{align}
and for the first term after the inequality, Assumptions \ref{ASSM: estimation first-stage}-\ref{ASSM: estimation basis functions} and \eqref{eq: rhohat Op1} imply
\begin{align*}
	&\int \big( \{q(d,\widehat{\nu}(d,z)) - q(d,\nu_0(d,z))\}'\widehat{\rho}\big)^2 dF_{DZ}(d,z)\\
	&\leq \max_{1\leq l\leq \dim(V)} \sup_{d,v}\bigg| \widehat{\rho}'\frac{\partial  q(d,v)}{\partial v_l} \bigg|^2 \int \bigg| \sum_{j=1}^{\dim(V)} \big(\widehat{\nu}_j(d,z) - \nu_{0,j}(d,z)\big) \bigg|^2dF_{DZ}(d,z) =O_{\mathbb{P}} \big( \omega_n^2 \big) .
\end{align*}	
For the other term, let $\|\cdot\|_{L^2}$ be the $L^2$ norm.  By the triangle inequality,
\begin{equation*}
	\|\widehat{\mu}-\mu_0\|_{L^2} \leq \|\widehat{\mu}-\rho_{*}'q\|_{L^2} + \|\rho_{*}'q - \bar{\rho}'q\|_{L^2} + \|\bar{\rho}'q-\mu_0\|_{L^2}
\end{equation*}
where $\|\bar{\rho}'q-\mu_0\|_{L^2}=o(n^{-1/4})$ by Assumption \ref{ASSM: estimation approximation error}.
Since the largest eigenvalue of $\Omega$ is uniformly bounded,
\begin{equation*}
	\|q'(\widehat{\rho}-\rho_{*})\|_{L^2}^2 = (\widehat{\rho}-\rho_{*})'\Omega (\widehat{\rho}-\rho_{*}) \leq C \|\widehat{\rho}-\rho_{*}\|_2^2 = O_{\mathbb{P}}\big( (\kappa_n/\omega_n) \omega_n^{2\xi/(2\xi+1)}\big)
\end{equation*}
where the last equality uses \eqref{eq: convergence rate rhohat}. By Lemma \ref{lemma: coefficient vectors L2 bounds}, $\|q'(\widehat{\rho}-\rho_{*})\|_{L^2}^2\leq C\omega_n^{2\xi/(2\xi+1)}$ holds. Then,
\begin{equation*}
	\|\widehat{\mu}-\mu_0\|_{L^2} = O_{\mathbb{P}}\Big(  (\kappa_n/\omega_n) \omega_n^{2\xi/(2\xi+1)}\Big) + o\big(n^{-1/4}\big).
\end{equation*}

For the derivative estimators, doing similar calculation to \eqref{eq: add and subtract},
\begin{align*}
	&\int \big( \widehat{\rho}' \partial\{q(d,\widehat{\nu}(d,z)) - q(d,\nu_0(d,z))\}/\partial v_l\big)^2 dF_{DZ}(d,z)\\
	&\leq \max_{1\leq l'\leq \dim(V)} \sup_{d,v}\bigg| \widehat{\rho}'\frac{\partial^2  q(d,v)}{\partial v_l \partial v_{l'}} \bigg|^2 \int \bigg| \sum_{j=1}^{\dim(V)} \big(\widehat{\nu}_j(d,z) - \nu_{0,j}(d,z)\big) \bigg|^2dF_{DZ}(d,z) =O_{\mathbb{P}} \big( \Xi_{2,n}^2\tilde{\omega}_n^2 \big) 
\end{align*}
and it remains to bound $\| \partial (\widehat{\mu}-\mu_0)/\partial V_l\|_{L^2}\leq \| \partial (\widehat{\mu}-q'\rho_{*})/\partial V_l\|_{L^2} + \| \partial (q'\rho_{*}-\mu_0)/\partial V_l\|_{L^2}$.
For the ``bias'' component,
\begin{align*}
	\|\partial (\mu_0 - \rho_{*}'q)/\partial V_l \|_{L^2} &\leq \|\partial (\mu_0 - \bar{\rho}'q)/\partial V_l \|_{L^2} + \| (\partial q /\partial V_l)'(\bar{\rho} - \rho_{*}) \|_{L^2}\\
	&\leq o(1) + \left| (\widetilde{\rho} - \rho_{*})'\mathbb{E}\left[\frac{\partial q(D,V)}{\partial V_l}\frac{\partial q(D,V)'}{\partial V_l}\right](\widetilde{\rho} - \rho_{*})\right|^{1/2}\\
	&\leq o(1) + C\Xi_{1,n}\|\widetilde{\rho} - \rho_{*}\|_2  = o(1) + \Xi_{1,n}\omega_n^{2\xi/(2\xi+1)}
\end{align*}
where I used Assumption \ref{ASSM: estimation approximation error} and Lemma \ref{lemma: coefficient vectors L2 bounds}. 
For the random component,
\begin{equation*}
	\Big\| \frac{\partial q}{\partial V_l}'(\widehat{\rho}-\rho_{*})\Big\|_{L^2}^2 = (\widehat{\rho} - \rho_{*})'\mathbb{E}\left[\frac{\partial q(D,V)}{\partial V_l}\frac{\partial q(D,V)'}{\partial V_l}\right](\widehat{\rho} - \rho_{*}) = O_{\mathbb{P}}\bigg(\Xi_{1,n}^2 \frac{\kappa_n^2}{\omega_n^2} \omega_n^{4\xi/(2\xi+1)} \bigg)
\end{equation*}
where I used Assumption \ref{ASSM: estimation approximation error} and \eqref{eq: convergence rate rhohat}. \hfill\qed 

\subsubsection{Remainder terms}
Write $\eta=B(X)- V$ where $\mathbb{E}[\eta|D,Z]=0$. In the sequel, $ a_n \lesssim b_n$ indicates that there exists a fixed constant $C>0$ such that $a_n\leq C b_n$ where $C$ does not dependent on $n$. Recall that all the expectations are computed with respect to $\{W_i:i\in I_l\}$ conditional on observations not in $I_l$. Thus, I treat $(\widehat{\mu}_l,\widehat{\nu}_l,\widehat{\alpha}_l)$ as if fixed when computing expectations. To save space, I write $\mathbb{E}[\,\cdot\,|I_l^c]=\mathbb{E}_l[\,\cdot\,]$ and drop the subscript $l$ from $\widehat{\mu}_l,\widehat{\nu}_l,\widehat{\alpha}_l, R_{jl}$. 
\begin{lemma}\label{remainder lemma R1 R2}
	$\mathbb{E}_l[R_{1}+R_{2}]=O_{\mathbb{P}}(\tilde{\omega}_n (\kappa_n/\omega_n) \omega_n^{2\xi/(2\xi+1)} +\Xi_{2,n}\tilde{\omega}_n^2)$, $\mathbb{E}_l[R_{1}^2]=o_{\mathbb{P}}(1)$, and $\mathbb{E}_l[R_{2}^2]=o_{\mathbb{P}}(1)$.
\end{lemma}
\begin{proof}
	Since $R_{1} = \widehat{\mu}_2(\widehat{V}) - \mu_{0,2}(V)$, Lemma \ref{lemma: convergence rate of muhat} implies $\mathbb{E}_l[R_{1}^2]=o_{\mathbb{P}}(1)$. For $R_2$, write $\partial \mu(D,V) =\partial \mu(D,V)/\partial V'$ and
	\begin{align*}
		R_2 &= \alpha_0(D,\widehat{V})[Y-\widehat{\mu}(D,\widehat{V})] - \alpha_0(D,V)\varepsilon \\
		&\quad + [\partial\widehat{\mu}_2(\widehat{V})-\alpha_0(D,\widehat{V}) \partial\widehat{\mu}(D,\widehat{V})][B(X)-\widehat{V}] - [\partial\mu_{0,2}(V)-\alpha_0(D,V)\partial\mu_0(D,V)]\eta\\
		&= [\alpha_0(D,\widehat{V}) -\alpha_0(D,V)]\varepsilon + \alpha_0(D,\widehat{V})[\mu_0(D,V)-\widehat{\mu}(D,\widehat{V})] \\
		&\quad + [\partial\widehat{\mu}_2(\widehat{V})-\partial\mu_{0,2}(V)-\alpha_0(D,\widehat{V}) \partial\widehat{\mu}(D,\widehat{V}) + \alpha_0(D,V)\partial\mu_0(D,V)]\eta \\
		&\quad + [\partial\widehat{\mu}_2(\widehat{V})-\alpha_0(D,\widehat{V}) \partial\widehat{\mu}(D,\widehat{V})][V-\widehat{V}].
	\end{align*}
	By the assumption, $\mathbb{E}[\varepsilon^2|D,Z]$, $\alpha_0$, $\partial \mu_0$, $\partial \mu_{0,2}$ and $\mathbb{E}[\|\eta\|_2^2|D,Z]$ are all bounded and $\alpha_0$ is Lipschitz continuous.
	\begin{align*}
		\mathbb{E}_l[R_2^2] &\lesssim \|\widehat{\nu}-\nu_0\|_{L^2}^2 + \int |\widehat{\mu}(d,\widehat{\nu}(d,z))-\mu_0(d,\nu_0(d,z))|^2 dF_{DZ}(d,z) \\
		&\quad + \int \|\partial \{\widehat{\mu}_2(\widehat{\nu}(d,z)) - \mu_{0,2}(\nu_0(d,z))\}\|_2^2 dF_{DZ}(d,z) + \Xi_{1,n}^2 \| \widehat{\nu}-\nu_0\|_{L^2}^2 \\
		&\quad + \int \|\partial \{\widehat{\mu}(d,\widehat{\nu}(d,z)) - \mu_{0}(d,\nu_0(d,z))\}\|_2^2 dF_{DZ}(d,z) = o_{\mathbb{P}}(1)
	\end{align*}
	where I used \eqref{eq: rhohat Op1} to deduce $|\partial \widehat{\mu}(d,v)/\partial v_l|=O_{\mathbb{P}}( \Xi_{1,n} )$.
	
	For $R_1+R_2$, note $\mathbb{E}[\varepsilon|D,Z]=0$ by Lemma \ref{lemma: E[Y|D,Z] measurable wrt V}. Then,
	\begin{align}
		\label{eq: term 1 R1+R2}
		\mathbb{E}_l[R_1+R_2] &=\mathbb{E}_l[-\mu_{0,2}(V)+\alpha_0(D,V)\mu_0(D,V)] + \mathbb{E}_l[\widehat{\mu}_2(V)-\alpha_0(D,V)\widehat{\mu}(D,V)]\\
		\label{eq: term 2 R1+R2}
		&\quad -\mathbb{E}_l[\widehat{\mu}_2(V) - \widehat{\mu}_2(\widehat{V}) - \partial \widehat{\mu}_2(\widehat{V})\{V-\widehat{V}\}] \\
		\label{eq: term 3 R1+R2}
		&\quad + \mathbb{E}_l[\{\alpha_0(D,\widehat{V})-\alpha_0(D,V)\}\{\mu_0(D,V)-\widehat{\mu}(D,\widehat{V})\}]\\
		\label{eq: term 4 R1+R2}
		&\quad - \mathbb{E}_l[\{\alpha_0(D,\widehat{V})-\alpha_0(D,V)\}\partial \widehat{\mu}(D,\widehat{V})\{V-\widehat{V}\} ] \\
		\label{eq: term 5 R1+R2}
		&\quad + \mathbb{E}_l[\alpha_0(D,V)\{\widehat{\mu}(D,V)-\widehat{\mu}(D,\widehat{V})-\partial\widehat{\mu}(D,\widehat{V}) (V-\widehat{V})\}].
	\end{align}
	The right-hand side of \eqref{eq: term 1 R1+R2} equals zero as $\mathbb{E}[\alpha_0(D,V)\mu(D,V)] = \mathbb{E}[m(W,\mu,\nu_0)]$ for any $\mu\in\Gamma$ as shown by CNS. For \eqref{eq: term 5 R1+R2}, by Taylor expansion, for $t \in [0,1]$ (dependent on $(d,z)$),
	{\small\begin{align*}
			\eqref{eq: term 5 R1+R2} &= \int \alpha_0(d,\nu_0(d,z)) (\nu_0(d,z) - \widehat{\nu}(d,z))'\frac{\partial^2 \widehat{\rho}'q(d,t\nu_0(d,z) + (1-t)\widehat{\nu}(d,z) )}{\partial v \partial v'} (\nu_0(d,z) - \widehat{\nu}(d,z)) dF_{DZ}(d,z)\\
			&\leq C \max_{1\leq l,l'\leq \dim(V)} \sup_{d,v} | \widehat{\rho}'\partial^2 q(d,v)/\partial v_l \partial v_{l'} | \|\widehat{\nu} -\nu_0\|_{L^2}^2 = O_{\mathbb{P}}(\Xi_{2,n}\tilde{\omega}_n^2).
	\end{align*}}%
	An analogous argument shows $\eqref{eq: term 2 R1+R2} = O_{\mathbb{P}}(\Xi_{2,n}\tilde{\omega}_n^2)$. For \eqref{eq: term 3 R1+R2}, the Cauchy-Schwartz inequality and Lemma \ref{lemma: convergence rate of muhat} imply it is $O_{\mathbb{P}}(\tilde{\omega}_n (\kappa_n/\omega_n) \omega_n^{2\xi/(2\xi+1)} )$. For \eqref{eq: term 4 R1+R2}, \eqref{eq: rhohat Op1} implies it is $O_{\mathbb{P}}(\Xi_{1,n}\tilde{\omega}_n^2)$.
\end{proof}

\begin{lemma}\label{remainder lemma R3}
	$\mathbb{E}_l[R_3]=0$ and $\mathbb{E}_l[R_3^2]=o_{\mathbb{P}}(1)$.
\end{lemma}
\begin{proof}
	Recall 
	\begin{align*}
		R_3 &= \widehat{\alpha}(D,V) \varepsilon + [\partial \mu_{0,2}(V) -\widehat{\alpha}(D,V)\partial \mu_0(D,V)]\eta \\
		&\quad - \alpha_0(D,V)\varepsilon + [\partial \mu_{0,2}(V)-\alpha_0(D,V)\partial \mu_0(D,V)]\eta.
	\end{align*}
	Using $\mathbb{E}[\varepsilon|D,V]=0$ and $\mathbb{E}[\eta|D,V]=0$, $\mathbb{E}_l[R_3]=0$ almost surely. Also,
	\begin{equation*}
		R_3^2 \lesssim  |\widehat{\alpha}(D,V)-\alpha(D,V)|^2\varepsilon^2 + |\widehat{\alpha}(D,V)-\alpha_0(D,V)|^2 |\partial\mu_0(D,V)\eta|^2
	\end{equation*}
	and $\mathbb{E}_l[R_3^2]=o_{\mathbb{P}}(1)$ follows from $\|\widehat{\alpha}-\alpha_0\|_{L^2}=o_{\mathbb{P}}(1)$ and boundedness of $\mathbb{E}[\varepsilon^2|D,Z]$, $\partial \mu_0$, and $\mathbb{E}[\|\eta\|_2^2|D,Z]$. 
\end{proof}

\begin{lemma}\label{remainder lemma R4}
	$\mathbb{E}_l[R_4]=O_{\mathbb{P}}(\Xi_{0,n}\Xi_{2,n}\tilde{\omega}_n^2+(\kappa_n/\omega_n)^2 \omega_n^{4\xi/(2\xi+1)})$ and $\mathbb{E}_l[R_4^2]=o_{\mathbb{P}}(1)$.
\end{lemma}
\begin{proof}
	Note
	\begin{equation*}
		R_4 = [\widehat{\alpha}(D,\widehat{V})-\alpha_0(D,\widehat{V}) ][Y-\widehat{\mu}(D,\widehat{V})] - [\widehat{\alpha}(D,\widehat{V}) - \alpha_0(D,\widehat{V})]\partial \widehat{\mu}(D,\widehat{V})[B(X)-\widehat{V}]- R_3 .
	\end{equation*}
	By \eqref{eq: rhohat Op1}, $\sup_{d,v}|\widehat{\mu}(d,v)|+\sup_{d,v}|\widehat{\alpha}(d,v)|=O_{\mathbb{P}}(\Xi_{0,n})$. Then, with probability arbitrarily close to one,
	\begin{align*}
		\mathbb{E}_l[(R_4+R_3)^2]&\lesssim \int |\widehat{\alpha}(d,\widehat{\nu}(d,z)) - \alpha_0(d,\nu_0(d,z))|^2 dF_{DZ}(d,z) [1+\Xi_{0,n}+\Xi_{1,n}]^2 \\
		&\quad  +  \|\widehat{\nu}-\nu_0\|_{L^2}^2 [1+\Xi_{0,n}+\Xi_{1,n}]^2.
	\end{align*}
	Thus, $\mathbb{E}_l[R_4^2]=o_{\mathbb{P}}(1)$ holds. For the expectation, note $\mathbb{E}_l[R_3]=0$ and
	\begin{align*}
		\mathbb{E}_l[R_4] &= \mathbb{E}_l\big[ \{\widehat{\alpha}(D,\widehat{V})-\alpha_0(D,\widehat{V})\}\{\widehat{\mu}(D,V)-\widehat{\mu}(D,\widehat{V})-\partial\widehat{\mu}(D,\widehat{V})(V-\widehat{V})\} \big] \\
		&\quad + \mathbb{E}_l\big[ \{\widehat{\alpha}(D,\widehat{V})-\alpha_0(D,\widehat{V})\}\{\mu_0(D,V)-\widehat{\mu}(D,V)\}\big]\\
		&\leq\big( \sup_{d,v} |\widehat{\alpha}(d,v)| +C\big) \mathbb{E}_l\big[|\widehat{\mu}(D,V)-\widehat{\mu}(D,\widehat{V})-\partial\widehat{\mu}(D,\widehat{V})(V-\widehat{V})|\big] \\
		&\quad + \sqrt{\mathbb{E}_l\big[ \{\widehat{\alpha}(D,\widehat{V})-\alpha_0(D,\widehat{V})\}^2\big] \mathbb{E}_l\big[\{\mu_0(D,V)-\widehat{\mu}(D,V)\}^2\big]} 
	\end{align*}
	where the first equality uses $\mathbb{E}[\varepsilon|D,Z]=0, \mathbb{E}[\eta|D,Z]=0$. 
	Using $\sup_{d,v}|\widehat{\alpha}(d,v)|=O_{\mathbb{P}}(\Xi_{0,n})$, calculations similar to those in Lemma \ref{remainder lemma R1 R2} imply $\mathbb{E}_l[R_4]=O_{\mathbb{P}}(\Xi_{0,n}\Xi_{2,n}\tilde{\omega}_n^2 + (\kappa_n/\omega_n)^2 \omega_n^{4\xi/(2\xi+1)} )$.
\end{proof}

\subsubsection{Lemmas}
I use some of lemmas in CNS and \cite{Bradicetal}. For ease of reference, I collect them here.
\begin{lemma}[Lemma A2 CNS]\label{lemma: CNS lemma A2}
	$(\bar{\rho}-\rho_{*})'\Omega(\bar{\rho}-\rho_{*})\leq C \omega_n^{4\xi/(2\xi+1)}$.
\end{lemma}
\begin{lemma}[Lemma A3 CNS]\label{lemma: CNS lemma A3}
	$\#|J_{*}|\leq C \omega_n^{-2/(2\xi+1)}$.
\end{lemma}
\begin{lemma}[Lemma C1 \cite{Bradicetal}] \label{lemma: Bradic lemma C1}
	For any $a\in\mathbb{R}^p$ such that $\|a-b_s\|_2\leq C s^{-r}$ for any $s\geq 0$, where $C,r>0$ are constants and $b_s=\argmin_{\|v\|_0}\|a-v\|_2$. If $r>1/2$ and $s\geq 2$, then $\|a-b_s\|_1\leq D s^{1/2-r}$ where $D>0$ is a constant depending only on $C$ and $r$.
\end{lemma}

I establish additional lemmas to characterize the convergence rates of the Lasso estimators.
\begin{lemma}\label{lemma: coefficient vectors L2 bounds}
	$\|\rho_{*}-\widetilde{\rho}\|_2\leq C \omega_n^{2\xi/(2\xi+1)}$ and $\|\rho_{*}-\bar{\rho}\|_2 \leq C \omega_n^{2\xi/(2\xi+1)}$.
\end{lemma}
\begin{proof}
	Let $J_1=J_0\cup J_{*}$ and note that $\|(\rho_{*})_{J_1}\|_2=\|(\rho_{*})\|_2$, $\|\bar{\rho}_{J_1}\|_2=\|\bar{\rho}\|_2$. Then, $0=\|(\rho_{*}-\widetilde{\rho})_{J_1^c}\|_2\leq 3\|(\rho_{*}-\widetilde{\rho})_{J_1}\|_2$ trivially holds. Lemma \ref{lemma: CNS lemma A3} implies $\#|J_1|\leq C \omega_n^{-2/(2\xi+1)}$. By Assumption \ref{ASSM: estimation basis functions},
	\begin{equation*}
		\|\rho_{*}-\widetilde{\rho}\|_2^2\leq C (\rho_{*}-\widetilde{\rho})'\Omega (\rho_{*}-\widetilde{\rho}) \leq 2 C(\rho_{*}-\bar{\rho})'\Omega (\rho_{*}-\bar{\rho}) +  2 C(\widetilde{\rho}-\bar{\rho})'\Omega (\widetilde{\rho}-\bar{\rho})
	\end{equation*}
	and the last two terms can be bounded by $C \omega_n^{4\xi/(2\xi+1)}$ by Lemma \ref{lemma: CNS lemma A2} and \eqref{eq: sparse approximation}.
	
	For the other conclusion, $\|\rho_{*}-\bar{\rho}\|_2\leq \|\rho_{*}-\widetilde{\rho}\|_2 + \|\widetilde{\rho}-\bar{\rho}\|_2 \leq  C \omega_n^{2\xi/(2\xi+1)}$.
\end{proof}

\begin{lemma}\label{lemma: Bradic lemma B4}
	If $\xi>1/2$, then $\|\rho_{*}-\bar{\rho}\|_1\leq C \omega_n^{(2\xi-1)/(2\xi+1)}$.
\end{lemma}
\begin{proof}
	By $\|\bar{\rho} - \widetilde{\rho}\|_2\leq C s^{-\xi}$ and using the argument of Lemma \ref{lemma: Bradic lemma C1},
	\begin{equation*}
		\|\bar{\rho}_{J_0^c}\|_1\leq C s_0^{1/2-\xi}\leq C \omega_n^{(2\xi-1)/(2\xi+1)}.
	\end{equation*}
	Let $J_1=J_0\cup J_{*}$ and note that $J_1^c\subset J_{*}^c$ and $J_1^{c}\subset J_0^c$. Then,
	\begin{equation*}
		\|(\rho_{*})_{J_1^c}-\bar{\rho}_{J_1^c}\|_1 = \|\bar{\rho}_{J_1^c}\|_1 \leq \|\bar{\rho}_{J_0^c}\|_1.
	\end{equation*}
	By Lemma \ref{lemma: CNS lemma A3}, $\#|J_1|\leq \#|J_{*}|+\#|J_0|\leq C\omega_n^{-2/(2\xi+1)} + s_0 \leq C \omega_n^{-2/(2\xi+1)}$.
	Therefore,
	\begin{align*}
		\|\rho_{*}-\bar{\rho}\|_1 & = \|(\rho_{*})_{J_1}-\bar{\rho}_{J_1}\|_1 + \|(\rho_{*})_{J_1^c}-\bar{\rho}_{J_1^c}\|_1 \leq  \|(\rho_{*})_{J_1}-\bar{\rho}_{J_1}\|_1 +\|\bar{\rho}_{J_0^c}\|_1\\
		&\leq \sqrt{\#|J_1|} \|(\rho_{*})_{J_1}-\bar{\rho}_{J_1}\|_2 + C \omega_n^{(2\xi-1)/(2\xi+1)}\\
		&\leq  C \omega_n^{-1/(2\xi+1)}\|\rho_{*}-\bar{\rho}\|_2+ C \omega_n^{(2\xi-1)/(2\xi+1)}\\
		&\leq C \omega_n^{(2\xi-1)/(2\xi+1)}
	\end{align*}
	where the last inequality follows from Lemma \ref{lemma: coefficient vectors L2 bounds}.
\end{proof}

\begin{lemma}\label{lemma: L1 boundedness coefficient vectors}
	If $\xi>1/2$, $\|\bar{\rho}\|_1\leq C$ and $\|\rho_{*}\|_1\leq C$.
\end{lemma}
\begin{proof}
	If $\max_{1\leq l\leq 2K_n} |\bar{\rho}_{l}|\leq C$ holds, then the conclusion follows by Lemma \ref{lemma: Bradic lemma C1} since $\|\bar{\rho}\|_1\leq \|\bar{\rho}-b_2\| + \|b_2\|_2 \leq C$ where $b_s$ is as defined in Lemma \ref{lemma: Bradic lemma C1}.
	
	Now to show $\max_{1\leq l\leq 2K_n} |\bar{\rho}_{l}|\leq C$, note $\|(\rho_{*})_{J_{*}}\|_2=\|\rho_{*}\|_2$, and thus, $\|(\rho_{*})_{J_{*}^c}\|_2\leq 3\|(\rho_{*})_{J_{*}}\|_2$. Using Assumption \ref{ASSM: estimation approximation error},
	\begin{equation*}
		\rho_{*}'\rho_{*}\leq C \rho_{*}'\Omega\rho_{*}\leq C
	\end{equation*}
	where the last equality holds because $\mathbb{E}[|q(D,V)'(\bar{\rho}-\rho_{*})|^2]=o(1)$ by Lemma \ref{lemma: CNS lemma A2}, and by $\bar{\rho}$ being coefficients of least squares projection, $\mathbb{E}[|q(D,V)'\bar{\rho}|^2]\leq \mathbb{E}[\mu_0(D,V)^2]<\infty$. Since $L^2$ norm of $\rho_{*}$ is uniformly bounded, $\max_{1\leq l\leq 2K_n} |\rho_{*l}|\leq C$.	
	By Lemma \ref{lemma: Bradic lemma B4}, $\|\rho_{*}-\bar{\rho}\|_1=o(1)$ and $\max_{1\leq l\leq 2K_n} |\bar{\rho}_{l}|\leq C$ follows.
	
	By the triangle inequality, $\|\rho_{*}\|_1\leq \|\rho_{*}-\bar{\rho}\|_1+\|\bar{\rho}\|_1\leq C$ also holds.
\end{proof}
Given a vector $A\in\mathbb{R}^L$, let $\|A\|_{\infty}= \max_{1\leq l\leq L}|a_{l}|$.
\begin{lemma}\label{lemma: convergence rate of Omegahat rho}
	$\|\widehat{\Omega}\rho_{*} - \Omega\rho_{*}\|_{\infty} = O_{\mathbb{P}}(\omega_n)$ and $\|\widehat{M}_n^{\mu} -M^{\mu}\|_{\infty} = O_{\mathbb{P}}(\omega_n)$.
\end{lemma}
\begin{proof}
	I first prove $\|\widehat{M}_n^{\mu} -M^{\mu}\|_{\infty} =  O_{\mathbb{P}}(\omega_n)$.
	Let $\widecheck{M}_n^{\mu} = n^{-1}\sum_{i=1}^n q(D_i,V_i)Y_i$. With $s=2$, Lemma \ref{lemma: unif convergence rate} implies 
	\begin{equation*}
		\|\widecheck{M}_n^{\mu}-M^{\mu}\|_{\infty} = O_{\mathbb{P}}\left(\frac{\Xi_{0,n}\log(K_n)}{\sqrt{n}}\right).
	\end{equation*}
	It remains to bound $\|\widehat{M}_n^{\mu}-\widecheck{M}_n^{\mu}\|_{\infty}$.
	\begin{align*}
		\|\widehat{M}_n^{\mu}-\widecheck{M}_n^{\mu}\|_{\infty}&= \max_{1\leq l\leq 2K_n}\bigg|\frac{1}{n}\sum_{i=1}^n  \big[q(D_i,\widehat{V}_i) - q(D_i,V_i)\big]Y_i\bigg| \\
		&\leq  C\max_{1\leq l\leq \dim(V)}\sup_v \left\|\frac{\partial p(v)}{\partial v_l}\right\|_{\infty} \frac{1}{n}\sum_{i=1}^n  \|\widehat{V}_i - V_i\|_1 |Y_i|.
	\end{align*}
	and 
	\begin{equation*}
		\mathbb{E}\Big[\|\widehat{M}_n^{\mu}-\widecheck{M}_n^{\mu}\|_{\infty}\Big]\leq C \Xi_{1,n} \bigg|\int \|\widehat{\nu}_n(d,z) - \nu_0(d,z)\|_2^2 dF_{DZ}(d,z) \mathbb{E}[Y^2]\bigg|^{1/2}  = O_{\mathbb{P}}\big(\Xi_{1,n}\tilde{\omega}_n\big)
	\end{equation*} 
	where the expectation is conditional on $\widehat{\nu}_n$ and observations used to estimate $\widehat{\nu}_n$ are independent of the observations used to form the sum.
	
	Now consider $\widehat{\Omega}_n\rho_{*}$.
	Let $\widecheck{\Omega}_n = n^{-1}\sum_{i=1}^n q(D_i,V_i)q(D_i,V_i)'$. By Lemma \ref{lemma: CNS lemma A2}, $\mathbb{E}[|q(D,V)'(\bar{\rho}-\rho_{*})|^2]=o(1)$, and by $\bar{\rho}$ being coefficients of least squares projection, $\mathbb{E}[|q(D,V)'\bar{\rho}|^2]\leq \mathbb{E}[\mu_0(D,V)^2]<\infty$. Thus, $\mathbb{E}[|q(D,V)'\rho_{*}|^2]\leq C$, and Lemma \ref{lemma: unif convergence rate} implies
	\begin{equation*}
		\|\widecheck{\Omega}_n\rho_{*} -\Omega \rho_{*}\|_{\infty} = O_{\mathbb{P}}\left(\frac{\Xi_{0,n}\log(K_n)}{\sqrt{n}}\right).
	\end{equation*}
	Now,
	\begin{align*}
		\|\widehat{\Omega}_n\rho_{*}-\widecheck{\Omega}_n\rho_{*}\|_{\infty}&\leq \max_{1\leq l\leq 2K_n}\bigg|\frac{1}{n}\sum_{i=1}^n  \big[q_l(D_i,\widehat{V}_i) - q_l(D_i,V_i)\big]q(D_i,\widehat{V}_i)'\rho_{*}\bigg| \\
		&\quad + \max_{1\leq l\leq 2K_n}\bigg|\frac{1}{n}\sum_{i=1}^n q_l(D_i,V_i)\big[q(D_i,\widehat{V}_i) -q(D_i,V_i)\big]'\rho_{*} \bigg|\\
		&\leq C \max_{1\leq \ell\leq \dim(V)}\sup_v \left\|\frac{\partial p(v)}{\partial v_{\ell}}\right\|_{\infty} \frac{1}{n}\sum_{i=1}^n  \|\widehat{V}_i - V_i\|_1 \big|q(D_i,\widehat{V}_i)'\rho_{*}\big|  \\
		&\quad + C \max_{1\leq \ell\leq \dim(V)}\sup_v \left|\frac{\partial p(v)}{\partial v_{\ell}}'\rho_{*}\right| \max_{1\leq l\leq 2K_n}\left|\frac{1}{n}\sum_{i=1}^n  \|\widehat{V}_i - V_i\|_1 |q_l(D_i,V_i)| \right|.
	\end{align*}
	The first term of the right-hand side is $O_{\mathbb{P}}(\omega_n)$ using the argument for $\widehat{M}_n^{\mu}$. For the other term, $\|\rho_{*}\|_1\leq C$ by Lemma \ref{lemma: L1 boundedness coefficient vectors} and
	\begin{align*}
		\max_{1\leq l\leq 2K_n}&\left|\frac{1}{n}\sum_{i=1}^n  \|\widehat{V}_i - V_i\|_1 |q_l(D_i,V_i)| \right| 	\leq \max_{1\leq l\leq 2K_n}\mathbb{E}[\|\widehat{V}_i - V_i\|_1 |q_l(D_i,V_i)|] \\
		&\qquad + \max_{1\leq l\leq 2K_n}\left|\frac{1}{n}\sum_{i=1}^n  \|\widehat{V}_i - V_i\|_1 |q_l(D_i,V_i)| - \mathbb{E}[\|\widehat{V}_i - V_i\|_1 |q_l(D_i,V_i)|] \right|
	\end{align*}
	where the expectations are conditional on the observations used to construct $\widehat{\nu}$, and the result follows from Lemma \ref{lemma: rate remainder} and $\mathbb{E}[\|\widehat{V}_i - V_i\|_1 |q_l(D_i,V_i)|]\leq C \sqrt{\mathbb{E}[\|\widehat{V}_i - V_i\|_1^2]  } = O_{\mathbb{P}}(\tilde{\omega}_n)$.
\end{proof}
Lemma \ref{lemma: convergence rate of Omegahat rho} establishes $\|\widehat{\Omega}\rho_{*} - \Omega\rho_{*}\|_{\infty} = O_{\mathbb{P}}(\omega_n)$, $\|\widehat{M}_n^{\mu} -M^{\mu}\|_{\infty} = O_{\mathbb{P}}(\omega_n)$, and by following the arguments of Lemma A6 of CNS and Lemma B7 of \cite{Bradicetal}, one can show
\begin{equation}\label{eq: convergence rate rhohat}
	\|\widehat{\rho} - \rho_{*}\|_1 =O_{\mathbb{P}}\left((\kappa_n/\omega_n) \omega_n^{(2\xi-1)/(2\xi+1)}\right),\qquad \|\widehat{\rho} - \rho_{*}\|_2 = O_{\mathbb{P}}\left( (\kappa_n/\omega_n) \omega_n^{2\xi/(2\xi+1)} \right).
\end{equation}
Also, this result and Lemma \ref{lemma: L1 boundedness coefficient vectors} imply 
\begin{equation}\label{eq: rhohat Op1}
	\|\widehat{\rho}\|_1 = O_{\mathbb{P}}(1).
\end{equation}

The following uniform convergence rate result borrows ideas from Lemma B.1 of \cite{CattaneoCrumpJansson2013}.
\begin{lemma}\label{lemma: unif convergence rate}
	Let $(X_{i1},\dots, X_{ip},Y_i)$ be i.i.d.\ across $i$ and I write $X_i =(X_{i1},\dots, X_{ip})'$. For some positive sequence $\Xi_n$, $\max_{1\leq i\leq n,1\leq j\leq p}|X_{ij}|\leq \Xi_n$, $\mathbb{E}[Y^2|X]\leq C$, $\mathbb{E}[Y^s]<\infty$ for some $s\geq 2$, and $\max_{1\leq j\leq p}\mathbb{E}[X_{ij}^2]\leq C$. Letting $Z_{ij}=X_{ij}Y_i - \mathbb{E}[X_{ij}Y_i]$,
	\begin{equation*}
		\max_{1\leq j\leq p}\bigg|\frac{1}{n}\sum_{i=1}^n Z_{ij} \bigg| = O_{\mathbb{P}}\bigg( \sqrt{\frac{\log (p)}{n}}\max\bigg\{1, \Xi_n\sqrt{\frac{\log (p)}{n^{1-2/s}}}\bigg\}  \bigg) + o\big( \Xi_n n^{-(s-1)/s}\big).
	\end{equation*}
\end{lemma}
\begin{proof}
	Let $\tau_n = n^{1/s}$ and $Y_{in} = Y_i \mathbbm{1}\{|Y_i|\leq \tau_n\}$.
	\begin{align*}
		\mathrm{Pr}\bigg[  \sum_{i=1}^n X_{ij}Y_i \neq \sum_{i=1}^n X_{ij}Y_{in} \text{ for some } j \bigg] &\leq \mathrm{Pr}\Big[ \max_{1\leq i\leq n} |Y_i| >\tau_n  \Big]\\
		&\leq \sum_{i=1}^n \mathrm{Pr}\big[|Y_i|>\tau_n \big]\\
		&\leq n \mathbb{E}\big[|Y|^s\mathbbm{1}\{|Y|>\tau_n\}\big] \tau_n^{-1/s} = o(1).
	\end{align*}
	For the difference in the expectations,
	\begin{equation*}
		\big|\mathbb{E}[X_{ij}Y_i] - \mathbb{E}[X_{ij}Y_{in}] \big|\leq \Xi_n \mathbb{E}[|Y|\mathbbm{1}\{|Y|>\tau_n \}]\leq \Xi_n \tau_n^{-(s-1)}\mathbb{E}[|Y|^s\mathbbm{1}\{|Y|>\tau_n \}] = o\big( \Xi_n n^{-(s-1)/s}\big).
	\end{equation*}
	Now let $Z_{ijn} = X_{ij}Y_{in} - \mathbb{E}[X_{ij}Y_{in}]$ and for any $c> 0$,
	\begin{align*}
		\mathrm{Pr}\bigg[\max_{1\leq j\leq p}\bigg|\frac{1}{n}\sum_{i=1}^n Z_{ij} \bigg| >c\bigg] &\leq \mathrm{Pr}\bigg[\max_{1\leq j\leq p}\bigg|\frac{1}{n}\sum_{i=1}^n Z_{ijn} \bigg| >c\bigg] + \mathrm{Pr}\bigg[  \sum_{i=1}^n X_{ij}Y_i \neq \sum_{i=1}^n X_{ij}Y_{in} \text{ for some } j \bigg]\\
		&\quad + \mathbbm{1}\Big\{ \max_{1\leq j\leq p}\big|\mathbb{E}[X_jY\mathbbm{1}\{|Y|>\tau_n\}]\big| > c \Big\}\\
		&\leq \mathrm{Pr}\bigg[\max_{1\leq j\leq p}\bigg|\frac{1}{n}\sum_{i=1}^n Z_{ijn} \bigg| >c\bigg] + o(1)\\
		&\quad  + \mathbbm{1}\Big\{ \Xi_n n^{-(s-1)/s}\mathbb{E}[|Y|^s\mathbbm{1}\{|Y|>\tau_n \}]   > c \Big\}.
	\end{align*}
	Thus, the desired result follows if $\max_{1\leq j\leq p}|\frac{1}{n}\sum_{i=1}^n Z_{ijn} | = O_{\mathbb{P}}(\varrho_n)$ with
	\begin{equation*}
		\varrho_n = \sqrt{\frac{\log (p)}{n}}\max\bigg\{1, \Xi_n\sqrt{\frac{\log (p)}{n^{1-2/s}}}\bigg\}.
	\end{equation*}	
	Note $\mathbb{E}[Z_{ijn}^2]\leq2\mathbb{E}[X_{ij}^2Y_i^2]=2\mathbb{E}[X_{ij}^2\mathbb{E}[Y_i^2|X_i]]\leq C\mathbb{E}[X_{ij}^2]\leq C$. Since $\max_{1\leq i\leq n}|Z_{ijn}|\leq 2\tau_n\Xi_n$, Bernstein's inequality implies that for any $M>0$,
	\begin{align*}
		\mathrm{Pr}\bigg(\max_{1\leq j\leq p}\bigg|\frac{1}{n}\sum_{i=1}^n Z_{ijn}\bigg|> \varrho_n M\bigg) &\leq \sum_{j=1}^p \mathrm{Pr}\bigg(\bigg|\sum_{i=1}^n Z_{ijn}\bigg|> n\varrho_n M\bigg)\\
		&\leq   2\exp\bigg( \log (p)- \frac{n(\varrho_n M)^2/16}{ C + \tau_n \Xi_nM \varrho_n /6}\bigg).
	\end{align*}
	Now, suppose $\limsup_{n}\frac{\Xi_n^2\log (p)}{n^{1-2/s}}  < \infty$. Then, $\varrho_n =O( \sqrt{\log (p)/n})$, and
	\begin{align*}
		\exp\bigg( - \frac{n(\varrho_nM) ^2/16}{ C + \tau_n \Xi_n \varrho_nM /6}\bigg) &\leq \exp\bigg(- \frac{M\log (p)/16}{CM^{-1} + \tau_n \Xi_n \sqrt{\log (p)/n}/6}\bigg) \\
		&\leq  \exp\Big(- M\log (p)/(C+1)\Big) 
	\end{align*}
	where the last inequality uses $\tau_n \Xi_n \sqrt{\log (p)/n}\leq C$. 
	
	Next, if $\liminf_{n}\frac{\Xi_n^2\log (p)}{n^{1-2/s}}  > 0$, then $\varrho_n =O( \Xi_n\log (p)/n^{1-1/s})$ and
	\begin{align*}
		\exp\bigg( - \frac{n(\varrho_nM) ^2/16}{ C + \tau_n \Xi_n \varrho_nM /6}\bigg) &\leq \exp\bigg(-\frac{M\Xi_n^2\log (p)^2 /16n^{1-2/s}}{ CM^{-1} +  \Xi_n^2 \log (p)/6n^{1-2/s}}\bigg)\\ 
		&= \exp\bigg(- \frac{M\log (p)/16}{Cn^{1-2/s}/(M\log (p) \Xi_n^2)+1/6 }\bigg)\\
		&\leq \exp\Big(- M\log (p)/(C+1)\Big) 
	\end{align*}
	where the last inequality uses $\limsup_{n} \frac{n^{1-2/s}}{\Xi_n^2\log (p)} = (\liminf_n\frac{\Xi_n^2\log (p)}{n^{1-2/s}})^{-1}  <\infty$. 
	Thus, for both cases, by taking $M$ large enough,
	\begin{equation*}
		\mathrm{Pr}\bigg(\max_{1\leq j\leq p}\bigg|\frac{1}{n}\sum_{i=1}^n Z_{ijn}\bigg|> \varrho_n M\bigg) = o(1).
	\end{equation*}
	For the case where $\limsup_{n}\frac{\Xi_n^2\log (p)}{n^{1-2/s}}  = \infty, \liminf_{n}\frac{\Xi_n^2\log (p)}{n^{1-2/s}}=0$, arguing along subsequences and using the above two cases indicates the desired result.
\end{proof}

\begin{lemma}\label{lemma: rate remainder}
	If $\tilde{\omega}_n (\log (K_n) +\Xi_{0,n}) =O(1)$, then
	\begin{equation*}
		\max_{1\leq l\leq 2K_n}\left|\frac{1}{n}\sum_{i=1}^n  \|\widehat{V}_i - V_i\|_1 |q_l(D_i,V_i)| - \mathbb{E}[\|\widehat{V}_i - V_i\|_1 |q_l(D_i,V_i)|] \right| = O_{\mathbb{P}}\bigg(\sqrt{\frac{\log (K_n)}{n}} +\Xi_{0,n} \tilde{\omega}_n \bigg) 
	\end{equation*}
	where the expectations are computed conditional on $\widehat{\nu}_n$.
\end{lemma}
\begin{proof}
	The argument is analogous to the one for Lemma \ref{lemma: unif convergence rate}. In the notation of the lemma, $Y_i = \|\widehat{V}_i -V_i\|_1$, $X_{ij}=|q_j(D_i,V_i)|$.
	Let $\tau_n = C n^{1/2} \tilde{\omega}_n$ for some large $C>0$, and $n \mathbb{E}[\|\widehat{V}_i -V_i\|_1^2] \tau_n^{-2}= C^{-2} O_{\mathbb{P}}(1)$. Also, $\mathbb{E}[|q_j(D_i,V_i)|\|\widehat{V}_i -V_i\|_1\mathbbm{1}\{\|\widehat{V}_i -V_i\|_1>\tau_n\}]\leq \Xi_{0,n} \tilde{\omega}_n$. Letting $Z_{ijn}=X_{ij}Y_i\mathbbm{1}\{|Y_i|\leq \tau_n\} - \mathbb{E}[X_{ij}Y_i\mathbbm{1}\{|Y_i|\leq \tau_n\}]$, $\mathbb{E}[Z_{ijn}^2]\leq C \Xi_{0,n}^2 O_{\mathbb{P}}(\tilde{\omega}_n^2)$. Then, with $C$ large enough, $\mathbb{E}[Z_{ijn}^2] \leq C \Xi_{0,n}^2\tilde{\omega}_n^2$ occurs with probability close to one. On this event,
	\begin{align*}
		\mathrm{Pr}\left(\max_{1\leq j\leq 2K_n}\bigg|\frac{1}{n}\sum_{i=1}^n Z_{ijn}\bigg| > M\sqrt{\frac{\log (K_n)}{n} } \right)&\leq 4 \exp\bigg( \log (K_n)- \frac{ M^2 \log (K_n)/16}{ C\Xi_{0,n}^2\tilde{\omega}_n^2 + \tau_n M \sqrt{\log (K_n)/n} /6}\bigg)\\
		&\leq 4 \exp\Big( - \log (K_n) \Big) = o(1)
	\end{align*}
	where the last equality holds if $M$ and $n$ are sufficiently large.
\end{proof}

\subsection{Flexible parametric estimation}
Recall the setup
\begin{align*}
	\vartheta(d) & = \mathbb{E}[\mathbb{E}[Y|D=d,V]]\\
	\mathbb{E}[Y|D,V] &= \Lambda\big( p(D,V)'\gamma_0 \big)\\
	\mathbb{E}[B(X)|D,Z] &= Q(D,Z)\delta_0.
\end{align*}
Define the regression residuals $\varepsilon= Y-\mathbb{E}[Y|D,V], \eta=B(X)-\mathbb{E}[B(X)|D,Z]$. I impose the following conditions.
\begin{assumption}\label{assm:paramteric}
	Let $\theta=(\delta,\gamma)$ and $\mathcal{N} = \{\theta :\|\delta-\delta_0\|\vee\|\gamma-\gamma_0\|\leq \eta\}$ for some $\eta >0$.
	\begin{enumerate}[(i)]
		\item $\{Y(d):d\in\mathcal{D}\} \inde (D,Z)|X^*$.
		\item  \label{assm:parametric_smoothness}
		$\Lambda$ is strictly monotone and twice continuously differentiable with bounded derivatives, and $p$ is continuously differentiable in $V$.
		
		\item \label{assm:parametric_nonsingular}
		The matrices $\Gamma_2=\mathbb{E}[\dot{\Lambda}(p(D,V)'\gamma_0)p(D,V)p(D,V)']$, $\Gamma_1=\mathbb{E}[Q(D,Z)'Q(D,Z)]$ are non-singular where $\dot{\Lambda}$ denotes the first derivative of $\Lambda$.

		\item  $\mathbb{E}[\sup_{\theta\in\mathcal{N}}\|\{Y-\Lambda(p(D,Q(D,Z)\delta)'\gamma)\}p(D,Q(D,Z)\delta)\|^2]$, $\mathbb{E}[\sup_{\theta\in\mathcal{N}}\|Q(D,Z)'(X-Q(D,Z)\delta)\|^2]$, $\mathbb{E}[\sup_{\theta\in\mathcal{N}}|\Lambda(p(D,Q(D,Z)\delta)'\gamma)|^2 ]$, $\mathbb{E}[\sup_{\theta\in\mathcal{N}}\|p(D,Q(D,Z)\delta)p(D,Q(D,Z)\delta)'\| ]$, and \\ $\mathbb{E}[\sup_{\theta\in\mathcal{N}}|Y-\Lambda(p(D,Q(D,Z)\delta)'\gamma)| \{\|p(D,Q(D,Z)\delta)\| +1\} \|\frac{\partial p(D,Q(D,Z)\delta)}{\partial V'}Q(D,Z)\| ]$  are all finite.
	\end{enumerate}
\end{assumption}
\begin{theorem}
	Under Assumption \ref{assm:paramteric}, the flexible parametric estimator satisfies
	\begin{equation*}
		\sqrt{n}\big(\widehat{\beta}_n(d) - \beta(d) \big)\leadsto \mathrm{Normal}(0,\Psi(d))
	\end{equation*}
	where
	\begin{equation*}
		\Psi(d) = \mathrm{Var}\big[\Lambda(p(d,V)'\gamma_0) -\beta(d) + c_1(d)\Gamma_2^{-1} P\dot{\Lambda}(P'\gamma_0)\varepsilon+ \{c_1(d)\Gamma_2^{-1}\Gamma_3 +c_2(d)\}\Gamma_1^{-1}Q'\zeta\big],
	\end{equation*}
	$P=p(D,V)$, $Q=Q(D,Z)$, and
	\begin{align*}
		c_1(d) &= \mathbb{E}[\dot{\Lambda}(p(d,V)'\gamma_0)  p(d,V)']\\
		c_2(d) &= \mathbb{E}\Big[\dot{\Lambda}(p(d,V)'\gamma_0)\gamma_0' \frac{\partial p(d,V)}{\partial V'}Q\Big]\\
		\Gamma_3 &= -\mathbb{E}\Big[ |\dot{\Lambda}(P'\gamma_0)|^2 P\gamma_0'\frac{\partial p(D,V)}{\partial V'}Q  \Big].
	\end{align*}
	In addition, the variance estimator in the main paper is consistent for $\Psi(d)$.
\end{theorem}

\subsubsection{Proof}
In the sequel, I refer to \cite{NeweyMcfadden1994} as NM. Note that $\widehat{\delta}_n\to_{\mathbb{P}}\delta_0$ follows from standard arguments. Using Lemma 2.4 of NM, one can show $\widehat{\gamma}_n\to_{\mathbb{P}}\gamma_0$ and $\widehat{\beta}_n(d)\to_{\mathbb{P}}\beta(d)$. Then, asymptotic normality follows from Theorem 6.1 of NM where the moment function is
\begin{equation*}
	\left[\begin{array}{c}
		Q(D,Z)'(X- Q(D,Z)\delta)\\
		\{Y-\Lambda(p(D,Q(D,Z)\delta)'\gamma) \}\dot{\Lambda} (p(D,Q(D,Z)\delta)'\gamma) p(D,Q(D,Z)\delta)\\
		\Lambda(p(d,Q(D,Z)\delta)'\gamma) -\beta
	\end{array} \right].
\end{equation*}
The derivative of this moment condition with respect to $(\delta, \gamma, \beta)$ is
\begin{equation*}
	\left[\begin{array}{ccc}
		-Q(D,Z)'Q(D,Z) & 0 & 0\\
		\varphi_1(Y,D,Z,\delta,\gamma)& \varphi_2(Y,D,Z,\delta,\gamma) & 0\\
		\dot{\Lambda}(p(d,Q(D,Z)\delta)'\gamma) \gamma'\frac{\partial p(d,Q(D,Z)\delta)}{\partial V'} Q(D,Z) & \dot{\Lambda}(p(d,Q(D,Z)\delta)'\gamma) p(d,Q(D,Z)\delta)' & -1
	\end{array} \right]
\end{equation*}
where
\begin{align*}
	\varphi_1(Y,D,Z,\delta,\gamma)=& \big[ \{Y-\Lambda( p(D,Q(D,Z)\delta)'\gamma)\} \ddot{\Lambda} (p(D,Q(D,Z)\delta)'\gamma)-\big|\dot{\Lambda} (p(D,Q(D,Z)\delta)'\gamma)\big|^2   \big]\\
	& \times p(D,Q(D,Z)\delta)\gamma'\frac{\partial p(D,Q(D,Z)\delta)}{\partial V'}Q(D,Z) \\
	&+  \{Y-\Lambda( p(D,Q(D,Z)\delta)'\gamma)\} \dot{\Lambda} (p(D,Q(D,Z)\delta)'\gamma) \frac{\partial p(D,Q(D,Z)\delta)}{\partial V'}Q(D,Z),
\end{align*}
\begin{align*}
	\varphi_2(Y,D,Z,\delta,\gamma) =& \big[ \{Y-\Lambda( p(D,Q(D,Z)\delta)'\gamma)\} \ddot{\Lambda} (p(D,Q(D,Z)\delta)'\gamma)-\big|\dot{\Lambda} (p(D,Q(D,Z)\delta)'\gamma)\big|^2   \big] \\
	&\times p(D,Q(D,Z)\delta)p(D,Q(D,Z)\delta)',
\end{align*}
and $\ddot{\Lambda}$ is the second derivatives of $\Lambda$.
By the formula given in NM, the asymptotic linear representation of $\widehat{\beta}_n(d)$ is
\begin{align*}
	\sqrt{n}\big(\widehat{\beta}_n(d) - \beta(d)\big) =& \frac{1}{\sqrt{n}} \sum_{i=1}^n \{ \Lambda(p(d,V)'\gamma_0) -\beta(d)\} \\
	&- \frac{1}{\sqrt{n}} \sum_{i=1}^n G_{12}G_{11}^{-1} \left[ \begin{array}{c} Q(D_i,Z_i)'\zeta_i \\ p(D_i,V_i) \dot{\Lambda}(p(D_i,V_i)'\gamma_0) \varepsilon_i  \end{array} \right] + o_{\mathbb{P}}(1)
\end{align*}
where 
\begin{equation*}
	G_{12} = \left[\begin{array}{cc} \mathbb{E}\big[\dot{\Lambda} (p(d,V)'\gamma_0) \gamma_0'\frac{\partial p(d,V)}{\partial V'} Q(D,Z)\big] & \mathbb{E}\big[\dot{\Lambda}(p(d,V)'\gamma_0)  p(d,V)'\big] \end{array}\right]
\end{equation*}
and
\begin{equation*}
	G_{11} = \left[\begin{array}{cc}
		-\mathbb{E}[Q(D,Z)'Q(D,Z)] & 0 \\
		\mathbb{E}[\varphi_1(Y,D,Z,\delta_0,\gamma_0)]& \mathbb{E}[\varphi_2(Y,D,Z,\delta_0,\gamma_0)] \end{array} \right].
\end{equation*}
Using the block matrix inverse formula,
\begin{align*}
	&-G_{12}G_{11}^{-1} \left[ \begin{array}{c} Q(D_i,Z_i)'\zeta_i \\ p(D_i,V_i) \dot{\Lambda}(p(D_i,V_i)'\gamma_0) \varepsilon_i  \end{array} \right]\\
	&=  \mathbb{E}\Big[\dot{\Lambda}(p(d,V)'\gamma_0)\gamma_0' \frac{\partial p(d,V)}{\partial V'}Q(D,Z)\Big] \mathbb{E}[Q(D,Z)'Q(D,Z)]^{-1}  Q(Z_i,D_i)'\zeta_i \\
	&\quad+ \mathbb{E}[\dot{\Lambda}(p(d,V)'\gamma_0)  p(d,V)']\mathbb{E}[  |\dot{\Lambda}(p(D,V)'\gamma_0)|^2 p(D,V)p(D,V)']^{-1} p(D_i,V_i)\dot{\Lambda}(p(D_i,V_i)'\gamma_0)\varepsilon_i\\
	&\quad + \mathbb{E}[\dot{\Lambda}(p(d,V)'\gamma_0)p(d,V)'] \mathbb{E}[  |\dot{\Lambda}(p(D,V)'\gamma_0)|^2 p(D,V)p(D,V)']^{-1} \mathbb{E}[\varphi_1(Y,D,Z,\delta_0,\gamma_0)]\\
	&\qquad\times \mathbb{E}[Q(D,Z)'Q(D,Z)]^{-1} Q(Z_i,D_i)'\zeta_i .
\end{align*}
Consistency of the variance estimator follows from repeated applications of Lemma 4.3 of NM.

\section{Comparison with the integral equation approach}

\subsection{Identifying assumptions}
For the integral equation approach, the main identifying assumptions are Assumptions \ref{ASSM: conditional independence outcome}, \ref{ASSM: proxy exclusion}, \ref{ASSM: L2 completeness}, and the following conditions (I omit regularity conditions for simplicity). 
\begin{assumptionIE}\label{ASSM: high-level condition integral equation}
	Given $d\in\CD$, define the operator
	\begin{equation*}
		\Pi_d:L_2(F_{X|D=d})\mapsto L_2(F_{Z|D=d}),\quad \Pi_d(h)(z) = \mathbb{E}[h(X)|D=d,Z=z]
	\end{equation*}
	Assume this linear operator $\Pi_d$ is compact, and denote the singular values by $\{\pi_{j,d}\}_{j\geq 1}$ and associated singular-value functions $\{v_{j,d}\}_{j\geq 1}\subset L_2(F_{X|D=d})$, $\{u_{j,d}\}_{j\geq 1}\subset L_2(F_{Z|D=d})$. See Theorem 15.16 of \cite{Kress2014}. Write $\pi_j(d)\equiv \pi_{j,d}$, $v_{j}(d,x)\equiv v_{j,d}(x)$, and $u_{j}(d,z)\equiv u_{j,d}(z)$. Assume
	\begin{equation*}
		\sum_{j=1}^{\infty}\frac{1}{\pi_j(d)^2}\mathbb{E}\big[ \mathbb{E}[Y|D,Z]u_j(D,Z)\big|D=d\big]^2 <\infty.
	\end{equation*}
\end{assumptionIE}
\begin{assumptionIE}\label{ASSM: IE2}
	For any $b\in L^2(F_{X^*})$ and $d\in\CD$, $\mathbb{P}[\mathbb{E}[b(X^*)|D,Z]=0|D=d]=1$ implies $\mathbb{E}[b(X^*)|D]=0$ with probability one.
\end{assumptionIE}
Assumption \ref{ASSM: IE2} is a weak version of a completeness condition. For reference, a standard completeness conditional on $D$ is 
\begin{assumptionC}
	For any $b\in L^2(F_{X^*})$ and $d\in\CD$, $\mathrm{Pr}[\mathbb{E}[b(X^*)|D,Z]=0|D=d]=1$ implies $b(X^*)=0$ with probability one.
\end{assumptionC}
Since this completeness implies Assumption \ref{ASSM: IE2}, I refer to the latter as a weak version of completeness.

\cite{Deaner2021} used a standard completeness condition rather than Assumption \ref{ASSM: IE2}. \cite{Miao-Geng-TchetgenTchetgen} similarly imposed a standard completeness condition on $X$ given $(D,Z)$. Yet, for the integral equation approach, the weak version of Assumption \ref{ASSM: IE2} suffices for the identification of the ASF (and its conditional version given $D$).\footnote{An anonymous referee pointed out this sufficiency of the weak version of completeness conditions. They also proved and shared a version of Lemma \ref{lemma: common support condition implies weak completeness} in a report. I thank them for very helpful feedback.}

The following result shows that Assumption \ref{ASSM: IE2} follows from the assumptions imposed by the control function method.
\begin{lemma}\label{lemma: common support condition implies weak completeness}
	Under Assumptions \ref{ASSM: proxy exclusion}-\ref{ASSM: support condition}, Assumption \ref{ASSM: IE2} holds.
\end{lemma}
\begin{proof}
	As shown in the proof of Theorem \ref{thm:conditional independence}, 
	\begin{equation*}
		\frac{f_{X^*|DZ}(x^*|D,Z)}{f_{X^*}(x^*)}=\Pi^{\dagger}\bigg(\frac{f_{X|DZ}(\cdot|D,Z)}{f_X(\cdot)}\bigg)(x^*)
	\end{equation*}
	for some fixed mapping $\Pi^{\dagger}$. By Assumption \ref{ASSM: support condition}, for each $d\in\mathcal{D}$, there exists a function $\phi_d$ such that
	\begin{equation*}
		\frac{f_{X|DZ}(x|D,Z)}{f_X(x)} = \frac{f_{X|DZ}(x|d,\phi_d(D,Z))}{f_X(x)}  \qquad \forall x \text{ s.t. } f_{X}(x)>0
	\end{equation*}
	holds with probability one. Then,
	\begin{equation*}
		f_{X^*|DZ}(x^*|D,Z)= f_{X^*|DZ}(x^*|d,\phi_d(D,Z)) \qquad \forall x^* \text{ s.t. } f_{X^*}(x^*)>0
	\end{equation*}
	with probability one. Now, suppose that with $F_{Z|D=d}$ probability one,
	\begin{equation}\label{eq: IE2 hypothesis}
		0=\mathbb{E}[b(X^*)|D=d,Z]=\int b(x^*)f_{X^*|DZ}(x^*|d,Z) dx^*.
	\end{equation}
	Then, $\int b(x^*)f_{X^*|DZ}(x^*|D,Z)dx^*=0$ holds with probability one; otherwise, 
	\begin{equation*}
		0\neq \int b(x^*)f_{X^*|DZ}(x^*|D,Z)dx^* = \int b(x^*)f_{X^*|DZ}(x^*|d,\phi_d(D,Z))dx^*
	\end{equation*}
	with some positive probability. That is, there exists a positive probability set $\CA\subseteq \CZ_d$ (recall that $\CZ_d$ is the conditional support of $Z$ given $D=d$) such that
	\begin{equation*}
		\int b(x^*)f_{X^*|DZ}(x^*|d,z)dx^*\neq 0 \qquad \forall z\in \CA
	\end{equation*}
	but this contradicts the hypothesis \eqref{eq: IE2 hypothesis}. Therefore, if $\int b(x^*)f_{X^*|DZ}(x^*|d,Z)dx^*=0$ holds with $F_{Z|D=d}$ probability one, then $\mathbb{E}[b(X^*)|D]=0$ almost surely follows.
\end{proof}
Given this result, the main difference in the control function approach and the integral equation approach is Assumption \ref{ASSM: support condition} versus Assumption \ref{ASSM: high-level condition integral equation}.
This is indeed the key distinction because of the two results in the next section.
First, I show that one can achieve the identification by replacing Assumption \ref{ASSM: support condition} with Assumption \ref{ASSM: IE2} and a high-level condition that is analogous to Assumption \ref{ASSM: high-level condition integral equation}.
Second, I show that in some model, Assumption \ref{ASSM: high-level condition integral equation} fails while Assumptions \ref{ASSM: conditional independence outcome}-\ref{ASSM: support condition} hold.
Before showing these results, I first elaborate on the relationship between the common support condition and completeness conditions.

\paragraph{Common support condition and completeness condition}
Lemma \ref{lemma: common support condition implies weak completeness} indicates that the common support condition implies a weak version of completeness. Yet, the common support condition does not imply standard completeness conditions. To see this claim, consider $D=\mathbbm{1}\{ZX^*>\eta \}$ where $X^*,\eta \sim\mathrm{Normal}(0,1)$, $\mathrm{Pr}[Z=1]=2/3=1-\mathrm{Pr}[Z=-1]$, and $(Z, X^*,\eta)$ are mutually independent. Also, $X=X^*+U$ with $U\inde (X^*,Z,\eta)$ and $U\sim\mathrm{Normal}(0,1)$. Then, we have
\begin{equation*}
	f_{X^*|DZ}(x^*|d,z) = \frac{\phi(x^*) \Phi(zx^*)^d \Phi(-zx^*)^{1-d} }{\mathbb{E}[\Phi(X^*)]}
\end{equation*}
where $\phi,\Phi$ are the pdf and CDF of standard normal, respectively.
The common support condition holds because $f_{X^*|D,Z}(x^*|1,z)=f_{X^*|D,Z}(x^*|0,-z)$ for all $x^*$ (and by the existence of a one-to-one mapping between $f_{X^*|D,Z}(\cdot|d,z)$ and $f_{X|DZ}(\cdot|d,z)$), but bounded completeness fails as $\mathbb{E}[\cos(X^*)|D,Z]$ equals a constant  almost surely.

Also, the common support condition is not implied by a completeness condition. This fact can be seen from the threshold crossing model discussed in the main text.

An implication of the above fact is that identifying assumptions of my method and those of \cite{HuSchennach2008} are non-nested since the approach of \cite{HuSchennach2008} requires a completeness condition of $(X^*,Z)$ conditional on $D$ with respect to $Z$.

\subsection{Replacing Assumption \ref{ASSM: support condition} with a high-level condition and Assumption \ref{ASSM: IE2}}
As in Lemma \ref{lemma: E[Y|D,Z] measurable wrt V}, under Assumptions \ref{ASSM: conditional independence outcome}-\ref{ASSM: regularity} and $\{Y(d):d\in\CD\}\inde (D,Z)|X^*$,
\begin{equation*}
	\mathbb{E}[Y|D=d,Z] = \sum_{j=1}^{\infty} \tau_j^{-1} \mathbb{E}[ \mathbb{E}[Y(d)|X^*]\phi_j(X^*) ] \int f_{X|DZ}(x|d,Z)\varphi_j(x) d\lambda(x).
\end{equation*}
Suppose we impose the following condition
\begin{equation}\label{eq: high-level condition on X*}
	\sum_{j=1}^{\infty} \frac{1}{\tau_j^2} \mathbb{E}[ \mathbb{E}[Y(d)|X^*]\phi_j(X^*) ]^2 < \infty,
\end{equation}
which is analogous to Assumption \ref{ASSM: high-level condition integral equation}.
With this condition imposed, the infinite sum and the integral with respect to $x$ in the above display can be interchanged to obtain
\begin{align*}
	\mathbb{E}[Y|D=d,Z] &= \int \bigg(\sum_{j=1}^{\infty} \tau_j^{-1}\mathbb{E}[ \mathbb{E}[Y(d)|X^*]\phi_j(X^*) ] \varphi_j(x) \bigg) f_{X|DZ}(x|d,Z)d\lambda(x) \\
	&\equiv \mathbb{E}[\mathfrak{B} (d,X) |D=d,Z]
\end{align*}
where the function $\mathfrak{B}$ is what \cite{Miao-Geng-TchetgenTchetgen} calls a bridge function. With Assumption \ref{ASSM: IE2}, the ASF can be identified via $\mathbb{E}[\mathfrak{B}(d,X)]$. Note that the above identification argument did not use Assumption \ref{ASSM: support condition} and thus \eqref{eq: high-level condition on X*} can replace Assumption \ref{ASSM: support condition} (along with Assumption \ref{ASSM: IE2}) to achieve the identification. Yet, as discussed in the main paper, verifying a high-level condition like \eqref{eq: high-level condition on X*} seems difficult in practice.

\subsection{Counterexample}
I show that the high-level condition of the integral equation (Assumption \ref{ASSM: high-level condition integral equation}) fails to hold in the following simple example.
\begin{equation*}
	Y = \mathbbm{1}\{\beta D   \geq X^* \},\qquad X = X^*+U_x ,\qquad Z = X^*+U_z 
\end{equation*}
where
\begin{equation*}
	\left[ \begin{array}{c} X^*\\ D  \end{array}\right] \sim \mathrm{Normal}\left(\left[\begin{array}{c} 0\\ 0\end{array}\right], \left[\begin{array}{cc} \sigma_{x}^2 & \sigma_{x}\sigma_d \rho  \\ \sigma_{x}\sigma_d \rho&\sigma_d^2	\end{array}\right] \right),
\end{equation*}
$(X^*,D)\inde (U_x,U_z)$, $U_x\inde U_z$, and $U_x,U_z$ each follows a standard normal distribution. We have
\begin{equation*}
	(X,Z) |D\sim \mathrm{Normal}\left( \left[\begin{array}{c} \frac{\sigma_x}{\sigma_d}\rho D\\ \frac{\sigma_x}{\sigma_d}\rho D \end{array} \right] , \left[\begin{array}{cc} (1-\rho^2)\sigma_{x}^2 + 1 &(1-\rho^2)\sigma_{x}^2 \\ (1-\rho^2)\sigma_{x}^2&(1-\rho^2)\sigma_{x}^2+1	\end{array}\right] \right).
\end{equation*}
In the notation of Assumption \ref{ASSM: high-level condition integral equation}, we can take \citep[see e.g.,][]{CarrascoFlorensRenault2007}
\begin{equation*}
	u_{j}(d,z) = \frac{1}{\sqrt{j!}} He_j\bigg(\frac{z-\frac{\sigma_x}{\sigma_d}\rho d}{(1-\rho^2)\sigma_{x}^2+1}\bigg),\qquad \pi_{j}(d)=\bigg( \frac{(1-\rho^2)\sigma_{x}^2}{(1-\rho^2)\sigma_{x}^2+1} \bigg)^j
\end{equation*}
where $He_j(\cdot)$ is the $j$th order Hermite polynomial.
For reference, let $\varsigma =  \frac{(1-\rho^2)\sigma_{x}^2}{(1-\rho^2)\sigma_{x}^2+1}$ so $\pi_j(d) = \varsigma^j$.

For the outcome, note 
\begin{equation*}
	\left[\begin{array}{c} X^*\\ D \\ Z	\end{array}\right] = \left[\begin{array}{ccc} 1 &0&0\\ 0&1&0\\ 1&0&1	\end{array}\right] \left[\begin{array}{c} X^* \\ D\\ U_z \end{array} \right] \sim \mathrm{Normal}\left( \left[ \begin{array}{c} 0 \\ 0\\ 0 \end{array}\right], \left[\begin{array}{ccc} \sigma_{x}^2 &\sigma_{x}\sigma_d \rho & \sigma_{x}^2\\ \sigma_{x}\sigma_d \rho&\sigma_d^2&\sigma_{x}\sigma_d \rho\\ \sigma_{x}^2&\sigma_{x}\sigma_d \rho&\sigma_{x}^2+1	\end{array}\right] \right)
\end{equation*}
and $X^*|D,Z\sim\mathrm{Normal}(\frac{\varsigma\rho}{(1-\rho^2)\sigma_x\sigma_d} D +\varsigma Z,\varsigma)$. Thus,
\begin{equation*}
	\mathbb{E}[Y|D,Z] = \Phi\bigg( \frac{1}{\sqrt{\varsigma}}\Big[\beta-\frac{\sigma_{x}\rho/\sigma_d}{(1-\rho^2)\sigma_{x}^2+1}\Big]D -  \sqrt{\varsigma} Z\bigg)\equiv \Phi\big(\bar{\beta}D-\sqrt{\varsigma} Z\big).
\end{equation*}

\paragraph{Computing inner products}
\begin{equation*}
	\int \Phi(a +b z) He_j\left(\frac{z-\mu}{\sigma}\right)\sigma^{-1}\phi\left(\frac{z-\mu}{\sigma}\right)dz = \int \Phi(a+b\mu + b \sigma t) He_j(t)\phi(t)dt.
\end{equation*}
For $j\geq 1$, using integration by part,
\begin{align*}
	\int \Phi(\tilde{a} + \tilde{b} t) He_j(t)\phi(t)dt &=  -\Phi(\tilde{a} + \tilde{b}t)He_{j-1}(t)\phi(t)\Big\vert_{-\infty}^{\infty}  + \tilde{b} \int \phi(\tilde{a}+\tilde{b} t)  He_{j-1}(t)\phi(t)dt\\
	&=  \frac{\tilde{b}}{2\pi}\exp\Big(\frac{-\tilde{a}^2}{2(\tilde{b}^2+1)}\Big) \int \exp\Big(-\frac{(\tilde{b}^2+1) }{2}\Big(t+\frac{\tilde{a}\tilde{b}}{\tilde{b}^2+1}\Big)^2\Big)  He_{j-1}(t)dt
\end{align*}
where I used that $\frac{d}{dx} ( He_{j-1}(x) \exp(-x^2/2) ) = (-1) He_{j}(x)\exp(-x^2/2)$.

Now for some constants $\gamma >1,\delta$,
\begin{align*}
	&\int \exp\Big(-\frac{\gamma}{2}(t+\delta)^2\Big) He_{j}(t)dt \\
	&= \frac{1}{\sqrt{\gamma}}\int \exp\Big(-\frac{s^2}{2}\Big)He_j\Big(\frac{s}{\sqrt{\gamma}} -\delta\Big)ds \\
	&= \frac{1}{\sqrt{\gamma}}\int \exp\Big(-\frac{s^2}{2}\Big)\sum_{k=0}^j \binom{j}{k} (-\delta)^{j-k} He_k\Big(\frac{s}{\sqrt{\gamma}} \Big)ds \\
	&= \frac{1}{\sqrt{\gamma}}\int \exp\Big(-\frac{s^2}{2}\Big)\sum_{k=0}^j \binom{j}{k} (-\delta)^{j-k} \sum_{i=0}^{\lfloor k/2\rfloor} \gamma^{-k/2+i}(\gamma^{-1}-1)^i \binom{k}{2i}\frac{(2i)!}{i!}2^{-i}He_{k-2i}(s)ds \\
	&= \sqrt{\frac{2\pi}{\gamma}} \sum_{k=0}^{\lfloor j/2\rfloor} \binom{j}{2k} (-\delta)^{j-2k} (\gamma^{-1}-1)^{k} \frac{(2k)!}{k!} 2^{-k} \\
	&= \sqrt{\frac{2\pi}{\gamma}} \sum_{k=0}^{\lfloor j/2\rfloor} \binom{j}{2k} (-1)^{j-k} \delta^{j-2k} \left(\frac{\gamma}{\gamma-1}\right)^{-k}\frac{(2k)!}{k!} 2^{-k}\\
	&= \sqrt{\frac{2\pi}{\gamma}} (-1)^j \left(\frac{\gamma-1}{\gamma}\right)^{j/2} j! \sum_{k=0}^{\lfloor j/2\rfloor} \frac{ (-1)^{k}}{k!(j-2k)!} \delta^{j-2k} \left(\frac{\gamma}{\gamma-1}\right)^{j/2-k} 2^{-k}\\
	&= \sqrt{\frac{2\pi}{\gamma}} (-1)^j \left(\frac{\gamma-1}{\gamma}\right)^{j/2} He_{j}\left(\delta \sqrt{\frac{\gamma}{\gamma-1}}\right).
\end{align*}
Putting all together, with $\mu_d= \frac{\sigma_x}{\sigma_d}\rho d$, $s^2=(1-\rho^2)\sigma_{x}^2+1$
\begin{align*}
	\mathbb{E}\big[ \mathbb{E}[Y|D,Z]u_{j+1}(D,Z)\big|D=d\big] &= \frac{1}{\sqrt{(j+1)!}} \int \Phi(\bar{\beta}d -\sqrt{\varsigma} z) He_{j+1}\bigg(\frac{z-\mu_d}{s}\bigg)s^{-1} \phi\left(\frac{z-\mu_d}{s}\right)dz\\
	&= \frac{\tilde{b}(-1)^j}{\sqrt{2\pi \gamma (j+1)!}} \exp\bigg(\frac{-\tilde{a}^2}{2(\tilde{b}^2+1)}\bigg)   \bigg(\frac{\gamma-1}{\gamma}\bigg)^{j/2} He_{j}\left(\delta \sqrt{\frac{\gamma}{\gamma-1}}\right).
\end{align*}
where $\tilde{a} = (\bar{\beta}-\sqrt{\varsigma}\frac{\sigma_{x}}{\sigma_d}\rho)d$, $\tilde{b} = -\sqrt{\varsigma(1-\rho^2)\sigma_x^2+\varsigma}$,
\begin{equation*}
	\gamma = \varsigma[(1-\rho^2)\sigma_x^2+1]+ 1=  (1-\rho^2)\sigma_x^2+1,\quad \text{ and }\quad \delta = -\frac{d[\bar{\beta} -\sqrt{\varsigma}\sigma_x\rho/\sigma_d]\sqrt{\varsigma}}{\sqrt{(1-\rho^2)\sigma_x^2+1} } .
\end{equation*}

\paragraph{Asymptotics}
Using (8.22.8) in \citet[p.200]{szeg1939orthogonal},
\begin{align*}
	\frac{\Gamma(\frac{j}{2}+1)}{\Gamma(j+1)} e^{-x^2/4} 2^{j/2}He_j(x) &= \cos\bigg(x\sqrt{j+1/2}  - \frac{j\pi}{2}\bigg) \\
	&\quad + \frac{x^3}{12\sqrt{2}} \big(2j+1)^{-1/2} \sin\bigg(x\sqrt{j+1/2} - \frac{j\pi}{2}\bigg) + O(j^{-1})
\end{align*}
for $j\in\mathbb{N}$.
Then, for infinitely many $n\in\mathbb{N}$, for some constants $c_1,c_2>0$,
\begin{align*}
	\mathbb{E}\big[ \mathbb{E}[Y|D,Z]u_{2n+1}(D,Z)\big|D=d\big]  &\geq c_1  \frac{(2n)!}{n!2^{n}} \frac{1}{\sqrt{(2n+1)!}} \bigg(\frac{\gamma-1}{\gamma}\bigg)^{n}\\
	&=   c_1(2n+1)^{-1/2} \frac{(2n-1)!!}{\sqrt{(2n)!}} \bigg(\frac{\gamma-1}{\gamma}\bigg)^{n}\\
	&= c_1(2n+1)^{-1/2} \sqrt{ \frac{(2n-1)!!}{(2n)!!}} \bigg(\frac{\gamma-1}{\gamma}\bigg)^{n}\\
	&\geq  c_2 (2n+1)^{-1/2} n^{-1/4} \bigg(\frac{\gamma-1}{\gamma}\bigg)^{n}
\end{align*}
where $n!!$ denotes the double factorial of $n$, the second and third equality use $(2n)!=(2n)!!(2n-1)!!=n!2^n(2n-1)!!$, and the last inequality follows from results on Wallis' integrals.
Then,
\begin{equation*}
	\pi_{2n+1}(d)^{-2}\mathbb{E}\big[ \mathbb{E}[Y|D,Z]u_{2n+1}(D,Z)\big|D=d\big]^2 \geq c  n^{-3/2} \frac{1}{\varsigma^{4n}} \bigg(\frac{\gamma-1}{\gamma}\bigg)^{2n}
\end{equation*}
and if $ (\gamma-1)/\gamma  > \varsigma^{2} $, Assumption \ref{ASSM: high-level condition integral equation} fails.
Since $ (\gamma-1)/\gamma  =\varsigma$ and by $\varsigma<1$, Assumption \ref{ASSM: high-level condition integral equation} fails.

\subsubsection{Verifying conditions for the control function method}
I verify Assumptions \ref{ASSM: conditional independence outcome}-\ref{ASSM: control function dimension} for the above model. Note $Y(d)=\mathbbm{1}\{\beta d\geq X^*\}$ so Assumption \ref{ASSM: conditional independence outcome} $Y(d)\inde (D,Z)|X^*$ trivially holds. Assumption \ref{ASSM: proxy exclusion} holds as $U_x\inde (X^*,D,U_z)$. Assumption \ref{ASSM: L2 completeness} follows from the completeness property of exponential families \cite[see e.g.,][]{NeweyPowell2003ECMA}. Assumption \ref{ASSM: regularity} can be verified using the normality assumption. For Assumption \ref{ASSM: support condition}, let $\beta = \frac{\varsigma\rho}{(1-\rho^2)\sigma_x\sigma_d}$ and $\varsigma=\frac{(1-\rho^2)\sigma_{x}^2}{(1-\rho^2)\sigma_{x}^2+1}$. Then, $X|D,Z \sim \mathrm{Normal}(\beta D +\varsigma Z,\varsigma+1)$. Then, $\phi_{\tilde{d}}(d,z) = z + \frac{1}{\varsigma}(d -\beta\tilde{d})$ satisfies the condition. Finally, for Assumption \ref{ASSM: control function dimension}, consider the set of functions $\{\mathbb{E}[X^l|D,Z]:l\in\mathbb{N}\}$. These functions are continuously differentiable and $\mathbb{E}[X|D,Z] = \beta D + \varsigma Z$ satisfies the ``maximal set'' as the variance does not vary with $(D,Z)$ and higher moments can be expressed as a function of the first and second moments.  

\end{appendices}
\setstretch{1}
\bibliographystyle{ecta}
\bibliography{covariate-measurement-error}

\end{document}